\documentclass[11pt,a4paper]{article}

\usepackage[a4paper,margin=1.7cm]{geometry}
\usepackage{enumitem}
\usepackage{comment}
\usepackage{graphicx}
\usepackage{pgfplots}
\pgfplotsset{compat = 1.18}
\usepackage{makecell}
\usepackage{booktabs}
\usepackage{colortbl}
\usepackage{authblk}
\usepackage[font=small,
            width=0.9\textwidth]{caption}
\usepackage{threeparttable}
\usepackage{amsmath, amsthm,amssymb, amsfonts, bm, mathtools}
\usepackage{cancel}
\usepackage{multirow}
\usepackage{tikz}
\usepackage{lscape}
\usepackage{longtable}
\usepackage{subcaption}
\usetikzlibrary{arrows.meta, positioning, calc, shadings, trees,backgrounds,fit}
\usepackage[style=authoryear-comp,citestyle=authoryear,maxcitenames=2,natbib=true,backend=bibtex,url=false]{biblatex}

\definecolor{iblue}{rgb}{0.1,0,0.9}
\definecolor{ired}{rgb}{0.9,0,0.1}

\newcommand{\fine}{\hfill $\Box$}
\newcommand{\Small}[1]{\textstyle #1 \displaystyle}
\newcommand{\comillas}[1]{``\,#1\,''}

\usepackage{amsthm}
\usepackage[colorlinks=true,
            linkcolor=blue,   
            citecolor=blue,   
            urlcolor=blue]    
           {hyperref}

\DeclareRobustCommand{\orcidicon}[1]{%
  \texorpdfstring{%
    \href{https://orcid.org/#1}{%
      \textsuperscript{%
        \tikz[baseline=-0.6ex]{
          \fill[green!60!black] (0,0) circle (1.2ex);
          \node[white, font=\tiny\sffamily] at (0,0) {iD};
        }%
      }%
    }%
  }{}%
}

\newtheorem{motivatingexample}{Motivating example}

\renewbibmacro{in:}{}
\AtEveryBibitem{%
  \clearfield{issn}%
  \clearfield{url}%
  \clearfield{urldate}%
  \clearfield{urlyear}%
  \clearfield{note}%
  \clearfield{month}%
  \clearfield{day}%
  \clearlist{language}%
}

\newcommand{\ind}{\mathbb{I}}

\newcommand{\logit}{\operatorname{logit}}
\newcommand{\sigm}{\operatorname{logit}^{-1}}

\newcommand{\Abar}{\overline{A}}

\usepackage{multicol}
\usepackage{float}
\usepackage[section]{placeins}

\usepackage{microtype}
\usepackage[ruled,vlined]{algorithm2e}
\usepackage{setspace}
\usepackage{xcolor}

\title{Ordering Stochastic Block Models via prior transitivity}

\author[1]{Lapo Santi\orcidicon{0009-0005-9363-3353}}
\author[1,2]{Nial Friel\orcidicon{0000-0003-4778-0254}}
\author[3,4]{Pierpaolo De Blasi\orcidicon{0000-0002-6944-2283}}

\affil[1]{School of Mathematics and Statistics, University College Dublin, Dublin, Ireland}
\affil[2]{Insight Centre for Data Analytics, University College Dublin, Dublin, Ireland}
\affil[3]{ESOMAS Department, University of Torino, Torino, Italy}
\affil[4]{Collegio Carlo Alberto, Torino, Italy}

\date{\today}

\begin{document}

\maketitle

\abstract{
In directed networks, nodes may form groups with similar interaction patterns, while these groups may themselves follow an ordered structure. Existing methods typically treat these features separately, either clustering nodes without enforcing a coherent block order, or ranking individual nodes without allowing for structurally equivalent groups. We introduce the \emph{Transitive Stochastic Block Model} (TSBM), a Bayesian model for directed weighted networks that uses transitivity-inducing priors to infer ordered blocks. The model separates the total volume of interaction between two nodes from the direction of interaction conditional on interaction occurring, so that hierarchy is imposed on directional imbalance rather than interaction frequency. We consider two order-restricted specifications: a flexible weak-stochastic-transitivity version, which excludes cyclic dominance patterns while allowing heterogeneous block-pair strengths, and a Toeplitz strong-stochastic-transitivity version, in which directional advantage increases with rank separation. Posterior inference is performed through a Gibbs sampler using P'olya-Gamma data augmentation. Since ordered block labels are not exchangeable, we introduce an \emph{age-ordered} partition prior to infer the number of blocks jointly with node allocation. Simulation studies show that order-constrained priors improve prediction and partition recovery, especially in sparse networks. Across six empirical directed networks, the TSBM improves predictive performance in four cases and yields partitions with clearer ordered structure. The results also identify cases, such as nearly deterministic dominance networks or non-transitive citation networks, where imposing ordered blocks can harm prediction. The TSBM therefore provides a probabilistic framework for estimating ordered groups and assessing when a transitive block structure is supported by the data.
}

\section{Introduction}

Many directed networks, including dominance encounters among animals, cross-citations between journals, and friendship choices in schools, exhibit two forms of structure at once. First, nodes may form groups whose members behave similarly, giving rise to a community or clustering structure. Second, these groups may interact in a way that is consistent with an ordering, with most edges pointing from higher-ranked nodes to lower-ranked ones. This phenomenon is well documented in sociology \citep{doreianSymmetricAcyclicDecompositions2000}, anthropology \citep{hsiehFindingHierarchicalSubgroups2010}, ecology \citep{williamsonMouseSocialNetwork2016}, and related disciplines.

In animal societies, repeated agonistic encounters often give rise to dominance hierarchies, also known as pecking orders\footnote{The expression is literal in origin: \citet{schjelderupEbbeBeitrageSozialpsychologie1922}'s work on domestic hens described an order of precedence sustained by asymmetric pecking relations, in which some birds could peck others without receiving the same treatment in return}. Higher-ranked individuals are more likely to defeat, displace, or receive deference from lower-ranked ones. In this setting, groupings and rankings are not separate phenomena, but two facets of the same social structure; existing literature, however, often treats grouping and ordering as distinct tasks. 


Clustering models for network data, most notably the Stochastic Block Model
\citep[SBM;][]{nowickiEstimationPredictionStochastic2001}, recover groups of
nodes with similar interaction patterns, but do not ensure that these groups can
be arranged in a coherent hierarchy. Conversely, ranking models
\citep{bradleyRankAnalysisIncomplete1952, thurstone1927} and ranking algorithms
\citep{elo, pagerank, springrank} order entities hierarchically, but do not
identify groups of entities that play similar structural roles. They therefore
risk enforcing a strict node-level ordering, even when not all nodes can be distinguished and the ordering sits at the group-level instead.

This line of work is motivated by situations in which a directed network may
contain both grouping and hierarchical structure. Let \(G=(V,E)\) be a weighted
directed network, where \(V=\{1,\ldots,n\}\) is a set of nodes and
\(E\subseteq V\times V\) is a set of directed edges. The network is represented
by a weighted adjacency matrix
\[
A=(A_{ij})_{i,j=1}^n,
\]
where \(A_{ij}\geq 0\) denotes the weight of the directed edge from node \(i\)
to node \(j\), with \(A_{ij}=0\) when no such directed relation is observed. In
the examples below, larger values of \(A_{ij}\) indicate stronger evidence that
node \(i\) dominates, is preferred to, or is more highly regarded than node
\(j\). The point is not only that a hierarchy may be present, but that the
hierarchy may operate between groups of nodes rather than between all nodes
separately. In this sense, it represents a relaxation of a complete order among
the nodes. We provide below three non-exhaustive examples that may warrant this
modelling approach.

\begin{motivatingexample}[Animal dominance encounters]\label{mot:animal}
Consider a weighted directed network $G$ whose nodes represent animals living in
the same social group. An edge from animal \(i\) to animal \(j\) records
dominance events in which \(i\) prevails over \(j\), for example by winning a
contest, displacing \(j\) from food, or forcing \(j\) to retreat. Repeated
events of this kind often give rise to a pecking order, which is precisely a
set of ordered groups of animals. Such groups consist of individuals with
similar behaviour and are characterised by the fact that members of
higher-ranked groups tend to send proportionally more edges to lower-ranked
groups than they receive from them.
\end{motivatingexample}

\begin{motivatingexample}[Citations between journals]\label{mot:citations}
Consider a weighted directed network $G^\prime$ whose nodes represent academic
journals observed over a fixed time window. Keeping the convention that the
prevalent edge direction should point along the hierarchy, in the present
context an edge is defined to point from journal \(i\) to journal \(j\) with a
given weight representing the number of citations received by \(i\) from
articles published in \(j\). We expect groups of journals to cross-cite each
other, either because of topic proximity or because of synergies in the
research area. However, we also expect such journals to have different levels
of prestige, and more prestigious journals to receive proportionally more
citations from less-prestigious ones than they produce for them.
\end{motivatingexample}

\begin{motivatingexample}[Friendship nominations in schools]\label{mot:friends}
Consider a weighted directed network $G^{\prime \prime}$ whose nodes represent
students in a school. Again taking edges to point in the hierarchical
direction, an edge from student \(i\) to student \(j\) records that \(i\) is
nominated by \(j\) as a friend, or, in a weighted version, how often this
nomination is observed across surveys. Also in this context, we expect friends
that belong to the same social circle to nominate each other as friends, thus
forming groups of nodes with reciprocal interactions. However, we also expect
more popular groups of students to have members who attract many friendship
nominations from other, less visible students, who do not necessarily receive
the same attention in return.
\end{motivatingexample}

The three examples above are representative of situations in which the joint
modelling of both grouping structure and an ordering among such groups may be
warranted. The central scientific question of this paper is to assess when this
joint modelling is supported by the data. More precisely, we ask whether
imposing an order-inducing prior on the interaction probabilities among the
blocks may provide both inferential benefits, such as the discovery of more
order-respecting partitions, and predictive gains over existing alternative
methods.

To address this question, we introduce the \emph{Transitive Stochastic Block
Model} (TSBM), a Bayesian model for directed weighted networks in which groups
are not only \emph{stochastically equivalent}, as in standard SBMs, but also
\emph{stochastically ordered} through \emph{transitivity relations}. The model is built
around the following methodological contributions.

First, we decompose the \emph{directed edge counts} into an \emph{undirected}
component, which captures the total \emph{volume} of interaction between two
nodes, and a \emph{directional} component, which captures the direction of
interaction conditional on such interaction having occurred. Intuitively, this distinction matters in settings such as Example~\ref{mot:animal}: two animals may encounter each other frequently even when the outcome of those encounters is only weakly directional.

Second, we impose a coherent \emph{ordering} on the blocks through a prior that induces stochastic transitivity on the block-level directed interaction probabilities. We consider two versions of transitivity. The first, weak stochastic transitivity, (\emph{WST}), rules out cyclic dominance patterns, such as block \(1\) prevailing over block \(2\), block \(2\) over block \(3\), and block \(3\) over block \(1\). The absence of cycles is the minimal requirement for an ordering to exist. The second, strong stochastic transitivity, \emph{SST},
is more restrictive: other than ruling out cycles like WST, it also assumes that the ordering is stricter when two groups are farther away in the social ranking; this is a typical feature of more unequal societies, where high-ranked groups have increasingly large advantages over lower-ranked ones.

Third, we impose these ordering restrictions without sacrificing \emph{computational tractability}. By working on the \emph{log-odds} scale and using \emph{Pólya--Gamma data augmentation} \citep{polson2013}, we derive a full Gibbs sampler for posterior inference on the directional parameters, the volume parameters, the latent allocation of nodes to ordered blocks, and also the number of occupied blocks. 

Fourth, because \emph{ordered block labels} are not exchangeable, as they indicate not only which nodes are partitioned together, but also the relative position they occupy in the hierarchy, we introduce an \emph{age-ordered partition prior} \citep{Don:Tav:86,Don:Joy:91} to learn the number of blocks in the network in a Bayesian nonparametric fashion, thus avoiding more computationally demanding approaches like reversible-jump methods \citep{green1995}. Although these ordered partition laws have been known for more than three decades, to the best of our knowledge they have not previously been used as nonparametric priors for latent block models with non-exchangeable labels. 

Fifth, we show that in simulated networks, ordered prior information improves both prediction and partition recovery relative to the standard degree-corrected SBM (DC--SBM), especially when a directional signal is present but statistically weak and the network is sparse. Across the empirical applications, the two TSBM specifications improve predictive performance over the DC--SBM in four of the six datasets considered. The two remaining cases are equally informative. In the macaques network, the hierarchy is already very sharp, almost compatible with a strict node-level ordering. Here the DC--SBM predicts slightly better, as the contribution of the prior to the prediction is less important. The citation network provides a different scenario: although a prestige scale is visible among journals, and journals can be grouped into different quality tiers, the interaction patterns are not transitive; therefore the DC--SBM retains a predictive advantage, as it can accommodate such non-transitive behaviour.

Finally, especially where the data signal is weak, the ordered prior helps a coherent hierarchical structure emerge from noisy interactions. In these cases, the TSBM not only improves prediction, but also yields partitions that are substantially more order-respecting than those obtained under the unconstrained DC--SBM.

\subsection{Related work}
\label{sec:related-work}

Our model lies at the intersection of two strands of literature that have, until recently, developed largely independently: clustering methods for network data, and preference learning models for pairwise interactions and ranking data. We review the contributions most relevant to the present work, and clarify where the proposed model departs from them.

\paragraph{Stochastic block models, with and without block order.}

Our clustering backbone is the directed degree-corrected stochastic block model (DC--SBM) of \citet{karrer2011stochastic}. Recall that for $G=(V,E)$ a weighted directed network on $n$ nodes, the adjacency matrix $A=(A_{ij})\in\mathbb{N}^{n\times n}$ yields via $A_{ij}$ the number of directed interactions from node $i$ to node $j$. Now assign nodes to $K$ latent blocks through labels $\mathbf z=(z_1,\ldots,z_n)$, with $z_i\in[K]=\{1,\ldots,K\}$. Conditional on these labels, the directed degree-corrected SBM specifies
\begin{equation}
A_{ij}\,\big|\, z_i,z_j,\bm\eta,\lambda
\;\sim\;
\mathrm{Poisson}\!\left(\eta_i\,\eta_j\,\lambda_{k\ell}\right),
\qquad z_i=k,\;z_j=\ell,
\label{eq:poisson-sbm}
\end{equation}
where $\bm\eta=(\eta_1,\ldots,\eta_n)$ is a vector of node-specific degree propensities with $\eta_i>0$, and $\lambda=(\lambda_{k\ell})$ is an asymmetric matrix of block-pair intensities. Nodes in the same block share the same average directed interaction pattern, up to node-specific degree effects, a property called \emph{stochastic equivalence}.

The SBM has been extended in several directions, including overlapping and mixed-membership variants \citep{latoucheOverlappingStochasticBlock2011,airoldiIntroductionMixedMembership2014}, covariate models \citep{Legramanti_2022}, and zero-inflated models \citep{Lu_2025}. For our purposes, a particularly relevant contribution is \citet{pengBayesianDegreeCorrectedStochastic2014}, who formulate the SBM as a logistic regression with degree correction and use Pólya--Gamma data augmentation for posterior sampling. We use the same augmentation later in the paper (see Section~\ref{sect:inference}).

A recent line of work has recognised that, in many directed networks, the blocks themselves may be ordered. \citet{letiziaResolutionRankingHierarchies2018} introduced a ranked SBM that recovers ordering through a penalty on backward edges, while \citet{iacovissi2022} combine mixed-membership community structure with SpringRank-type scores, using a latent node-level mechanism that distinguishes interactions driven by community affinity from those driven by hierarchy. \citet{peixotoOrderedCommunityDetection2022} develops nonparametric Bayesian community detection for directed networks with ordered blocks.
The modelling perspective, however, is different from ours, as it hinges on the microcanonical formulation of SBM and Minimum Description Length (MDL) methods.
In particular, the hierarchy is useful when it yields a shorter description of the inter-block connectivity matrix, and so the number of blocks is estimated via the MDL criterion. 
The TSBM instead imposes the ordering using the stochastic transitivity assumptions inherited from preference theory; these are encoded via a prior on the block-level interaction probabilities. 

\paragraph{Connections to preference theory: from pairs to rankings.}

The connection with preference theory and pairwise-comparison data emerges by viewing the nodes of the directed network as the items, players, or actors being compared. Under this interpretation, an unordered dyad \(\{i,j\}\) represents the opportunity for comparison between two items, while the direction of an observed edge records the comparison outcome: an edge from \(i\) to \(j\) provides evidence that \(i\) prevails over, dominates, cites, defeats, or otherwise points towards \(j\), depending on the application. Classical paired-comparison models focus on this directional component, namely the probability that one item is preferred to another conditional on a comparison being observed.

Models for pairwise-comparison data have a long history, with classical examples including the Bradley--Terry model \citep{bradleyRankAnalysisIncomplete1952} and the Thurstone model \citep{thurstone1927}. Consider a collection of \(n\) items, which in the network setting correspond to the nodes \(V=\{1,\ldots,n\}\). In these models, the probability that item \(i\) is preferred to item \(j\), for \(i\neq j\), depends on a one-dimensional latent strength vector \(\bm u=(u_1,\ldots,u_n)\):
\begin{equation}\label{eq:LST}
\Pr(i\succ j)=F(u_i-u_j),
\end{equation}
where \(F\) is the logistic CDF in the Bradley--Terry model and the Gaussian CDF in the Thurstone model. A single latent scale induces a complete probabilistic ordering of the items, but requires all pairwise probabilities to be explained through differences in \(\bm u\).

Recent work has extended Bradley--Terry and Plackett--Luce models to include a block structure \citep{pearceErosheva2025,santifriel2025,abodi}, thereby producing groups of items with similar preference patterns and relative strengths. See also an antecedent of these papers \citep{basiniAssessingCompetitiveBalance2023} in the context of football modelling. 
These models are close in spirit to our goal, since they produce ordered blocks of similar strength. However, they typically model who wins, or who is preferred, conditional on the observed comparison. A generative model for network data should also model the comparison (the edge) presence, other than the comparison ultimate winner (the direction of the edge). 

A second limitation is that models of the form \eqref{eq:LST} impose linear stochastic transitivity (LST) \citep{oliveiraStochasticTransitivityAxioms2018}. LST is convenient and tractable, but it is also the most restrictive of the usual stochastic-transitivity assumptions. As argued by \citet{shah2016}, it may fit comparison data poorly when preferences are ordered but not well described by a single latent utility scale. SST and WST provide more flexible alternatives that progressively relax LST. 

The remainder of the paper is organised as follows. Section~\ref{sec:hierarchical-decomposition} introduces the TSBM, including the volume/direction decomposition, the prior on ordered partitions, and the order-restricted priors on directional parameters. Section~\ref{sect:inference} derives the Pólya--Gamma augmentation and the resulting Gibbs sampler. Section~\ref{sec:sim-strong} reports simulation studies, Section~\ref{sec:applications} presents six empirical applications, and the final section concludes.
\section{The Transitive Stochastic Block Model}
\label{sec:hierarchical-decomposition}

In this section, we introduce a novel Bayesian model for generating and analysing networks that exhibit both grouping and ordering structures. The TSBM produces groups that can be arranged 
such that nodes in higher-ranked blocks tend, on average, to send edges to nodes in lower-ranked blocks more often than they receive edges from them. The order of the blocks is therefore inferred from systematic asymmetries in the direction of edges. We define this model through the following three key steps:

\begin{enumerate}[leftmargin=2em]
\item \textbf{The directed counts are decomposed into 
a volume component and a direction component.}
In Section~\ref{sec:likelihood-decomp}, we reinterpret the standard DC--SBM model in~\eqref{eq:poisson-sbm} as a two-step process. First we generate the total undirected \emph{volume} of edges between nodes, and then 
we model the probability that an edge has a certain \emph{direction}.
This two-step formulation is equivalent to 
the DC--SBM likelihood, but it is the volume/direction decomposition that makes it possible to impose an order on the blocks.

\item \textbf{Order-respecting priors on the directional probabilities.} In Section~\ref{sec:psi_prior}, we show that having ordered blocks is equivalent to having directional probabilities that satisfy transitivity. We then show how to account for transitivity
without sacrificing computational tractability: we first re-express interaction probabilities as log-odds, and then we place Gaussian priors on those log-odds to obtain weak and strong stochastic transitivity (WST and SST).

\item \textbf{An ordered partition of the nodes.} 
Once WST and SST restrictions are in place, the likelihood is no longer invariant to permutations of the block labels, and since standard Bayesian priors on partitions are invariant to group labels permutations \citep{deblasi2015}, we need to adopt a different class of priors over partitions, namely \emph{age-ordered} partitions \citep{Don:Tav:86,Gne:Pit:05aop}, as detailed in Section~\ref{sec:ocrp-prior}.
\end{enumerate}

The full generative process is summarised in Algorithm~\ref{alg:TSBM-generative}.

\subsection{Decomposing the DC--SBM likelihood}
\label{sec:likelihood-decomp}

We begin by noting that the ordering of the blocks should not be imposed on the total volume of edges between two blocks. In the animal dominance setting of Example~\ref{mot:animal}, an animal at the top of the hierarchy need not fight every low-ranking animal often; some individuals may rarely meet, or may avoid one another altogether. What makes the relationship hierarchical is that, conditional on an encounter being observed, the higher-ranked animal is more likely to win, displace, or receive deference. Therefore, what matters for the order is the \emph{net directional imbalance}: conditional on interaction occurring between blocks $k$ and $\ell$, do edges tend to point from the higher-ranked block to the lower-ranked one?

To disentangle the direction of the interaction from its occurrence, we express the directed Poisson DC--SBM in~\eqref{eq:poisson-sbm} using the standard Poisson--multinomial conditioning identity: independent Poisson counts, conditional on their sum, are multinomial, and hence binomial in the two-direction case. Hence, we parametrise the directed DC--SBM intensity matrix $\lambda=(\lambda_{k\ell})$ in terms of
\[
\kappa_{k\ell}\coloneqq \lambda_{k\ell}+\lambda_{\ell k},
\qquad
\rho_{k\ell}\coloneqq 
\frac{\lambda_{k\ell}}{\lambda_{k\ell}+\lambda_{\ell k}},
\]
and factorize the DC--SBM distribution of the pair $(A_{ij},A_{ji})$ 
as
\begin{align}
p(A_{ij},A_{ji}\mid z_i=k,z_j=\ell,\bm\eta,\kappa,\rho)
&=
p(A_{ij}\mid N_{ij},z_i=k,z_j=\ell,\rho)
\;
p(N_{ij}\mid z_i=k,z_j=\ell,\bm\eta,\kappa)\nonumber\\
&=
\underbrace{\mathrm{Binomial}\bigl(A_{ij};N_{ij},\rho_{k\ell}\bigr)}_{\text{Direction}}
\cdot
\underbrace{\mathrm{Poisson}\bigl(N_{ij};\eta_i\eta_j\kappa_{k\ell}\bigr)}_{\text{Volume}},
\label{eq:dyadic-volume-direction}
\end{align}
where $N_{ij} = A_{ij} + A_{ji}$.
Thus, \eqref{eq:poisson-sbm} can be reinterpreted as a two-step process, in which first weighted undirected edges are sampled according to a symmetric intensity matrix $\kappa=(\kappa_{k\ell})$ and degree correction parameters $\bm\eta=(\eta_1,\ldots,\eta_n)$; subsequently, these edges are given a direction from $k$ to $\ell$ with probability $\rho_{k\ell}\in(0,1)$, that is according to a probability matrix $\rho=(\rho_{k\ell})$ that satisfies $\rho_{k\ell}+\rho_{\ell k}=1$ for every $k,\ell$ (in particular $\rho_{kk}=1/2$), alternatively $\rho+\rho^\top={\rm 1}_K{\rm 1}_K^\top$. The probabilities $\rho_{k\ell}$ are the parameters on which we will place order-respecting priors in the following section.

\subsection{Transitive-inducing priors on the directional log-odds}
\label{sec:psi_prior}

We now formalise what it means for the directional probabilities $\rho_{k\ell}$ introduced in Section~\ref{sec:likelihood-decomp} to induce an order among the blocks. 
Recall that $\rho_{k\ell}$ is the probability that, given an interaction between blocks $k$ and $\ell$, it flows from $k$ to $\ell$, with $\rho_{k\ell}+\rho_{\ell k}=1$. When $\rho_{k\ell}>1/2$, we say informally that block $k$ probabilistically ``beats" block $\ell$, and define the \emph{stochastic preference relation} \citep{ryanRepresentationBinaryChoice2016,fishburn1973,oliveiraStochasticTransitivityAxioms2018} as
\[
k \succ_{\rho} \ell
\quad\Longleftrightarrow\quad
\rho_{k\ell}>\tfrac12 .
\]

If ties are excluded, namely if \(\rho_{k\ell}\neq 1/2\) for all \(k\neq \ell\), then for every unordered pair of blocks exactly one of \(k\succ_{\rho}\ell\) and \(\ell\succ_{\rho}k\) holds. The relation \(\succ_{\rho}\) is then complete and asymmetric. In graph-theoretic language, it defines a tournament on the block set \([K]\): every pair of blocks is joined by one directed comparison.

A tournament, however, is not necessarily ordered. It may contain cycles. For instance, one may have
\[
1\succ_{\rho}2,\qquad
2\succ_{\rho}3,\qquad
3\succ_{\rho}1.
\Longleftrightarrow  
\rho_{12}>\tfrac12,\qquad
\rho_{23}>\tfrac12,\qquad
\rho_{31}>\tfrac12.
\]
Such a pattern cannot be represented by any complete ordering of the blocks, in the sense that no permutation exists of the three blocks such that all three pairwise directions point from higher to lower rank. A complete order requires that the stochastic preference relation induced by \(\rho\) must be transitive, that is, absent of cycles. More formally, stochastic transitivity requires the existence of a permutation $\pi$ of $[K]$ such that $\rho_{\pi(r)\pi(s)}>1/2$ for every $r<s$.
This is precisely \emph{Weak Stochastic Transitivity (WST)} \citep{oliveiraStochasticTransitivityAxioms2018}:
\[
\rho_{k\ell}\ge\tfrac12
\quad\text{and}\quad
\rho_{\ell m}\ge\tfrac12
\quad\Longrightarrow\quad
\rho_{k m}\ge\tfrac12.
\]
Without loss of generality, we fix a \emph{canonical ordering} of the block labels so that $1\succ_\rho 2\succ_\rho\cdots\succ_\rho K$, which implies WST is satisfied if the upper triangular entries of $\rho$ are larger than \(1/2\), as illustrated in Figure~\ref{fig:wst-sst-side-by-side}(a).
The previous condition is the minimal requirement for the block probabilities to induce an order. On the other hand, \emph{Strong Stochastic Transitivity} (SST) imposes an additional constraint \citep{fishburn1973,oliveiraStochasticTransitivityAxioms2018}: 
\[
\rho_{km}
\ge
\max\{\rho_{k\ell},\rho_{\ell m}\},
\qquad k<\ell<m, \; \rho_{k\ell},\rho_{\ell m} \geq \tfrac12
\]
Thus SST requires not only that all dyads point in the hierarchy direction, but also that higher-ranked blocks gain a larger directional advantage the further down they interact. In matrix terms, and under the canonical ordering above, SST is satisfied if the upper-triangular entries of $\rho$ not only are positive, but also increase along rows and decrease down columns.

\paragraph{Formalising WST and SST conditions with conjugate priors on the log-odds scale.}

Stochastic transitivity such as WST and SST have been defined on the probability scale; however, for prior elicitation purposes, it is convenient to work on the log-odds scale. In fact this makes posterior tractable, as it yields conditionally Gaussian updates under the Pólya--Gamma augmentation (more details later in Section~\ref{sect:inference}). For an upper-triangular block pair \(k<\ell\), define
\[
\psi_{k\ell}
\coloneqq
\logit(\rho_{k\ell})
=
\log\frac{\lambda_{k\ell}}{\lambda_{\ell k}} .
\]
Thus \(\psi_{k\ell}\) is the log-odds that, conditional on an interaction
between blocks \(k\) and \(\ell\), the interaction is directed from the higher
ranked block \(k\) to the lower ranked block \(\ell\). The matrix $\psi=(\psi_{k\ell})$ is \emph{anti-symmetric}: from $\logit(\rho_{\ell k})=-\psi_{k\ell}$ and $\rho_{\ell k}=1-\rho_{k\ell}$ it follows that $\psi=-\psi^\top$. Equivalently, for any \(k\neq \ell\), write
\[
\rho_{k\ell}
=
\sigm\bigl(s_{k\ell}\psi_{k\wedge\ell,k\vee\ell}\bigr),\quad s_{k\ell}\coloneqq \operatorname{sgn}(\ell-k)
\label{eq:rho-logit}
\]
where \(k\wedge\ell\) and \(k\vee\ell\) denote the smaller and larger of the
two indices. For \(k<\ell\), the following are equivalent:
\[
\psi_{k\ell}>0
\quad\Longleftrightarrow\quad
\rho_{k\ell}>\tfrac12
\quad\Longleftrightarrow\quad
k\succ_\rho \ell .
\]

We are now ready to define the order-inducing priors on the log-odds directional parameters $\psi_{k\ell}$.

\paragraph{WST prior.}

To have WST, the only requirement the prior needs to satisfy is the sign: \(\psi_{k\ell}\ge 0\) for all \(k<\ell\). We therefore place
independent Gaussian priors, truncated to the non-negative half-line, on the
upper-triangular entries  of the $K\times K$ matrix $\psi=(\psi_{k\ell})$, bearing in mind that $\psi_{kk}=0$ and the anti-symmetry relation $\psi_{k\ell}+\psi_{\ell k}=0$:
\begin{equation}
\psi_{k\ell}\overset{\text{i.i.d.}}{\sim}\mathcal N^{+}(\mu_0,\sigma_0^2),\quad k<\ell
\label{eq:wst-prior}
\end{equation}
We refer to the support of the WST prior as
\[
\mathcal C_{\mathrm{WST}}
=
\left\{
\psi\in\mathbb R^{K\times K}: \psi+\psi^\top=0,
\psi_{k\ell}\ge0
\ \text{for all } k<\ell
\right\}.
\]
An illustration of the induced structure on the probability matrix $(\rho_{k\ell})$ is in the left panel of Figure~\ref{fig:wst-sst-side-by-side}.

\begin{figure}[t]
\centering
\hspace*{2cm}
\begin{minipage}{0.42\textwidth}
\centering
\begin{tikzpicture}[scale=0.75, every node/.style={font=\small}]
  \foreach \i in {1,...,4} {
    \foreach \j in {1,...,4} {
      \pgfmathtruncatemacro{\y}{5-\i}
      \pgfmathtruncatemacro{\x}{\j}
      \ifnum\i=\j
        \def\cellcolor{gray!20}
      \else
        \ifnum\i<\j
          \def\cellcolor{blue!35}
        \else
          \def\cellcolor{blue!5}
        \fi
      \fi
      \draw[fill=\cellcolor] (\x-0.5,\y-0.5) rectangle (\x+0.5,\y+0.5);
      \ifnum\i=\j
        \node at (\x,\y) {$\tfrac12$};
      \else
        \ifnum\i<\j
          \node at (\x,\y) {\scriptsize $\rho_{k\ell}$};
        \fi
      \fi
    }
  }

  \node[anchor=east] at (0.5,4.5) {$k$};
  \foreach \i in {1,...,4} {
    \pgfmathtruncatemacro{\y}{5-\i}
    \node[anchor=east] at (0.4,\y) {$\i$};
  }
  \node[anchor=north] at (2.5,0.2) {$\ell$};
  \foreach \j in {1,...,4} {
    \node[anchor=north] at (\j,0.4) {$\j$};
  }

  \begin{scope}[shift={(4.85,2)}]
    \draw[fill=blue!35] (0,0.8) rectangle (0.5,1.4);
    \node[anchor=west, text width=3.0cm, align=left] at (0.65,1.1)
      {\scriptsize $\rho_{k\ell}{>}1/2$ ($k{<}\ell$)};
    \draw[fill=gray!20] (0,0.1) rectangle (0.5,0.7);
    \node[anchor=west, text width=3.0cm, align=left] at (0.65,0.4)
      {\scriptsize $\rho_{kk}=\tfrac12$};
    \draw[fill=blue!5] (0,-0.6) rectangle (0.5,0.0);
    \node[anchor=west, text width=3.0cm, align=left] at (0.65,-0.3)
      {\scriptsize $\rho_{\ell k}=1{-}\rho_{k\ell}$ ($k{>}\ell$)};
  \end{scope}
\end{tikzpicture}
\caption*{\parbox{\textwidth}{\centering\footnotesize \textbf{(a)} WST ($K=4$)}}
\end{minipage}
\hspace*{0.1cm}
\begin{minipage}{0.42\textwidth}
\centering
\begin{tikzpicture}[scale=0.75, every node/.style={font=\small}]
  \foreach \i in {1,...,4} {
    \foreach \j in {1,...,4} {
      \pgfmathtruncatemacro{\y}{5-\i}
      \pgfmathtruncatemacro{\x}{\j}
      \ifnum\i=\j
        \def\cellcolor{gray!20}
      \else
        \pgfmathtruncatemacro{\d}{\j-\i}
        \ifnum\d>0
          \ifnum\d=1
            \def\cellcolor{orange!30}
          \else
            \ifnum\d=2
              \def\cellcolor{orange!50}
            \else
              \def\cellcolor{orange!70}
            \fi
          \fi
        \else
          \pgfmathtruncatemacro{\revd}{\i-\j}
          \ifnum\revd=1
            \def\cellcolor{orange!14}
          \else
            \ifnum\revd=2
              \def\cellcolor{orange!9}
            \else
              \def\cellcolor{orange!5}
            \fi
          \fi
        \fi
      \fi
      \draw[fill=\cellcolor] (\x-0.5,\y-0.5) rectangle (\x+0.5,\y+0.5);

      \ifnum\i=\j
        \node at (\x,\y) {$\tfrac12$};
      \else
        \ifnum\d>0
          \node at (\x,\y) {\scriptsize $\rho_{\d}$};
        \fi
      \fi
    }
  }

  \node[anchor=east] at (0.5,4.5) {$k$};
  \foreach \i in {1,...,4} {
    \pgfmathtruncatemacro{\y}{5-\i}
    \node[anchor=east] at (0.4,\y) {$\i$};
  }
  \node[anchor=north] at (2.5,0.2) {$\ell$};
  \foreach \j in {1,...,4} {
    \node[anchor=north] at (\j,0.4) {$\j$};
  }

  \begin{scope}[shift={(4.85,2.0)}]
    \draw[fill=orange!30] (0,1.0) rectangle (0.5,1.6);
    \node[anchor=west, text width=3.0cm, align=left] at (0.65,1.3)
      {\scriptsize distance $d=1$};
    \draw[fill=orange!50] (0,0.2) rectangle (0.5,0.8);
    \node[anchor=west, text width=3.0cm, align=left] at (0.65,0.5)
      {\scriptsize distance $d=2$};
    \draw[fill=orange!70] (0,-0.6) rectangle (0.5,0.0);
    \node[anchor=west, text width=3.0cm, align=left] at (0.65,-0.3)
      {\scriptsize distance $d=3$};
  \end{scope}
\end{tikzpicture}
\caption*{\parbox{\textwidth}{\centering\footnotesize \textbf{(b)} Toeplitz SST ($K=4$)}}
\end{minipage}

\caption{Side-by-side comparison of (a) the WST and (b) the Toeplitz SST. 
In panel (b), lower-triangular cells use lighter shades of the same distance class to encode the reverse direction, with $d=3$ the lightest.
}
\label{fig:wst-sst-side-by-side}
\end{figure}

\paragraph{SST prior.}

Recall that SST imposes $K(K-1)/2$ inequality constraints on the probability matrix $\rho=(\rho_{k\ell})$. Specifically, under the convention that $1\succ_\rho 2 \ldots \succ_\rho K$, together with anti-symmetry $\rho+\rho^\top={\rm 1}_K{\rm 1}_K^\top$ we have entries increasing in the rows and decreasing in the columns \citep{shah2016}. When $K=4$, for example, the constraints are
  $$\rho_{12},\rho_{23},\rho_{34}\geq \Small{\frac12},\quad \rho_{13}\geq \rho_{12}\vee\rho_{23}, \quad 
  \rho_{24}\geq \rho_{23}\vee\rho_{34}, \quad \rho_{14}\geq \rho_{13}\vee\rho_{24}$$
To have a prior that fulfils all of them, while retaining tractability for posterior inference, is not straightforward. We therefore work with a more restrictive but tractable specification, which consists of assuming that pairs of blocks that share the same distance in the ranking, like $\rho_{12},\rho_{23},\rho_{34}$ in the example above, feature the same directional probability. Let \(d=|\ell-k|\) denote the distance between two blocks in the order. Instead of assigning a separate log-odds to every upper-triangular dyad, as in WST, we
set
\begin{equation}
\rho_{k\ell}
=
\sigm\bigl(s_{k\ell}\psi_d\bigr),
\qquad
s_{k\ell}=\operatorname{sgn}(\ell-k),
\qquad
d=|\ell-k|.
\label{eq:sst-rho-distance}
\end{equation}
The free parameters are the log-odds $\psi_1,\ldots,\psi_{K-1}$, where $\psi_d$ is the forward log-odds for block pairs at distance $d$. To encode SST, we impose
\begin{equation}
0\le\psi_1\le\psi_2\le\cdots\le\psi_{K-1}.
\label{eq:sst-monotone-psi}
\end{equation}
Since $\sigm$ is monotone and $m-k\ge\max\{\ell-k,m-\ell\}$ for $k<\ell<m$, the constraint \eqref{eq:sst-monotone-psi} implies $\rho_{km}=\sigm(\psi_{m-k})\ge\max\{\rho_{k\ell},\rho_{\ell m}\}$, i.e., SST holds. 
The resulting matrix has a \emph{skew-symmetric Toeplitz} structure: all block pairs at the same rank distance share the same forward log-odds, and hence the same forward probability; see Figure~\ref{fig:wst-sst-side-by-side}(b). This is a strict sub-model of the full SST class for \(K\ge 3\). To distinguish it from unrestricted full SST, we refer to it as \emph{Toeplitz SST} until the end of the current Section. Thereafter, whenever no ambiguity can arise, we simply call it SST.
To define it algebraically, let \(D_j\in\mathbb R^{K\times K}\) be the shift matrix with entries equal to \(1\) on the \(j\)-th superdiagonal and \(0\) elsewhere, for \(j=1,\ldots,K-1\). Define
\[
\operatorname{toeplitz}(c_1,\ldots,c_{K-1})
:=
\sum_{j=1}^{K-1} c_j\left(D_j-D_j^\top\right).
\]
Then the support of the Toeplitz SST prior is
\[
\mathcal C_{\mathrm{Toep}}
=
\left\{
\psi \in\mathbb R^{K\times K}:
\psi=\operatorname{toeplitz}(\psi_1,\ldots,\psi_{K-1}),\;
0\le\psi_1\le\psi_2\le\cdots\le\psi_{K-1}
\right\}.
\]
So we are left with $K-1$ inequality constraints. We enforce the monotonicity in \eqref{eq:sst-monotone-psi} through positive
increments. We define for $k\leq K-1$,
\begin{equation}\label{eq:sst-prior1}
 \psi_k=\delta_1+\cdots+\delta_k,
\end{equation} 
or, in vector form, $(\psi_1,\ldots,\psi_{K-1})=S\delta$, for $\delta=(\delta_1,\ldots,\delta_{K-1})$ and \(S\) the $(K{-}1)\times(K{-}1)$ lower-triangular cumulative-sum matrix. So the support of the prior is
\[
\mathcal C_{\mathrm{Toep}}
=
\left\{
\psi \in\mathbb R^{K\times K}:
\psi=\textrm{toeplitz}(S\delta),\;
\delta_k\ge 0,\; \text{for all }k
\right\}.
\]
All increments share the same i.i.d.\ truncated-Gaussian prior:
\begin{equation}\label{eq:sst-prior2}
\delta_k\overset{\mathrm{iid}}{\sim}\mathcal N^{+}(0,\tau_0^2),
\quad k=1,\ldots,K-1.
\end{equation}
Each increment $\delta_k=\psi_k-\psi_{k-1}$ measures the extra forward bias
gained when block distance grows from $k-1$ to $k$. If the posterior
concentrates near $\delta_k=0$, distance adds little beyond a general baseline tilt in the directional probabilities; large increments signal sharply increasing separation with distance.

The Toeplitz SST prior induced on \((\psi_1,\ldots,\psi_{K-1})\) is a constrained multivariate distribution on the ordered cone \(0\leq \psi_1\leq\cdots\leq\psi_{K-1}\). Equivalently, it is obtained by writing \(\psi_d=\sum_{r=1}^d\delta_r\), where the increments \(\delta_1,\ldots,\delta_{K-1}\) are independently distributed on the positive half-line. The cumulative-sum matrix $S$ induces dependence among the cumulative log-odds \((\psi_1,\ldots,\psi_{K-1})\); however, conditionally on the other increments, the full conditional distribution of each \(\delta_r\) has a Gaussian kernel truncated to \((0,\infty)\). This conditional tractability is the feature exploited in Section~\ref{sect:inference}.


The relation between the three relevant supports is
\[
\mathcal C_{\mathrm{Toep}}
\subset
\mathcal C_{\mathrm{SST}}
\subset
\mathcal C_{\mathrm{WST}}.
\]
The first inclusion is strict in general because full SST allows heterogeneous
matrices that are not Toeplitz. The second inclusion is the usual implication
from strong to weak stochastic transitivity.

\subsubsection{Geometry and interpretation of the transitivity constraints}
\label{subsubsec:lst-comparison}

The Toeplitz SST prior has \(K-1\) free parameters, as in \eqref{eq:sst-monotone-psi}. Full SST and WST are both embedded in the ambient space of dimension \(K(K-1)/2\), with full SST further restricted by the strong-transitivity inequalities.

\begin{figure}[t]
\centering
\includegraphics[width=0.50\textwidth]{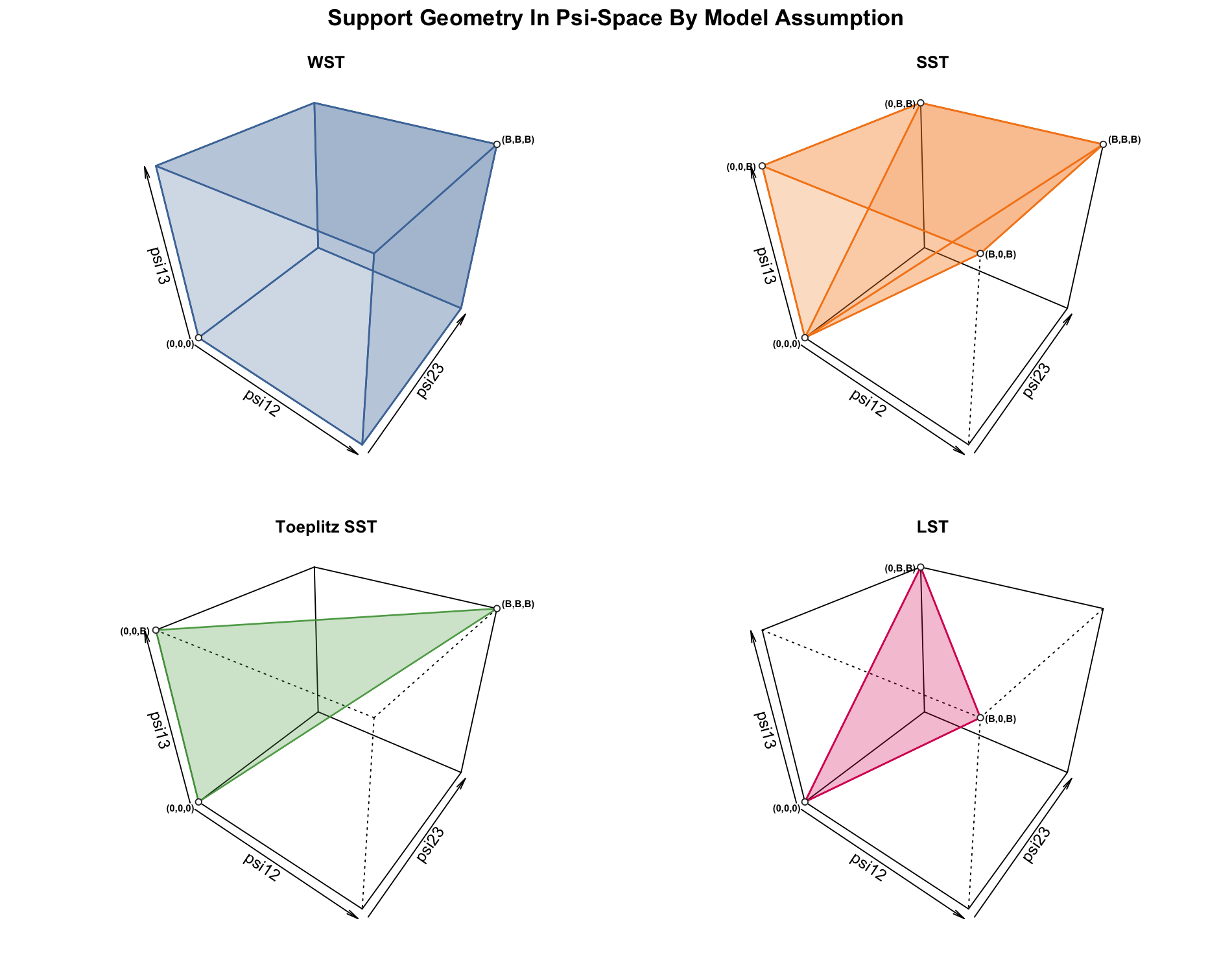}
\caption[Geometry of support restrictions for WST, SST, Toeplitz SST, and LST at K=3]{Geometry of support restrictions in \(\psi\)-space for WST, SST, Toeplitz SST, and LST at \(K=3\), shown as shaded support surfaces with annotated vertices and a common viewing angle. Setting \((x,y,z)=(\psi_{12},\psi_{23},\psi_{13})\), the coordinate \(z=\psi_{13}\) is the long-range comparison and sits on the vertical axis. $\mathcal{C}_\mathrm{WST}$ consists of 
\(x,y,z\ge0\), which fills the cube. $\mathcal{C}_\mathrm{SST}$ has \(x,y\ge0,\; z\ge x,\; z\ge y\), which is a pyramid inside the cube  \([0,B]^3\) with  volume \(B^3/3\), because \(\psi_{13}\) must be the largest of three exchangeable coordinates.
$\mathcal{C}_\mathrm{Toep}$ ($x=y,\;0\le x\le z$) and $\mathcal{C}_\mathrm{LST}$ ($z=x+y,\;x,y\ge0$) are two-dimensional surfaces; their three-dimensional volumes vanish, but their areas inside $[0,B]^3$ are $B^2/\sqrt{2}$ and $\sqrt{3}\,B^2/2$ respectively, with $\mathcal{C}_\mathrm{LST}$ larger by a factor of $\sqrt{3/2}\approx 1.22$ (see Appendix~\ref{app:geometry-supports}). 
}
\label{fig:psi-geometries}
\end{figure}

Figure~\ref{fig:psi-geometries} compares these restrictions for $K=3$, with coordinates $(x,y,z)=(\psi_{12},\psi_{23},\psi_{13})$ inside $[0,B]^3$. 
WST fills the entire positive cube, allowing for instance $\psi_{12}=3$, $\psi_{23}=3$, $\psi_{13}=0.2$. For a citation network such as Example~\ref{mot:citations}, citations may be strongly biased between adjacent-tier journals while the top and bottom tiers remain only mildly asymmetric. SST rules out this behaviour. The condition \(\psi_{13}\ge\psi_{12}\vee\psi_{23}\) says that moving further down the hierarchy cannot reduce the directional advantage. In an animal dominance hierarchy such as Example~\ref{mot:animal}, contests between neighbouring status classes may still be close to balanced, whereas contests between a clearly dominant adult and a young subordinate may be almost deterministic. 

It is also useful to compare Toeplitz SST with block-level linear stochastic transitivity (LST), recalled in \eqref{eq:LST}. This is the structure underlying Bradley--Terry and Plackett--Luce-type models \citep{bradleyRankAnalysisIncomplete1952, plackett1975, lucePossiblePsychophysicalLaws1959}: pairwise log-odds are generated by differences on a one-dimensional latent strength scale, \(\psi_{k\ell}=u_k-u_\ell\), for \(k<\ell\). Its defining property is additivity along the order. For every triple \(k<\ell<m\), one has \(\psi_{km}=\psi_{k\ell}+\psi_{\ell m}\).
This additivity condition is equivalent to the existence of latent block strengths $u_1,\ldots,u_K$ such that \(\psi_{k\ell}=u_k-u_\ell\) for all \(k<\ell\).
Equivalently, writing the adjacent utility gaps as \(a_r=u_r-u_{r+1}\ge0\), LST gives \(\psi_{k\ell}=\sum_{r=k}^{\ell-1}a_r\):
\[
\mathcal{C}_{\mathrm{LST}}
=
\left\{
\psi\in\mathbb R^{K\times K}: \psi+\psi^\top=0,
\psi_{k\ell}=\Small{\sum_{r=k}^{\ell-1}}a_r
\ \text{for }k<\ell,\text{ and }a_1,\ldots,a_{K-1}\in\mathbb R^+
\right\}.
\]

Although Toeplitz SST may look restrictive, it has the same parameter count as LST, which is the baseline assumption of many models for preference data analysis. Neither model contains the other; instead, they impose different restrictions that intersect only when the increments are equally spaced:
\[
\mathcal{C}_{\mathrm{LST}}
\cap
\mathcal{C}_{\mathrm{Toep}}
=
\left\{
\psi:\psi_{k\ell}=c(\ell-k),\; c\ge 0
\right\}.
\]

See Appendix~\ref{app:geometry-supports} for detailed derivation and interpretation of the surface areas of $\mathcal{C}_{\mathrm{Toep}}$ and $\mathcal{C}_{\mathrm{LST}}$. In the remainder of the paper, unless otherwise stated, SST refers to the Toeplitz SST prior defined in \eqref{eq:sst-prior1}--\eqref{eq:sst-prior2}; the term full SST is reserved for the unrestricted strong-stochastic-transitivity cone.

\subsection{A prior on ordered partitions}
\label{sec:ocrp-prior}

The number of blocks $K$ in the TSBM \eqref{eq:dyadic-volume-direction} is a key quantity that determines the dimension of block parameters $\kappa$ and $\rho$. We want to be agnostic and estimate $K$ together with the allocation of nodes into ordered blocks, as driven by the latent allocation vector \(\mathbf z=(z_1,\ldots,z_n)\). Here \(z_i=k\), with \(k\in[K]\), assigns node \(i\) to block \(k\), and records its position in the hierarchy \(1\succ 2\succ\cdots\succ K\). It is important to distinguish the number of blocks, $K$, from the number of \emph{occupied blocks}, henceforth denoted by $K_n$, that is the number 
of blocks of the partition of $[n]$ induced by $\mathbf z$. This distinction is important when $K$ is set to a sufficiently large value, deemed larger than the true number of blocks: this is called the \emph{overfitted} case when one estimates the number of blocks through $K_n$.

An alternative way of handling an unknown number of blocks is to compare the fit of models with different values of $K$, for instance
using marginal likelihoods or information criteria such as BIC or ICL \citep{celeux2006deviance,biernacki2000assessing}. A third alternative consists in putting a prior on the number of components in a finite mixture model \citep{nobileBayesianFiniteMixtures2007}, and use trans-dimensional algorithms
such as reversible-jump MCMC \citep{green1995}, that update the number of components and the allocation jointly. Recently, more efficient sampling schemes of conditional Gibbs type have been introduced in \citet{fruhwirthSchnatter2021telescoping,Deb:Gil:23}.

A different route, common in Bayesian nonparametrics, is to put a prior directly on the partition $\Pi_n$ of $[n]$ and use \(\mathbf z\) only as a convenient encoder, the labelling being immaterial. In the usual setting the random partition is exchangeable in that its distribution is invariant under permutations of node indexes. Equivalently, for \(\{A_1,\ldots,A_K\}\) a partition of \([n]\), then
\[
\Pr(\Pi_n=\{A_1,\ldots,A_K\})
=
p(|A_1|,\ldots,|A_K|),
\]
for some symmetric function \(p\) of compositions $(n_1,\ldots,n_K)$ of $n$. This function $p$ is called the \emph{exchangeable partition probability
function}, or EPPF \citep{pitman2006}. Of particular importance for posterior inference is that the EPPF is subject to an addition rule expressing the consistency as $n$ varies of the sequence of random partitions $(\Pi_n)$. This is equivalent to the existence of predictive laws, typically described via a Chinese restaurant process (CRP) urn scheme \citep{pitman2006}, that expresses the probability, given the current partition, that a new statistical unit is assigned to one of the existing blocks or to a new one. Dirichlet process, Pitman--Yor, and, more in general, Gibbs-type priors stand out for their mathematical tractability \citep{gnedin2006exchangeable,deblasi2015}. Their predictive probabilities are key ingredients in devising \emph{collapsed} or \emph{marginal} Gibbs samplers allowing for estimation of the number of blocks via the number of occupied blocks \(K_n\), hinging on the seminal work of \citet{neal2000} for the Dirichlet process case.
Applications of these priors in SBM settings are growing rapidly; we mention here the infinite relational model of \citet{kemp2006irm} for the Dirichlet process case, \citet{Gen:Bha:Pat:19}, and \citet{Legramanti_2022} for Gibbs-type priors.

Using these priors directly would be inappropriate, because they are invariant
to block labelling. In the TSBM, block labels carry the nodes' positions in the
hierarchy, as elicited by the WST or SST prior on the probability matrix
\(\rho\). A permutation of the block labels changes the ranking of the blocks,
and therefore changes the likelihood.
We need to pair the WST and SST priors on block parameters with a prior on {\it ordered} partitions with tractable predictive probabilities, namely those of assigning a node to an existing ordered block or of creating a new block in a specific position in the hierarchy. 

Ordered random partitions have been studied in population genetics. \citet{Don:Tav:86,Don:Joy:91} investigated the allelic types of a sample from a population where types are ordered by their ages, namely the times since they first appeared in the sample. They derived the ordered counterpart of the Dirichlet process EPPF, which notably satisfies consistency between different sample sizes. \citet{Gne:Pit:05aop} later developed an elegant theory of regenerative composition structures, and derived the ordered counterpart of the EPPF of the Pitman--Yor process. See \citet{balocchi2024ordering} for a recent application in species-sampling problems.

In the sequel we will resort to the \emph{age-ordered} random partition of \citet{Don:Tav:86,Don:Joy:91}, which stands out for analytical tractability. To the best of our knowledge, it has never been used in latent blocks models such as ours. In their original setting, the hierarchy is determined by the \emph{age}: blocks of higher age are those that appear first in the sampling from the population. \citet{Don:Joy:91} explicitly noted that different orderings might be of interest, including a ``pecking order''. We retain the same age-ordering convention, but interpret it in the TSBM orientation: lower-ranked blocks, appearing later in the hierarchy, may be larger than the earlier higher-ranked blocks. And so our labelling $k=1,2,\ldots$ is coherent with labelling by age. In fact, the relative sizes of the blocks in order of appearance converge to the distribution of the Dirichlet process probability weights in size-biased order: $\pi_i\overset{d}{=}w_i\prod_{j<i}(1-w_j)$ for $w_i\overset{\text{i.i.d.}}{\sim}\text{beta}(1,\vartheta)$ and $\vartheta>0$ the concentration parameter \citep{pitman2006}.


We recall next the ordered EPPF of the Dirichlet process, which yields predictive probabilities that correspond to an ordered version of the CRP. 
Consider an ordered partition $\{A_1,\ldots,A_K\}$ of $[n]$ with $K$ blocks of sizes $n_1,\ldots,n_K$ and let $m_k=\sum_{\ell=1}^k n_\ell$, $k=1,\ldots,K$, that is \(m_k\) is the number of nodes in the first \(k\) ranks of the hierarchy. The ordered EPPF is given by
\begin{equation}
\label{eq:ocrp-allocation}
p(n_1,\ldots,n_K)
=
\frac{\vartheta^K}{\vartheta_{(n)}}\,
\frac{\prod_{k=1}^K n_k!}{m_1m_2\cdots m_K},
\end{equation}
see equation (4.1) in \citet{Don:Tav:86}. Here $\vartheta_{(n)}=\Gamma(\vartheta+n)/\Gamma(\vartheta)$ is the ascending factorial. Summing over all permutations of $[K]$ and exploiting Proposition 2.1 in \citet{Don:Tav:86} one recovers the EPPF of the Dirichlet process, cf. equation (2.19) in \citet{pitman2006}. 
To derive the predictive probabilities, consider a new node, $n+1$, that can either join one of the \(K\) existing blocks or create a new
singleton block in one of \(K+1\) possible insertion positions. We express these probabilities in terms of the law of the allocation variable $z_{n+1}$ given the first $n$, collected into the vector $\mathbf z=(z_1,\ldots,z_n)$. We distinguish the values of $z_{n+1}$ among \comillas{old} as $k^{\text old}$ for $k\in [K]$, and \comillas{new} as $r^{\mathrm{new}}$ for $r\in [K+1]$. The predictive probabilities are recovered as ratios of ordered EPPF, for example
  $$\Pr(z_{n+1}=2^{\text old}\mid \mathbf z)=\frac{p(n_1,n_2+1,\ldots,n_K)}{p(n_1,n_2,\ldots,n_K)},\quad
  \Pr(z_{n+1}=2^{\mathrm{new}}\mid \mathbf z)=\frac{p(n_1,1,n_2,\ldots,n_K)}{p(n_1,n_2,\ldots,n_K)}$$
One finds
\begin{align}
\label{eq:ocrp-old}
\Pr(z_{n+1}=k^{\text old}\mid \mathbf z)
&=
\frac{n_k+1}{\vartheta+n}
\prod_{\ell=k}^{K}\frac{m_\ell}{m_\ell+1},
\qquad k=1,\ldots,K,\\
\label{eq:ocrp-new}
\Pr(z_{n+1}=r^{\mathrm{new}}\mid \mathbf z)
&=
\frac{\vartheta}{\vartheta+n}\,
\frac{1}{m_{r-1}+1}
\prod_{\ell=r}^{K}\frac{m_\ell}{m_\ell+1},
\qquad r=1,\ldots,K+1.
\end{align}
with the conventions \(m_0=0\) and \(\prod_{\ell=K+1}^{K}(\cdot)=1\). In the Gibbs sampler, the same predictive law is applied after temporarily removing the node being updated. Thus, for node \(i\), the quantities entering the predictive probabilities are computed from \(z_{-i}\): any empty rank created by deleting node \(i\) is removed, the remaining occupied ranks are relabelled according to their induced order, and \(n\) is replaced by \(n-1\).

After forgetting the order of the blocks, the induced unordered partition recovers the Dirichlet process EPPF and the ordinary CRP urn scheme. Consequently, the number of occupied blocks \(K_n\) among \(n\) nodes has the same logarithmic growth controlled by the concentration parameter \(\vartheta\):
\begin{equation}
\label{eq:ocrp_prior_K}
\mathbb E[K_n]
=
\sum_{i=1}^{n}\frac{\vartheta}{\vartheta+i-1}
=
\vartheta\big(\operatorname{digamma}(\vartheta+n)-\operatorname{digamma}(\vartheta)\big)
\approx \vartheta\log n,\ \text{as }n\to\infty
\end{equation}
Therefore small values of $\vartheta$ favour fewer occupied blocks, and vice versa. What changes is not the distribution of \(K_n\), but the way occupied block sizes are arranged along the hierarchy. Below, we refer to the ordered EPPF in \eqref{eq:ocrp-allocation} also as a distribution over the vector of labels $\mathbf z$, and write $p(\mathbf z)$ accordingly. Specifically, we are referring to the whole sequence of laws indexed by the sample size $n$ with predictive probabilities \eqref{eq:ocrp-old} and \eqref{eq:ocrp-new}.
Appendix~\ref{app:ocrp-derivation} gives the derivation of these predictive probabilities and the induced prior diagnostics.

\subsection{Generative process and posterior}
\label{sec:generative-process}

The full TSBM generative process is collected in Algorithm~\ref{alg:TSBM-generative}. This summary comes after the three ingredients have been introduced, so that the ordered allocation, the volume--direction factorisation, and the two priors on directional parameters are all defined before they are used. We use the shape--rate parametrisation, with Gamma prior $\eta_i\overset{\mathrm{iid}}{\sim}\mathrm{Gamma}(a_\eta,b_\eta)$ and $\kappa_{k\ell}\overset{\mathrm{ind}}{\sim}\mathrm{Gamma}(a_\kappa,b_\kappa)$ for $k\le \ell$.

\newlength{\TSBMpanelheight}
\setlength{\TSBMpanelheight}{0.31\textheight} 

\begin{figure}[t]
\centering

\begin{minipage}[t][\TSBMpanelheight][t]{0.41\textwidth}
\vspace{0pt}
\centering

\begin{tikzpicture}[
   scale=0.95,
   transform shape,
   latent/.style = {draw, circle, minimum size=20pt, inner sep=0pt},
   obs/.style    = {latent, fill=gray!20},
   const/.style  = {rectangle, inner sep=0pt},
   plate/.style  = {rounded corners, draw, dashed, inner sep=5pt},
   >=Stealth,
   node distance=1.15cm
]

\node[const, align=center] (hyper)
{$\varphi \coloneqq \{ a_\kappa,b_\kappa,\ a_\eta,b_\eta,$\\
 $\text{hyper}(\psi),\ \vartheta \}$};

\node[latent] (kappa) [below left= 1.7 cm and 1 of hyper] {$\kappa_{k\ell}$};
\node[latent] (psi)   [below right=1.7cm and 1 cm of hyper]
{$\psi_{k\ell}/\psi_d$};

\node[latent] (eta) [below right= 1 cm and 0.8 cm of kappa] {$\eta_i$};
\node[latent] (z)   [right= 1.5cm of eta] {$z_i$};

\node[obs] (N) [below=2.2cm of eta] {$N_{ij}$};
\node[obs] (A) [right=1.55cm of N] {$A_{ij}$};

\draw[->] (hyper) -- (kappa);
\draw[->] (hyper) -- (psi);
\draw[->] (hyper) -- (eta);
\draw[->] (hyper) -- (z);

\draw[->] (kappa) -- (N);
\draw[->] (eta)   -- (N);
\draw[->] (z)     -- (N);
\draw[->] (z)     -- (A);
\draw[->] (psi)   -- (A);
\draw[->] (N)     -- (A);

\begin{scope}[on background layer]
  \node[plate, fit={(eta) (z)}, label=below:{\small $i\in [n]$}] {};
  \node[plate, fit={(N)  (A)}, label=below:{\small $i<j$}] {};
  \node[plate, fit={(kappa)}, label=right:{\small $k\le\ell$}] {};
  \node[plate, fit={(psi)}, 
      label= right:{\small \shortstack[l]{\footnotesize $k<\ell$ \\\footnotesize (WST) \\ 
       \footnotesize  $d\in [K-1]$ \\ 
       \footnotesize ($\mathrm{SST}$)}}] {};
\end{scope}

\end{tikzpicture}

\end{minipage}
\hfill
\begin{minipage}[t][\TSBMpanelheight][t]{0.48\textwidth}
\vspace{0pt}

\begin{algorithm}[H]
\footnotesize
\DontPrintSemicolon
\caption{Generative process}
\label{alg:TSBM-generative}

\KwIn{$n$, $\varphi$, and model WST/SST}
\BlankLine
Draw
\(
\mathbf z \sim p(\mathbf z), \eta_i\sim\mathrm{Gamma}(a_\eta,b_\eta)\)
 and set \(
K_n =\#\mathrm{unique}(\mathbf z).
\)
\BlankLine
For $ 1\le k\le \ell\le K_n$, draw:
\(
\kappa_{k\ell}\sim\mathrm{Gamma}(a_\kappa,b_\kappa),
\)
\; and set \(\kappa_{\ell k}=\kappa_{k\ell}\).
\BlankLine

If model = WST, for $ 1\le k<\ell\le K_n$, draw: $\psi_{k\ell}\sim \eqref{eq:wst-prior}$
\BlankLine
If model = SST, for $d \in [K_n - 1]$, draw: $\delta_d\sim$\eqref{eq:sst-prior2}. Then, set 
$\psi_d=\sum_{r=1}^d\delta_r$
\BlankLine

For $k\neq\ell$, define \(\rho_{k\ell}\) from \eqref{eq:phi_def}; for a dyad with \(z_i=k\) and \(z_j=\ell\), this means \(\rho_{z_i z_j}=\sigm(\phi_{ij})\).
\BlankLine

\For{$i<j$, with $z_i=k$ and $z_j=\ell$}{
\[
\begin{aligned}
N_{ij}&\sim\operatorname{Poisson}(\eta_i\eta_j\kappa_{k\wedge\ell,\;k\vee\ell}),\\
A_{ij}&\sim\operatorname{Binomial}(N_{ij},\rho_{k\ell}).
\end{aligned}
\]
Set $A_{ji}=N_{ij}-A_{ij}$.\;
}

\end{algorithm}
\end{minipage}

\caption{DAG and generative process for the TSBM. The $\psi$ node denotes block-level log-odds $\psi_{k\ell}$ under WST and distance-level log-odds $\psi_d$ under SST. Shaded nodes are observed.}
\label{fig:dag-TSBM-generative}

\end{figure}

Let $\mathcal I=\{(i,j):1\le i<j\le n\}$ and define, for each $(i,j)\in\mathcal I$,
\begin{equation}\label{eq:phi_def}
\phi_{ij}
=
\begin{cases}
0, & z_i=z_j,\\[3pt]
s_{ij}\psi_{z_i\wedge z_j,z_i\vee z_j}, & z_i\neq z_j,\; (\mathrm{WST}),\\[3pt]
s_{ij}\psi_{|z_i-z_j|}, & z_i\neq z_j,\; (\mathrm{SST}),
\end{cases}
\qquad
s_{ij}=\operatorname{sgn}(z_j-z_i).
\end{equation}
Thus $\rho_{z_i z_j}=\sigm(\phi_{ij})$. The notation $x\wedge y=\min(x,y)$ and $x\vee y=\max(x,y)$ is used to keep the log-odds indexed by the upper-triangular block pair, while the sign $s_{ij}$ records the direction induced by the current ordering.

Up to constants not depending on the unknowns, the posterior distribution is
\begin{align}
\label{eq:full-post}
p(\bm\eta,\kappa,\psi,\mathbf z\mid A,N;\varphi)
\propto
\left\{
\prod_{(i,j)\in\mathcal I}
p(N_{ij}\mid \bm\eta,\kappa,\mathbf z)\,
p(A_{ij}\mid N_{ij},\psi,\mathbf z)
\right\}
p(\bm\eta)\,p(\kappa)\,p(\psi)\,p(\mathbf z).
\end{align}
Here \(p(\psi)\) is the WST or SST prior of Section~\ref{sec:psi_prior}, and \(\varphi\) collects the fixed hyperparameters of the priors on \(\bm\eta\), \(\kappa\), \(\psi\), and the ordered partition. The factorisation in~\eqref{eq:full-post} also makes clear where the computational difficulty lies. Conditional on the allocation \(\mathbf z\), the volume component retains the usual Poisson--Gamma structure. The direction component is less immediate: the block-level parameter \(\psi\) enters through a logistic likelihood, and its support depends on the chosen ordered model, either WST or SST. The inference scheme therefore has to keep the directional parameters explicit, respect the corresponding order constraints, and still provide tractable updates. The next section shows that this can be achieved by introducing Pólya--Gamma latent variables for the directional likelihood.

\section{Inference}\label{sect:inference}

We fit the posterior in~\eqref{eq:full-post} with a partially collapsed Gibbs
sampler. The sampler returns draws of the ordered allocation \(\mathbf z\), the
volume parameters \((\bm\eta,\kappa)\), and the directional hierarchy parameter
\(\psi\). The number of occupied blocks \(K_n\) is
then recovered from \(\mathbf z\). The Gaussian prior on \(\psi\) is
non-conjugate to the directional component of the likelihood, preventing a
direct Gibbs update. Following
\citet{pengBayesianDegreeCorrectedStochastic2014}, we therefore use
Pólya--Gamma augmentation, as detailed in the next section.

\subsection{Pólya--Gamma data augmentation}\label{subsec:PG-augment}

For an unordered dyad $(i,j)$, the directional likelihood $p(A_{ij}\mid N_{ij},\psi,\mathbf z)$ in~\eqref{eq:full-post} 
features the ratio
$\exp\{A_{ij}\phi_{ij}\}/(1+\exp\{\phi_{ij}\})^{N_{ij}}$ where $\phi_{ij}$ is the directional log-odds with different formulation in terms of the directional parameter $\psi$ depending on the transitivity models in use, see \eqref{eq:phi_def}.
This term is not conjugate to the Gaussian priors on $\psi$, therefore we introduce one Pólya--Gamma distributed auxiliary variable per unordered dyad,
\[
\omega_{ij}\mid N_{ij},\phi_{ij}
\sim
\mathrm{PG}(N_{ij},\phi_{ij}),
\] 
where $\mathrm{PG}(\cdot,\cdot)$ denotes the Pólya--Gamma distribution with density $p_{\mathrm{PG}}(\cdot\mid N_{ij},\phi_{ij})$. Cf. \citet{polson2013}. 
The augmented directional likelihood is
\begin{equation}
\label{eq:aug-dyad}
p(A_{ij},\omega_{ij}\mid N_{ij},\phi_{ij})
\propto
\prod_{(i,j) \in \mathcal I}\exp\!\left\{
\Abar_{ij}\phi_{ij}
-\frac{1}{2}\omega_{ij}\phi_{ij}^2
\right\}
p_{\mathrm{PG}}(\omega_{ij}\mid N_{ij},0).
\end{equation}
where $\Abar_{ij}\coloneqq A_{ij}-\frac{N_{ij}}{2}$ 
corresponds to the counts $A_{ij}$ centered to have zero mean.
This term is now conjugate to the Gaussian priors on $\psi$, in both the WST and the SST case. The variables $\omega_{ij}$ are discarded after sampling, as they carry no inferential meaning. Finally, 
\begin{align}
\label{eq:full-post-regime}
p(\bm\eta,\kappa,\psi,\mathbf z,\omega \mid A,N)
&\propto
\prod_{(i,j)\in\mathcal I}
\exp\!\Big\{
-\eta_i\eta_j\,\kappa_{z_i z_j}
+N_{ij}\big(\log\eta_i+\log\eta_j+\log\kappa_{z_i z_j}\big)
\Big\}
\nonumber\\
&\quad\times
\exp\!\Big\{
\Abar_{ij}\phi_{ij}
-\frac{1}{2}\omega_{ij}\phi_{ij}^{2}
\Big\}
p_{\mathrm{PG}}(\omega_{ij}\mid N_{ij},0)
\nonumber\\
&\quad\times
\Bigl[\prod_{k\le\ell}\kappa_{k\ell}^{a_\kappa-1}
e^{-b_\kappa\kappa_{k\ell}}\Bigr]
\Bigl[\prod_{i=1}^n\eta_i^{a_\eta-1}
e^{-b_\eta\eta_i}\Bigr]
p(\psi)
p(\mathbf z),
\end{align}
where $p(\psi)$ is the WST or SST prior of Section~\ref{sec:psi_prior}, and $p(\mathbf z)$ is the prior corresponding to age-ordered random partition~\eqref{eq:ocrp-allocation}. Note that log-odds $\phi_{ij}$ is a deterministic function of $(\psi,z_i,z_j)$, so the random variable transformation introduces no additional Jacobian term.

\subsection{One Gibbs sweep}
\label{subsec:gibbs-sweep-overview}

A sweep of the sampler updates the volume parameters \(\kappa\), the directional parameters \(\psi\), the Pólya--Gamma variables \(\omega_{ij}\), the degree corrections \(\eta_i\), and the allocation variables \(z_i\). Algorithm~\ref{alg:TSBM-gibbs-compact} presents the overview of the Gibbs sampler, while below we inspect each single update.

\begin{algorithm}[htpb]
\DontPrintSemicolon
\setstretch{0.95}
\setlength{\algomargin}{2.0em}
\SetInd{0.55em}{0.75em}

\KwIn{Initial state
\((\mathbf z^{(0)},\bm\eta^{(0)},\kappa^{(0)},\psi^{(0)})\);
number of iterations \(T\).}
\KwOut{Posterior draws
\(\{(\mathbf z^{(t)},\bm\eta^{(t)},\kappa^{(t)},\psi^{(t)})\}_{t=1}^T\).}

\For{\(t=1,\ldots,T\)}{

  Draw \(\omega^{(t)}_{ij} \sim \mathrm{PG}\{N_{ij},\phi_{ij}^{(t-1)}\}\)
  for all \((i,j)\in\mathcal I\).\;

  \BlankLine

  Draw \(\psi^{(t)} \sim p(\psi\mid \mathbf z^{(t-1)},\omega^{(t)},A,N)\)
  from the WST or SST truncated-normal full conditional.\;
  \Indp
  Determine \(\phi^{(t)}\) from \(\psi^{(t)}\) and \(\mathbf z^{(t-1)}\).\;
  \Indm

  \BlankLine

  Draw \(z_i^{(t)}\) from its ordered-allocation full conditional, with
  \(\kappa\) collapsed, for \(i=1,\ldots,n\).\;
  \Indp
  If removing node \(i\) leaves its block empty, set
  \(K_n\leftarrow K_n-1\), and adjust \(\kappa\) and \(\psi\).\;
  If a birth move is selected, insert a new block, set
  \(K_n\leftarrow K_n+1\), and adjust \(\kappa\) and \(\psi\).\;
  \Indm

  \BlankLine

  Draw \(\kappa^{(t)} \sim
  p(\kappa\mid \mathbf z^{(t)},\bm\eta^{(t-1)},A,N)\)
  from its Gamma full conditionals.\;

  \BlankLine

  Draw \(\eta_i^{(t)} \sim
  p(\eta_i\mid \mathbf z^{(t)},\kappa^{(t)},\eta_{-i}^{(t)},A,N)\)
  from its Gamma full conditional, for \(i=1,\ldots,n\).\;
}

\caption{TSBM partially collapsed Gibbs sampler.}
\label{alg:TSBM-gibbs-compact}
\end{algorithm}
\begin{enumerate}[leftmargin=2em]

\item \textbf{Update the volume parameters \texorpdfstring{($\kappa,\mathbf \eta)$}{(kappa, eta)}.}
The symmetric matrix $\kappa$ update results from standard Poisson-Gamma conjugacy:
\[
\kappa_{k\ell}\mid\text{rest}
\sim
\mathrm{Gamma}\Big(a_\kappa+\sum\nolimits_{(i,j)\in\mathcal I_{k\ell}}N_{ij},\,
b_\kappa+\sum\nolimits_{(i,j)\in\mathcal I_{k\ell}}\eta_i\eta_j\Big),
\quad k\le\ell, 
\]
The degree corrections $\eta_i$ are then updated sequentially following Gamma-Gamma conjugacy:
\[
\eta_i\mid\text{rest}
\sim
\mathrm{Gamma}\!\left(
a_\eta+\sum\nolimits_{j\ne i}N_{ij},\,
b_\eta+\sum\nolimits_{j\ne i}\eta_j\kappa_{z_i z_j}
\right).
\]
\item \textbf{Update the directional parameter \texorpdfstring{$\psi$}{(psi)}.}
Conditionally on Pólya--Gamma auxiliary variables, we obtain Gaussian full conditionals for $\psi$. Under WST, for each block pair $k<\ell$ the full conditional is
\[
\psi_{k\ell}\mid\text{rest},
\sim
\mathcal N\!\left(
\frac{\bar y_{k\ell}+\sigma_0^{-2}\mu_0}
     {\bar\omega_{k\ell}+\sigma_0^{-2}},
\,
(\bar\omega_{k\ell}+\sigma_0^{-2})^{-1}
\right) \mathbb I(\psi_{k\ell}>0), \, \text{ where} \, \begin{cases}
    \bar y_{k\ell}
&=
\sum_{(i,j)\in\mathcal I_{k\ell}}s_{ij}\Abar_{ij}, \\
\bar\omega_{k\ell}
&=
\sum_{(i,j)\in\mathcal I_{k\ell}}\omega_{ij}.
\end{cases}
\]
Under the SST prior, we have Gaussian updates for the i.i.d. positive increments $\delta=(\delta_1,\ldots,\delta_{K_n-1})$, which we then combine using $\psi_d=\sum_{r\le d}\delta_r$. Expanding the quadratic form in $\delta$ gives truncated--normal updates
\[
\delta_u\mid\text{rest}
\sim
\mathcal N (\mu_u,c_u^{-1})  \mathbb I(\delta_u>0),
\qquad u=1,\ldots,K_n-1,
\]
with $(\mu_u,c_u)$ full expressions reported in Appendix~\ref{app:toep-directional-update}.

\item \textbf{Update the Pólya--Gamma \texorpdfstring{$\omega$}{omega}.}
For each unordered dyad,
\[
\omega_{ij}\mid\text{rest}
\sim
\mathrm{PG}\bigl(N_{ij},\phi_{ij}\bigr).
\]

where $\mathrm{PG}(\cdot, \cdot)$ is, again, given in \citet{polson2013}.

\end{enumerate}

We conclude the outline of the Gibbs sampler by describing how to compute the ordered-allocation full conditional and, jointly, how to update \(K_n\), the number of occupied blocks.

\subsubsection{Updating the ordered allocation}
\label{subsec:z-update-unified}

In the sequel we refer to the urn scheme induced by the predictive probabilities
\eqref{eq:ocrp-old} and \eqref{eq:ocrp-new} as the \emph{ordered CRP}.\footnote{This urn scheme can be applied in the single-site Gibbs update after temporarily removing node \(i\), since the underlying ordered partition law is exchangeable in the node labels and coherent as the sample size \(n\) varies.}
For the update of node \(i\), this urn scheme gives the probability
\(\Pr(z_i=k^{\mathrm{old}}\mid z_{-i})\)
of assigning node \(i\) to an existing block, for \(k\in[K_{-i}]\), and the probability
\(\Pr(z_i=r^{\mathrm{new}}\mid z_{-i})\)
of creating a new block at slot \(r\in[K_{-i}+1]\). Here \(z_{-i}\) is the vector of labels \(\mathbf z\) without \(z_i\) and, throughout this subsection, \(K_{-i}\) denotes the number of occupied blocks after removing node \(i\); the relabelling required when that removal empties a block is described in Appendix~\ref{app:z-full-conditional}. These \emph{prior} predictive probabilities represent the prior contribution to the Gibbs update of the allocation \(\mathbf z\) that respects the block ordering.

We compute the probability of assigning node \(i\), conditional on \(z_{-i}\), the model parameters, and the data. The candidate values are the \(K_{-i}\) existing blocks and the \(K_{-i}+1\) possible insertion slots. By Bayes' rule,
\begin{align}
\Pr(z_i=t\mid z_{-i},\text{rest})
&=
\frac{
\Pr(z_i=t,A\mid z_{-i},\text{rest})
}{
p(A\mid z_{-i},\text{rest})
}
\nonumber\\
&\propto
\Pr(z_i=t\mid z_{-i})\,
\frac{
p(A\mid z_i=t,z_{-i},\text{rest})
}{
p(A_{-i}\mid z_{-i},\text{rest})
}
\nonumber\\
&=
\Pr(z_i=t\mid z_{-i})\,
\exp\!\{\ell_i(t;\mathcal M)\},
\label{eq:z-update-full-conditional}
\end{align}
where \(t\) denotes either an existing block or a new block, and \(A_{-i}\) denotes the data after removing all dyads involving node \(i\). The ratio in the second line is evaluated on the log scale, i.e., \(\ell_i(t;\mathcal M)\), for computational convenience, with the volume parameter \(\kappa\) integrated out. We now describe the two cases, old vs new block, leaving the exact formulae for \(\ell_i(t;\mathcal M)\) to Appendix~\ref{app:z-full-conditional}.

\paragraph{1) Existing-block candidates}

For \(k\in[K_{-i}]\), all parameters are already defined after temporarily removing node \(i\). The corresponding full conditional is
\begin{align}
\Pr(z_i=k^{\mathrm{old}}\mid z_{-i},\text{rest})
&\propto
\Pr(z_i=k^{\mathrm{old}}\mid z_{-i})
\exp\!\left\{\ell_i^{\mathrm{old}}(k;\mathcal M)\right\},
\label{eq:z-update-existing-weight}
\displaybreak[0]\\
\ell_i^{\mathrm{old}}(k;\mathcal M)
&=
\ell_{i,\mathrm{vol}}^{\mathrm{old}}(k)
+
\ell_{i,\mathrm{dir}}^{\mathrm{old}}(k;\mathcal M).
\nonumber
\end{align}
The volume term $\ell_{i,\mathrm{vol}}^{\mathrm{old}}(k)$ is evaluated in partially collapsed form, by integrating out the affected \(\kappa\) parameters and thus it features Poisson--Gamma marginal likelihood ratios \citep{vanDyk01062008}. The directional term $\ell_{i,\mathrm{dir}}^{\mathrm{old}}(k;\mathcal M)$ is then computed with the current WST or SST parameters, using pooled node-to-block quantities. Appendix~\ref{app:z-full-conditional} gives the exact formulae, interprets the pooled terms, and shows which factors cancel out in likelihood ratios.

\paragraph{2) New-block candidates}

For \(r\in[K_{-i}+1]\), the candidate move creates a new singleton block at insertion slot \(r\). The corresponding full conditional is
\begin{align}
\Pr(z_i=r^{\mathrm{new}}\mid z_{-i},\text{rest})
&\propto
\Pr(z_i=r^{\mathrm{new}}\mid z_{-i})
\exp\!\left\{\ell_i^{\mathrm{new}}(r;\mathcal M)\right\},
\label{eq:z-update-new-weight}
\displaybreak[0]\\
\ell_i^{\mathrm{new}}(r;\mathcal M)
&=
\ell_{i,\mathrm{vol}}^{\mathrm{new}}
+
\ell_{i,\mathrm{dir}}^{\mathrm{new}}(r;\mathcal M).
\nonumber
\end{align}
If node \(i\) is assigned to a new block at slot \(r\), then existing blocks with rank \(\ell<r\) keep their rank, whereas existing blocks with rank \(\ell\ge r\) are shifted to rank \(\ell+1\). The volume contribution \(\ell_{i,\mathrm{vol}}^{\mathrm{new}}\) is evaluated in partially collapsed form, by integrating out the volume parameters involving the new block, and does not depend on \(r\). It therefore needs to be computed only once for the new-block case.

The dependence on \(r\) enters through the prior term \(\Pr(z_i=r^{\mathrm{new}}\mid z_{-i})\) and through the directional term \(\ell_{i,\mathrm{dir}}^{\mathrm{new}}(r;\mathcal M)\). Under WST, only the signs of the comparisons between node \(i\) and the existing blocks depend on the insertion slot. Under SST, the insertion also changes the distances between existing block pairs that straddle slot \(r\), so the corresponding old--old directional contributions must be corrected as well. See Appendix~\ref{app:z-full-conditional} for details.

\subsubsection{Sampling the ordered allocation}

After computing the \(K_{-i}\) terms 
  $\Pr(z_i=k^{\mathrm{old}}\mid z_{-i})\exp\!\left\{\ell_i^{\mathrm{old}}(k;\mathcal M)\right\}$
in \eqref{eq:z-update-existing-weight} and the \(K_{-i}+1\) terms
  $\Pr(z_i=r^{\mathrm{new}}\mid z_{-i})\exp\!\left\{\ell_i^{\mathrm{new}}(r;\mathcal M)\right\}$
in \eqref{eq:z-update-new-weight}, they are divided by their sum so to form the \(2K_{-i}+1\) probabilities to be used to sample the allocation $z_i$. 

If a new block $r^{\text{new}}$ is selected, the old blocks \(r,\ldots,K_{-i}\) shift to \(r+1,\ldots,K_{-i}+1\), node \(i\) becomes the singleton block at slot \(r\), and the parameters integrated out in the partially collapsed probability calculation are subsequently instantiated from their full conditionals: the new volume parameters \(\kappa_{r\ell}\) and, depending on the model, either the new WST log-odds or the new SST increment $\delta_{K_{-i}}$ corresponding to the new largest-distance log-odds $\psi_{K_{-i}}$. 

Following \citet{vanDyk01062008}, we place this partially collapsed update of the allocation \(\mathbf z\) right before the $\kappa$ update, so that the integrated-out parameters are refreshed immediately. This completes the single-site update of the ordered allocation; we next describe how the resulting MCMC output is summarised into point estimates.

\subsubsection{Identifiability and hyperparameter choice}

The volume likelihood $p(N_{ij}\mid \bm\eta,\kappa,\mathbf z)$ in~\eqref{eq:full-post} 
depends on
\(\bm\eta\) and \(\kappa)\) only through the products
\(\eta_i\eta_j\kappa_{z_i z_j}\), cf. \eqref{eq:dyadic-volume-direction}. Hence, for any positive constants
\(c_1,\dots,c_{K_n}\), the transformation
\begin{equation}
\eta'_i = c_{z_i}\,\eta_i, \qquad
\kappa'_{k\ell} = \frac{\kappa_{k\ell}}{c_k c_\ell}
\label{eq:tsbm-scale-transform}
\end{equation}
leaves the volume likelihood unchanged. We fix
this scale non-identifiability by rescaling \(\bm\eta\) and \(\kappa\) so that
\begin{equation}
\sum_{i:z_i=k}\eta_i=n_k,
\qquad
k=1,\ldots,K_n,
\label{eq:tsbm-eta-normalisation}
\end{equation}
where \(n_k=\#\{i:z_i=k\}\). That is, after an unconstrained update of \(\bm\eta\), set
\[
c_k=\frac{n_k}{\sum_{h:z_h=k}\eta_h},
\qquad
\eta_i \leftarrow c_{z_i}\eta_i,
\qquad
\kappa_{k\ell}\leftarrow \frac{\kappa_{k\ell}}{c_k c_\ell}.
\]
This deterministic rescaling leaves the volume likelihood unchanged and fixes
one representative from each likelihood-equivalent parametrisation. 

Regarding hyperparameters, the main tuning knob is \(\vartheta\), the
concentration parameter of the age-ordered partition prior, since
\(\mathbb E[K_n]\approx \vartheta\log n\). We set \(\vartheta=0.5\) throughout,
fix the remaining Gamma hyperparameters for \(\kappa\) and \(\eta\) at one, and show in Appendix~\ref{app:sim-sensitivity} that moderate perturbations
mainly shift the posterior block count rather than the qualitative model
comparison or predictive conclusions.


\section{Model choice, point estimates and uncertainty quantification}
\label{sec:summaries}

The sampler produces posterior draws \(\{(\psi^{(t)},\bm \eta^{(t)},\mathbf z^{(t)}, \kappa^{(t)})\}_{t=1}^T\). After discarding the first \(B\) samples as burn-in, we summarise the posterior distribution and perform model-choice comparisons. We compare our TSBM with the DC--SBM introduced in \eqref{eq:poisson-sbm}. Its full specification is relegated to Appendix~\ref{app:DCSBM_gnedin}.

\subsection{Model choice via leave-one-out cross-validation}
\label{subsec:predictive}

We use leave-one-out predictive accuracy for model choice. For each unordered dyad $(i,j) \in \mathcal I$, let \(y_{ij}=(A_{ij},N_{ij})\) and 
  $y_{-(ij)}=(A_{i'j'},N_{i'j'})_{(i'j') \neq(ij)}.$ 
For model \(\mathcal M\), the leave-one-out expected log predictive density is
\[
\mathrm{ELPD}_{\mathcal M}
=
\sum_{(i,j) \in \mathcal I}
\log
\int
p_{\mathcal M}(y_{ij}\mid \theta_{\mathcal M})\,
p_{\mathcal M}(\theta_{\mathcal M}\mid y_{-(ij)})\,d\theta_{\mathcal M},
\]
where \(\theta_{\mathcal M}\) denotes the parameters and latent variables of model \(\mathcal M\). Larger values mean better estimated out-of-sample prediction. The same dyadwise contributions \(y_{ij}\) are used for WST, SST, and the DC--SBM, so the comparison is made on a common predictive scale.

The integral above is not computed by refitting the model \( |\mathcal I| \) times. Instead, following \citet{vehtari2017practical}, we approximate it by Pareto-smoothed importance sampling, using the R package \texttt{loo}. For each dataset, we report
\[
\mathrm{LOOIC}_{\mathcal M}=-2\widehat{\mathrm{ELPD}}_{\mathrm{LOO},\mathcal M},
\]
because it is easier to interpret, together with the difference from the selected model\footnote{Thus, the selected model has \(\Delta\mathrm{ELPD}=0\), denoted and all other models have negative values.}, denoted \(\Delta_{\mathrm{ELPD},\mathcal M}\), and the corresponding standard error.

\subsection{Posterior summary of the partition and number of blocks}
\label{subsec:partition-summary}

Let 
\(\{\mathbf z_{\mathcal M}^{(t)}\}_{t=1}^T\) be the posterior draws under model $\mathcal M$. Let also $\hat{\mathbf{z}}_{\mathcal M}$ be the VI point estimate, that is the partition minimising posterior expected Variation of Information loss \citep{wade2018, Meil__2007, rastelli2018}.
For WST and SST, the VI optimisation is restricted to the sampled ordered partitions. The selected estimate is therefore one of the posterior draws and inherits the block ordering encoded by the model labels.\footnote{This is implemented by setting \texttt{method = "draws"} in the \texttt{mcclust.ext::minVI} function in \texttt{R}.} Instead, for the DC--SBM, the VI point estimate is obtained using the standard unconstrained VI optimization. We therefore relabel the estimated partition ex post, with labels in the order of the mean empirical out-degree relative to their in-degree, so that smaller labels correspond to higher-ranked blocks and the resulting partition is comparable to the TSBM estimates. In simulation studies, where the true partition is known, the VI distance is used to measure the agreement between the point estimate and the true partition, with lower VI values indicating better recovery.

Finally, we summarise the posterior distribution of the number of occupied blocks,
\[
K_n^{(t)} = \#\{ \text{unique labels in } \mathbf z_{\mathcal M}^{(t)}\},
\qquad t=1,\ldots,T.
\]
One caveat is that the VI point estimate tends to be parsimonious. Hence the number of blocks in the displayed partition \(\hat{\mathbf z}_{\mathcal M}\) need not coincide with the posterior mean reported for \(K_n\).

\subsection{Order posterior summaries}
\label{subsec:order-quality}

The VI point estimate $\hat{\mathbf{z}}$ also lets us measure how well the observed edge directions agree with its order. In the TSBM, since smaller labels correspond to stronger blocks, edges with \(k<\ell\) run forward, while edges with \(k>\ell\) run backwards. The \emph{violation} rate is
\begin{equation}
\zeta^{\mathrm{viol}}_{\hat{\bm{z}}}
= 1-
\frac{\sum_{k<\ell}C_{k\ell}(\hat{\mathbf z})}
  {\sum_{k\ne \ell}C_{k\ell}(\hat{\mathbf z})}, \qquad C_{k\ell}(\hat{\mathbf z}) = \sum_{i:\hat z_i=k}\sum_{j:\hat z_j=\ell} A_{ij}
\label{eq:backward-flow-rate}
\end{equation}
where \(C_{k\ell}(\hat{\mathbf z})\) is the total volume of edges from \(k\) to \(\ell\). It is the fraction \comillas{back/cross} of the number of cross-block edges that points \emph{backwards}, that is from a weaker block to a stronger block, over the total. Values near zero correspond to a clean empirical hierarchy; values near \(1/2\) indicate little directional signal in the displayed order.

When the displayed estimate has a single block, the denominator in
\eqref{eq:backward-flow-rate} is zero, so \(\zeta^{\mathrm{viol}}_{\hat{\bm{z}}}\) is
not defined and is not reported. For WST and SST, \(\zeta^{\mathrm{viol}}_{\hat{\bm{z}}}\) is computed on an ordered point estimate. For the DC--SBM, we need to relabel the point estimate ex post to make sure it is aligned with the TSBM.

\section{Simulation study}
\label{sec:sim-strong}

The simulation study has three aims. First, it assesses whether and under which conditions the sampler presented above in Alg.~\ref{alg:TSBM-gibbs-compact} recovers the relevant model quantities when data are generated from the TSBM. Second, it compares the TSBM models with the directed degree-corrected SBM (DC--SBM), which is the natural unconstrained baseline: it is equivalent from a likelihood standpoint, but does not impose an ordering on the directional probabilities via the prior. Third, it performs some various checks that are reported in Appendix~\ref{app:simulation-supplement}. Finally, the code to reproduce the simulation study, the analyses, and the application section that follows is available at
\href{https://github.com/laposanti/Reproducibility-support-for-Ordered-SBMs}
{\texttt{GitHub Reproducibility Support}}.

\subsection{Design}
\label{subsec:sim-design}

We compare three models throughout: the WST and SST variants of TSBM, and the DC--SBM. The two ordered models use the age-ordered partition prior of Section~\ref{sec:ocrp-prior}, with \(\vartheta=0.5\), \(a_\kappa=b_\kappa=a_\eta=b_\eta=1\), while the mean and variance hyperparameters of the log-odds truncated--normals are fixed to \(\mu_0 = 0, \sigma_0 = 2.5\) for WST. For SST, we set the increment prior scale to \(\tau_0= 0.5\). The DC--SBM uses a standard Dirichlet process partition prior with concentration \(\vartheta=0.5\), and hyperparameter \(a_\lambda = b_\lambda = 1\) for the Gamma prior on intensity parameters $\lambda_{k\ell}$, cf. equation \eqref{eq:poisson-sbm}. In all cases, the number of occupied blocks \(K_n\) is inferred from the data. These defaults are deliberately mild: the appendix sensitivity check shows that moderate perturbations mainly affect the inferred block count, not the broad ordering or the model choice. Additional simulation settings, convergence diagnostics \citep{Gelman_Rubin_1992}, and robustness checks are reported in Appendix~\ref{app:simulation-supplement}.

To generate the data, we use a controlled version of
Algorithm~\ref{alg:TSBM-generative}. Rather than drawing the ordered partition,
volume parameters, and directional log-odds from their priors, we fix them at
interpretable values, so that sparsity, number of ranks, and directional signal
can be varied separately and create distinct scenarios. For each of such scenarios, we generate \(n_{\mathrm{rep}}=5\) networks of \(n=60\) nodes assigned to \(K^\star\) equal-sized ordered blocks. We vary
\[
K^\star\in\{3,5,8\},\qquad
\kappa^\star\in\{0.75,1.50,3.00,6.00\},\qquad
\psi^\star\in\{0.2,0.7,1.3\}.
\]

The parameter \(K^\star\) is the true number of ordered blocks: larger values
make inference harder, since more ranks and more block-pair relations must be
recovered. 

The parameter \(\kappa^\star\) controls \emph{sparsity}: we generate $\kappa^\star_{k\ell}$ from a Gamma distribution with mean $\kappa^\star$. As for degree corrections, we draw \(\eta_i^\star\sim\mathrm{Unif}(0.8,1.2)\) to introduce moderate degree
heterogeneity. Hence, for two nodes \(i\) and \(j\) in blocks \(k\) and
\(\ell\), the total number of edges \(N_{ij}\) is sampled from a
Poisson distribution with mean
\(\eta_i^\star\eta_j^\star
\kappa^\star_{z_i^\star\wedge z_j^\star,z_i^\star\vee z_j^\star}\).
The parameter \(\psi^\star\) controls the \emph{strength of the signal}x, that is the directional
separation on the log-odds scale. Under WST, we generate $\psi^\star_{k \ell}$ from a truncated Normal with mean parameter $\psi^\star$; under SST we generate $\psi_d^\star$ as sum of independent positive increments from truncated Normal with mean parameter $\psi^\star/(K^\star-1)$. See Appendix~\ref{app:sim-design-details} for details. 

We retrieve the directional probabilities as
\(\rho^\star_{k\ell}
=\sigma\{s_{k\ell}\psi^\star_{k\wedge\ell,k\vee\ell}\}\), where
\(s_{k\ell}=\operatorname{sgn}(\ell-k)\), with the corresponding
distance-based version used under SST.
We then sample the direction of each dyad as
\[
A_{ij}\mid N_{ij}
\sim
\operatorname{Binomial}
\left(N_{ij},\rho^\star_{z_i^\star z_j^\star}\right),
\qquad
A_{ji}=N_{ij}-A_{ij}.
\]
The probability \(\rho^\star_{k\ell}\) therefore controls the directional bias,
conditional on the dyadic volume. Larger values of \(\psi^\star\) make the
ordered structure easier to recover, because they push the forward
probabilities farther away from \(1/2\). When \(\rho^\star_{k\ell}\) is close
to \(1/2\), interactions between two ordered blocks look nearly symmetric and
the hierarchy is weakly identified; when \(\rho^\star_{k\ell}\) is closer to
one, most interactions point from the higher-ranked block to the lower-ranked
block, making the directional signal clearer.

This gives \(3\times4\times3=36\) scenarios
for each transitivity regime; with \(n_{\mathrm{rep}}=5\) we analyse \(360\) datasets, on which TSBM--WST, TSBM--SST, and DC--SBM are each fitted once, for a total of \(1080\) posterior fits. Each run uses \(10{,}000\) MCMC iterations (burn-in \(2{,}000\)). We report the posterior mean \(\hat K_n\) with empirical \(95\%\) posterior credible intervals, the VI point estimate \(\hat{\mathbf{z}}_{\mathcal M}\), and predictive performance via mean LOOIC together with paired \(\Delta\)LOOIC values (standard errors in parentheses). Convergence diagnostics and full scenario tables are in Appendix~\ref{app:simulation-supplement}.

\subsection{Results}
\label{subsec:sim-results}

We report the full extent of the simulation study in Appendix~\ref{app:simulation-supplement}. Here we keep only two representative scenarios with $K^\star=8$; the first scenario has generating parameters $(\kappa^\star=0.75,\ \psi^\star=0.2)$. This means the networks are quite sparse, and we have a weak signal, i.e. little separation directional bias among the blocks. We see that in this scenario both WST and SST achieve a significantly better result than DC--SBM. The second scenario $(\kappa^\star=6.0,\ \psi^\star=1.3)$ corresponds to dense network with strong signal: all the three models achieve perfect recovery both of the latent partition and of the latent number of blocks. See Table \ref{Table:1}.

\begin{table}[htpb]
\centering
\caption{Simulation summary for two representative scenarios with \(K^\star=8\). \(\hat K_n\) is the posterior mean of the occupied block count, reported together with the empirical 95\% posterior credible interval over 5 replicates. Mean VI is averaged over the same 5 replicates, with lower values indicating better partition recovery. \textbf{Bold}: lowest VI within each row.}\label{Table:1}
\label{tab:sim-main-two-scenarios}
\resizebox{\textwidth}{!}{%
\begin{tabular}{llcccccc}
\toprule
& & \multicolumn{3}{c}{\(\hat K_{n, [2.5\%,97.5\%]}\)} & \multicolumn{3}{c}{Mean VI} \\
\cmidrule(l{3pt}r{3pt}){3-5}
\cmidrule(l{3pt}r{3pt}){6-8}
Scenario & Gen. model & WST & SST & DC--SBM & WST & SST & DC--SBM \\
\midrule \\
\multirow[t]{2}{*}{\makecell[l]{Sparse, weak signal\\[-1pt]{\footnotesize \((\kappa^\star=0.75,\ \psi^\star=0.2)\)}}}
  & SST & \(5.75_{[2.30,7.92]}\) & \(5.25_{[1.30,7.92]}\) & \(3.50_{[1.15,6.70]}\) & \textbf{1.57} & 1.61 & 2.15 \\
  & WST & \(5.50_{[4.08,6.92]}\) & \(5.50_{[3.22,6.92]}\) & \(3.50_{[2.00,5.85]}\) & 1.70 & \textbf{1.64} & 2.25 \\
\addlinespace
\multirow[t]{2}{*}{\makecell[l]{Dense, strong signal\\[-1pt]{\footnotesize \((\kappa^\star=6.0,\ \psi^\star=1.3)\)}}}
  & SST & \(8.00_{[8.00,8.00]}\) & \(8.00_{[8.00,8.00]}\) & \(7.50_{[7.00,8.00]}\) & \textbf{0.00} & \textbf{0.00} & 0.12 \\
  & WST & \(8.00_{[8.00,8.00]}\) & \(8.00_{[8.00,8.00]}\) & \(7.75_{[7.08,8.00]}\) & \textbf{0.00} & \textbf{0.00} & 0.06 \\
\bottomrule
\end{tabular}%
}
\end{table}

\begin{table}[htpb]
\centering
\caption{Compact predictive comparison for two representative scenarios with \(K^\star=8\). Entries are mean LOOIC with paired \(\Delta\pm\mathrm{SE}\) in parentheses, computed relative to the best model within each row. Lower values are better; the best value in each row is in \textbf{bold}.}
\label{tab:sim-main-two-scenarios-loo}
\resizebox{\textwidth}{!}{%
\begin{tabular}{llccc}
\toprule
Scenario & Gen. model & WST & SST & DC--SBM \\
\midrule \\
\multirow[t]{2}{*}{\makecell[l]{Sparse, weak signal\\[-1pt]{\footnotesize \((\kappa^\star=0.75,\ \psi^\star=0.2)\)}}}
  & SST & \(5431\ (5.8\pm4.0)\) & \(\mathbf{5425\ (0.0\pm0.0)}\) & \(5515\ (90.1\pm54.1)\) \\
  & WST & \(\mathbf{5348\ (0.0\pm0.0)}\) & \(5349\ (0.6\pm5.5)\) & \(5403\ (55.2\pm28.8)\) \\
\addlinespace
\multirow[t]{2}{*}{\makecell[l]{Dense, strong signal\\[-1pt]{\footnotesize \((\kappa^\star=6.0,\ \psi^\star=1.3)\)}}}
  & SST & \(13101\ (18.2\pm10.2)\) & \(\mathbf{13083\ (0.0\pm0.0)}\) & \(13263\ (180.0\pm88.8)\) \\
  & WST & \(\mathbf{12103\ (0.0\pm0.0)}\) & \(12224\ (122.0\pm34.6)\) & \(12154\ (51.6\pm63.7)\) \\
\bottomrule
\end{tabular}%
}
\end{table}

\begin{figure}[!htbp]
\centering
\begin{subfigure}{0.49\textwidth}
\centering
\includegraphics[width=\textwidth]{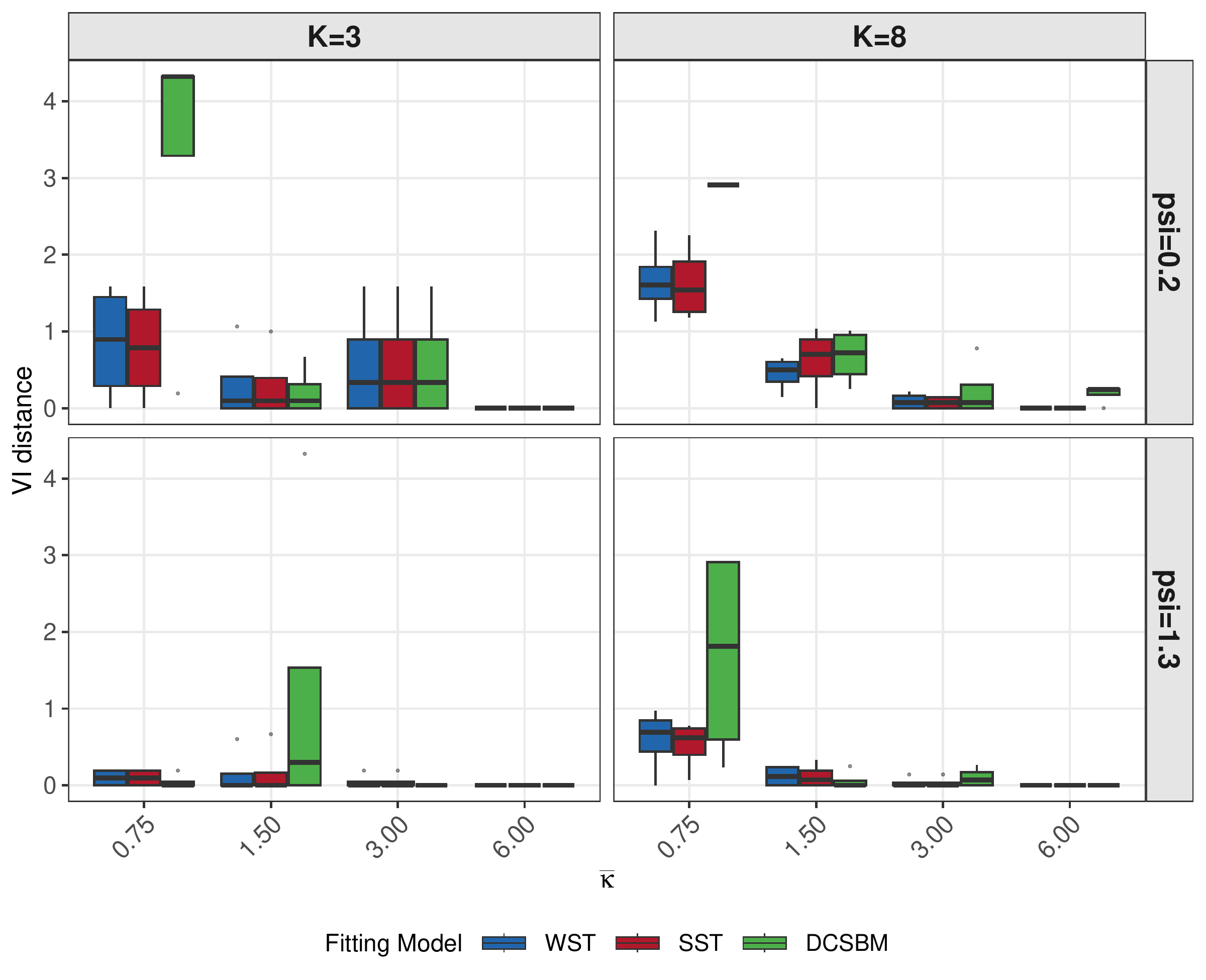}
\end{subfigure}\hfill
\begin{subfigure}{0.49\textwidth}
\centering
\includegraphics[width=\textwidth]{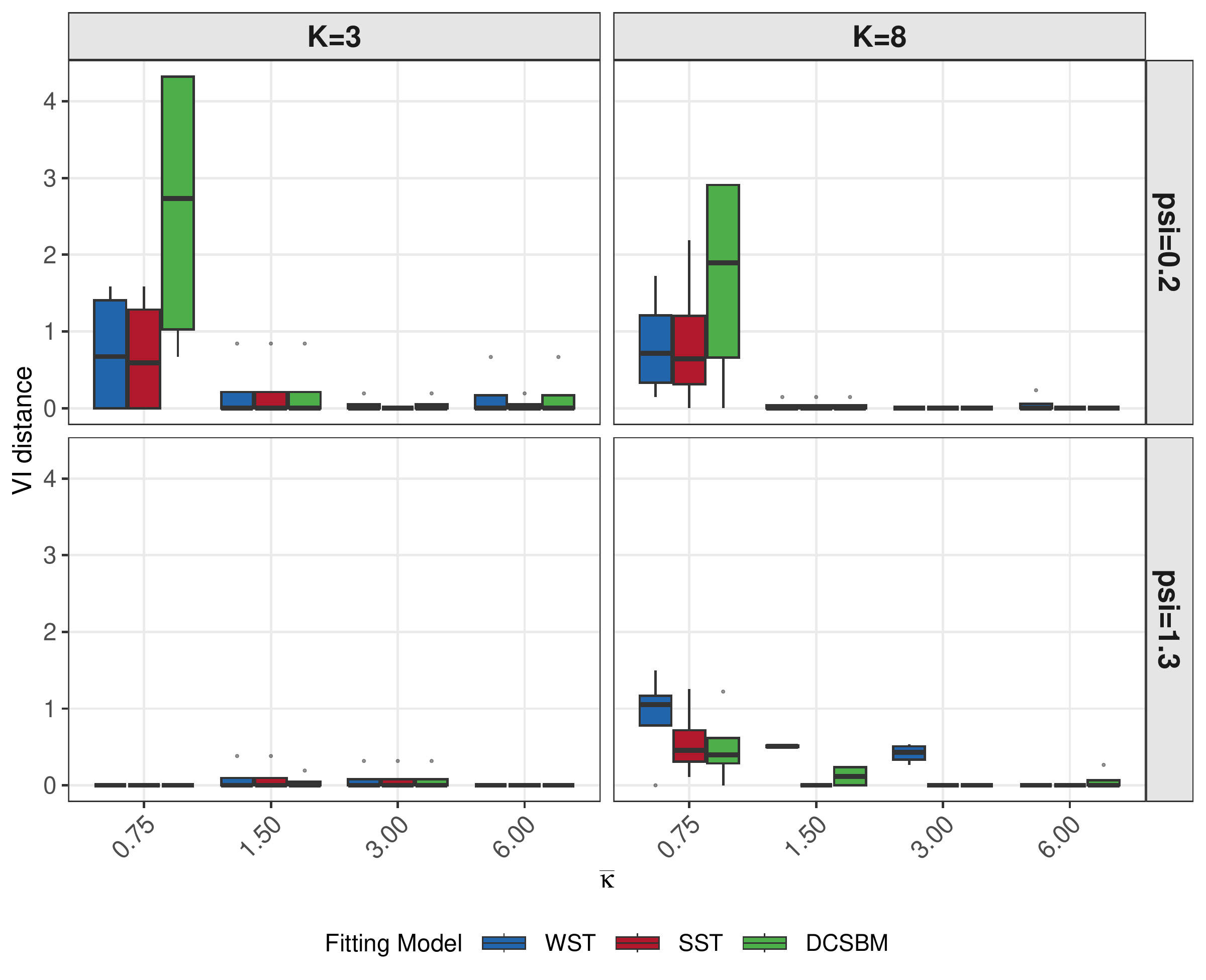}
\end{subfigure}
\caption{VI distance between the \texttt{minVI} partition estimate and the true partition for WST-generated (left) and SST-generated (right) scenarios. Each panel is a combination of $\psi^\star\in\{0.2,1.3\}$ rows and $K^\star\in\{3,8\}$ columns. Increasing $\kappa^\star$ values on the $x$-axis lead to more edges and thus, better signal. On the $y-$axis we have the VI distance from the true partition, where lower is better. Boxplots span $5$ replicates over the same scenario.}\label{fig:vi-boxplot-wst}
\end{figure}

The VI box plots in Figure~\ref{fig:vi-boxplot-wst} convey the same pattern. The largest gains of TSBM models over DC--SBM occur in sparse, weak-signal settings at higher $K^\star$. For example, when $K^\star=8$ under SST-generated data with sparse/weak signal, VI drops from $2.15$ (DC--SBM) to $1.57$ (best ordered model), a reduction of $0.58$ (about $27\%$). Under WST-generated data with sparse/weak signal, we can notice how in Table~\ref{tab:sim-main-two-scenarios} VI drops from $2.25$ to $1.64$ (reduction $0.61$, about $27\%$). The $K_n$ summaries tell the same story: in the sparse scenarios, TSBM models mildly underestimate $\hat{K}_n \approx5.3$--$5.8$, while DC--SBM does worse underestimating at $\hat{K}_n \approx3.5$ ($K^\star=8$ here). Predictive performance in Table~\ref{tab:sim-main-two-scenarios-loo} is consistent with the VI findings. 

Therefore, the simulation study shows that: (i) all three considered models, TSBM-WST, TSBM-$\mathrm{SST}$, and DC--SBM, perform similarly when the network is dense and the directional signal is strong; (ii) when the network is sparse and the signal is weak, the TSBM improves both predictive performance and partition recovery relative to the DC--SBM; and (iii) Appendix~\ref{app:sim-misspec} reports an additional study in which networks are generated with no global hierarchy at the block level, meaning with cycles built in. This is a misspecified scenario for the TSBM. The DC--SBM has a clear predictive advantage there and recovers the latent partition perfectly, while the two TSBM variants fail. This is expected and constitutes a useful diagnostic check.

\section{Applications}
\label{sec:applications}

We fit the DC--SBM, the WST--TSBM and the SST--TSBM to the six directed weighted networks in Table~\ref{tab:datasets}. Each edge is a count. In the animal dominance networks, as in Example~\ref{mot:animal}, $A_{ij}$ records how often an animal $i$ defeats or displaces $j$; in the citation and friendship networks, as in Examples~\ref{mot:citations} and~\ref{mot:friends}, it records the number of citations \emph{received} or the number of times $i$ is indicated as a friend from $j$, to maintain that the direction of the edge conveys a preference relation. 

\begin{table}[ht]
\centering
\small
\resizebox{\textwidth}{!}{%
\begin{tabular}{@{}p{3.2cm}p{6.2cm}cl@{}}
\toprule
Dataset & Edge meaning ($A_{ij}$) & $n$ & Reference \\
\midrule
Bighorn sheep dominance
& No.\ of dominance interactions in which sheep~$i$ beats sheep~$j$
& 28 & \citep{hass1991sheep,kunegisKONECTKoblenzNetwork2013} \\
\addlinespace
Spotted hyenas dominance
& No.\ of agonistic interactions won by hyena~$i$ against hyena~$j$
& 35 & \citep{strauss2022domarchive} \\
\addlinespace
Mountain goats dominance
& No.\ of agonistic encounters in which goat~$i$ dominates goat~$j$
& 45 & \citep{cote2000} \\
\addlinespace
Statistics journal citations
& No.\ of times journal~$i$ is cited by journal~$j$
& 47 & \citep{varinStatisticalModellingCitation2016} \\
\addlinespace
Japanese macaques dominance
& No.\ of dominance displays directed from macaque~$i$ to macaque~$j$
& 62 & \citep{fedigan1991,kutsukake2000} \\
\addlinespace
Illinois high-school friendships
& No.\ of times student~$i$ is indicated as a friend by student~$j$
& 70 & \citep{kunegisKONECTKoblenzNetwork2013, coleman1964introduction} \\
\bottomrule
\end{tabular}}
\caption{Directed weighted networks analysed in the application study.}
\label{tab:datasets}
\end{table}

All models use $10{,}000$ MCMC iterations, with the first $2{,}000$ discarded as burn-in. All hyperparameters are set like in the simulation study.

\begin{figure}[htbp]
\centering
\begin{subfigure}{.48\textwidth}
\centering
\includegraphics[width = \textwidth]{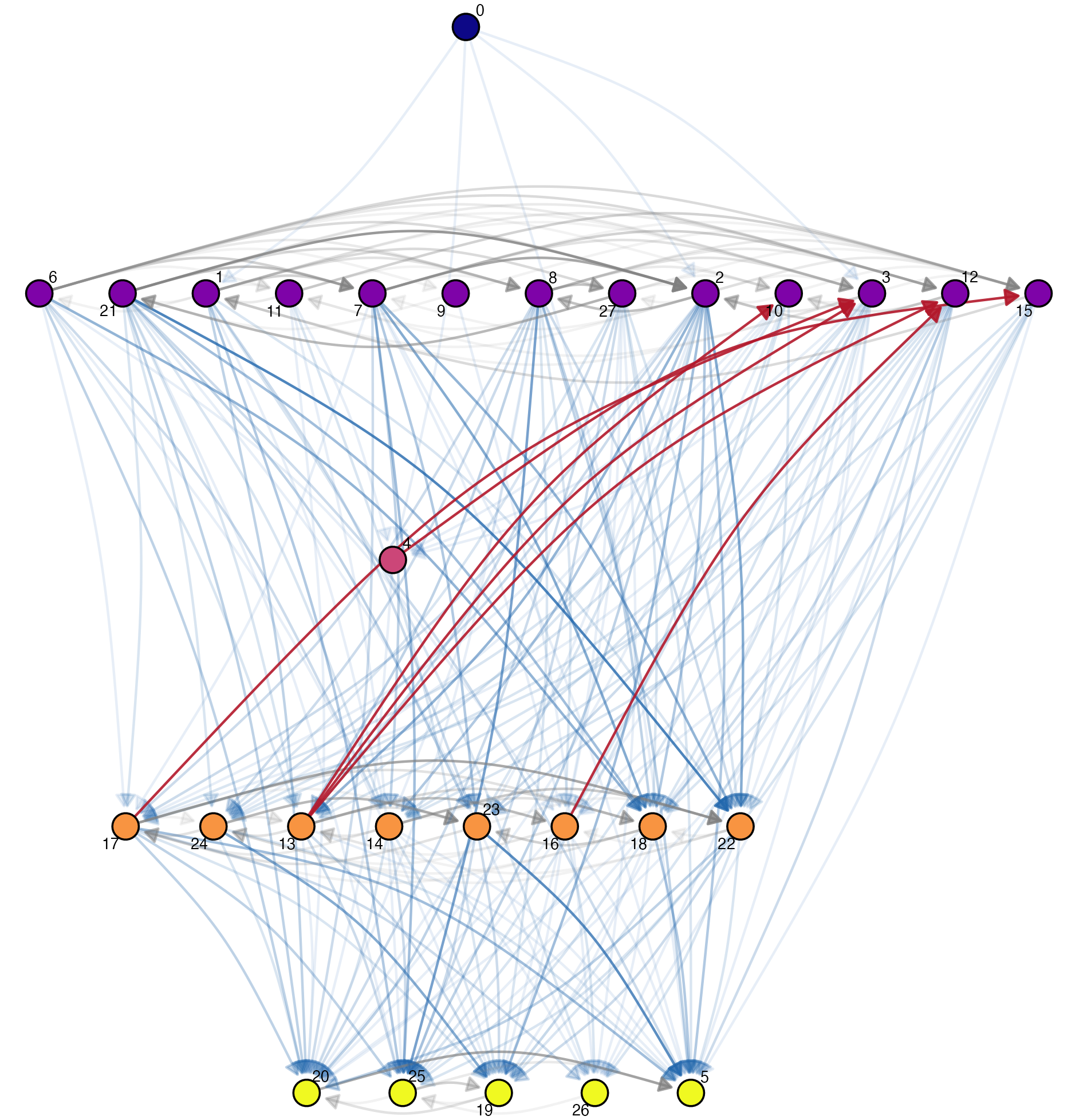}
\caption{$\mathrm{SST}$}
\label{fig:SST_sheep_tier_lineplot}
\end{subfigure}
\begin{subfigure}{.48\textwidth}
\centering
\includegraphics[width=\textwidth]{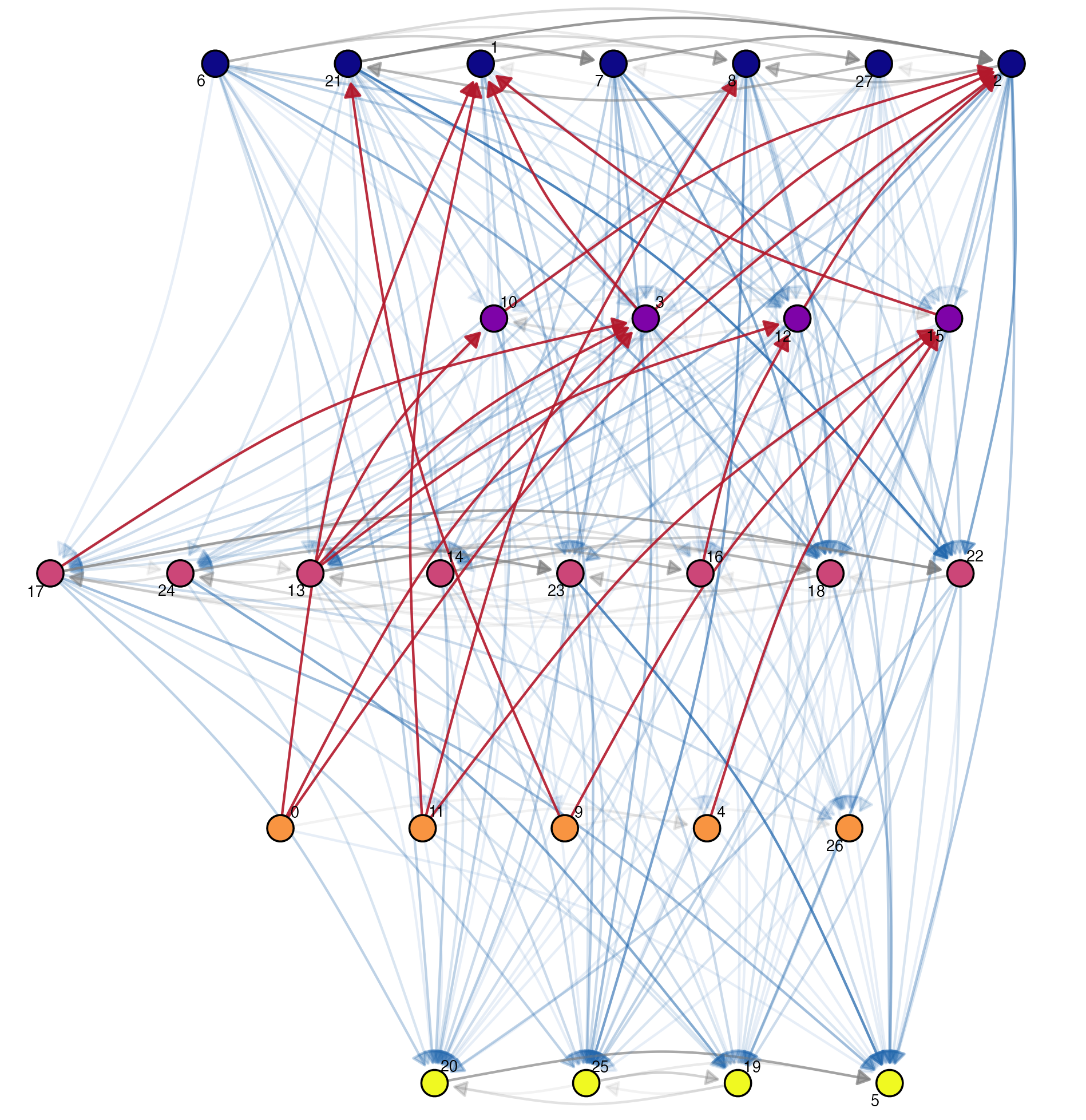}
\caption{DC--SBM}
\label{fig:DC-SBM_sheep_tier_lineplot}
\end{subfigure}
\caption{Partition point estimates for the bighorn sheep network \citep{hass1991sheep} under $\mathrm{SST}$ and DC--SBM. Notice that the number of blocks  does not coincide with the posterior mean \(\hat K_n\) reported in Table \ref{tab:model-selection}. Nodes are coloured by block, with the hierarchy running from top to bottom. Blue edges respect the ordering, while red edges point against it. The DC--SBM partition is relabelled so that layer 1 is the top block, and subsequent layers follow the induced ordering. The $\mathrm{SST}$ estimate has only 9 backward edges, $2.2\%$ of cross-block interactions, whereas DC--SBM gives a substantially less ordered partition, with 35 backward edges, $7.5\%$.}
\label{fig:tier_line_plot}
\end{figure}

\begin{table}[htpb]
\centering
\small
\caption[Posterior model-comparison summary]{Posterior summary of the occupied block count \(\hat K_n\), where \(\hat K_n\) denotes the posterior mean and the bracketed interval is the empirical \(95\%\) posterior credible interval, together with LOOIC computed on each dataset, model combination, and violation rate \eqref{eq:backward-flow-rate}. The winning model (in \textbf{bold}) has \(\Delta\mathrm{ELPD}=0\) by construction; the other two show \(\Delta\mathrm{ELPD}\) relative to the winner together with its standard error. A dagger marks rows for which \(|\Delta\mathrm{ELPD}|\ge 2\,\mathrm{SE}(\Delta\mathrm{ELPD})\), using the number of pointwise LOO terms for that dataset.}
\label{tab:model-selection}
\begingroup
\setlength{\tabcolsep}{3.5pt}
\begin{tabular}{llcrrrrr}
\toprule
Dataset & Model & $\hat{ K}_n\;[95\%\text{ CrI}]$ & LOOIC & $\Delta\text{ELPD}$ & SE($\Delta\text{ELPD}$)  & $\zeta_{\hat{\mathbf{z}}}^{\mathrm{viol}}$(back/cross)\\
\midrule
\multirow{3}{*}{Bighorn sheep} & \textbf{SST--TSBM} & $6\;[4,7]$ & $\mathbf{1{,}769.8}$ & $0$ & ---  &  2.2\% (9/400)\\
 & WST--TSBM & $6\;[4,7]$ & $1{,}826.2$ & $-28.2$ & $14.9$ & 4.8\% (19/393) \\
 & DC--SBM & $6\;[5,8]$ & $1{,}855.8$ & $\textbf{-43.0}^\dagger $ & $17.9$  &  7.5\% (35/466)\\
\addlinespace
\multirow{3}{*}{Spotted hyenas} & \textbf{WST--TSBM} & $14\;[13,15]$ & $\mathbf{3{,}885.9}$ & $0$ & ---  & 11.4\% (194/1{,}703)\\
 & SST--TSBM & $8\;[7,8]$ & $4{,}399.7$ & $\textbf{-256.9}^\dagger$ & $66.5$ & 11.4\% (166/1{,}461) \\
 & DC--SBM & $6\;[6,8]$ & $4{,}681.4$ & $\textbf{-397.8}^\dagger$ & $84.9$ & 10.4\% (135/1{,}299) \\
\addlinespace
\multirow{3}{*}{Mountain goats} & \textbf{SST--TSBM} & $8\;[6,8]$ & $\mathbf{2{,}515.2}$ & $0$ & ---   & 0.3\% (2/597) \\
 & DC--SBM & $4\;[4,6]$ & $2{,}590.8$ & $\textbf{-37.8}^\dagger$ & $8.3$   & 0.0\% (0/535)\\
 & WST--TSBM & $4\;[3,6]$ & $2{,}699.0$ & $\textbf{-91.9}^\dagger$ & $11.2$  & 0.4\% (2/547)\\
\addlinespace
\multirow{3}{*}{Stat.\ journals} & \textbf{DC--SBM} & $11\;[11,11]$ & $\mathbf{12{,}184.2}$ & $0$ & ---   & 31.5\% (3{,}914/12{,}416)\\
 & WST--TSBM & $14\;[12,14]$ & $12{,}496.3$ & $-156.1$ & $106.2$  & 29.6\% (3{,}608/12{,}188) \\
 & SST--TSBM & $5\;[5,5]$ & $17{,}460.5$ & $\textbf{-2{,}638.2}^\dagger$ & $313.5$ & 34.2\% (3{,}062/8{,}954)\\
\addlinespace
\multirow{3}{*}{Macaques} & \textbf{DC--SBM} & $11\;[11,12]$ & $\mathbf{6{,}187.1}$ & $0$ & ---  & 2.0\% (41/2026)\\
 & SST--TSBM & $10\;[9,11]$ & $6{,}354.1$ & $-83.5$ & $48.1$  & 0.6\% (11/1942)\\
 & WST--TSBM & $10\;[8,11]$ & $6{,}672.5$ & $\textbf{-242.7}^\dagger$ & $17.2$  & 0.9\% (17/1930) \\
\addlinespace
\multirow{3}{*}{High school} & \textbf{SST--TSBM} & $11\;[8,12]$ & $\mathbf{2{,}439.4}$ & $0$ & ---  & 18.8\% (29/154)\\
 & WST--TSBM & $9\;[7,11]$ & $2{,}456.5$ & $-8.5$ & $8.7$  & 29.4\% (40/136) \\
 & DC--SBM & $8\;[8,10]$ & $2{,}740.3$ & $\textbf{-150.4}^\dagger$ & $29.3$   & 46.1\% (59/128)\\
\bottomrule
\end{tabular}
\endgroup

\end{table}

Before detailing the results of our analysis, Figure~\ref{fig:tier_line_plot} shows an intuitive output for the bighorn sheep network, an instance of Example~\ref{mot:animal}. Panel~\ref{fig:SST_sheep_tier_lineplot} shows the SST--TSBM point estimate, while panel~\ref{fig:DC-SBM_sheep_tier_lineplot} shows the DC--SBM point estimate reordered by its induced hierarchy. Colours denote blocks; gray edges are within-block, blue edges respect the ordering, and red edges point against it.

The SST--TSBM estimate is visibly more ordered. Most remaining backward edges concentrate between blocks 2 and 4, which would be adjacent if the singleton block 3 were removed. This highlights a key difference between the posteriors under the ordered model and the DC--SBM: the former one concentrates reversals in a small local region, whereas the latter one produces a less coherent block ordering.

Table~\ref{tab:model-selection} reports the predictive comparison among the three fitted models. The TSBMs perform best on four networks: SST on sheep, goats and high school, and WST on hyenas. The DC--SBM performs best on citations and macaques, although in the citation network it is not statistically different from WST. In the macaque dataset, another instance of Example~\ref{mot:animal}, the number of reversals is so small that the hierarchy is almost perfect. As shown in the simulation study above, in this setting the DC--SBM can already recover the ordering without requiring additional prior regularisation.

The citation network in Example~\ref{mot:citations} illustrates a different situation, in which the TSBM is not selected. Here, stochastic transitivity is not strongly supported by cross-journal citation patterns, which explains why the TSBMs are less competitive despite producing partitions with substantially fewer backward edges.

\begin{figure}[htbp]
\centering
\begin{subfigure}{\textwidth}
\centering
\includegraphics[width=\textwidth]{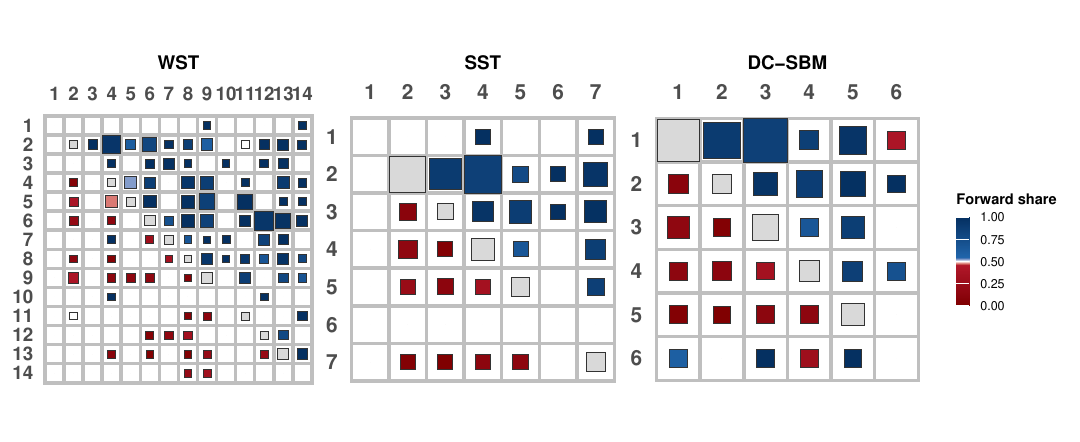}
\label{fig:block-flow-hyenas}
\end{subfigure}
\par\medskip
\caption{Empirical forward share for the spotted hyenas network ($n=35$) \citep{strauss2022domarchive}. The DC--SBM matrix is reordered so that block 1 is the strongest block, as in WST and $\mathrm{SST}$. Each square corresponds to a directed block pair $(k,\ell)$: its size is proportional to the directed edge count, and its colour gives the empirical forward share
$\tilde\rho_{k\ell}=C_{k\ell}/(C_{k\ell}+C_{\ell k})$,
where $C_{k\ell}$ is the numerator of the ratio in \eqref{eq:backward-flow-rate}, namely the number of edges from block $k$ to block $\ell$. The DC--SBM shows a clear cycle: the weakest block, block 6, sends more edges to block 1 than it receives from it, making the DC--SBM point-estimate partition incompatible with a global ordering.}
\label{fig:block-flow-animal}
\end{figure}

\begin{figure}[htbp]
\centering
\includegraphics[width=\textwidth]{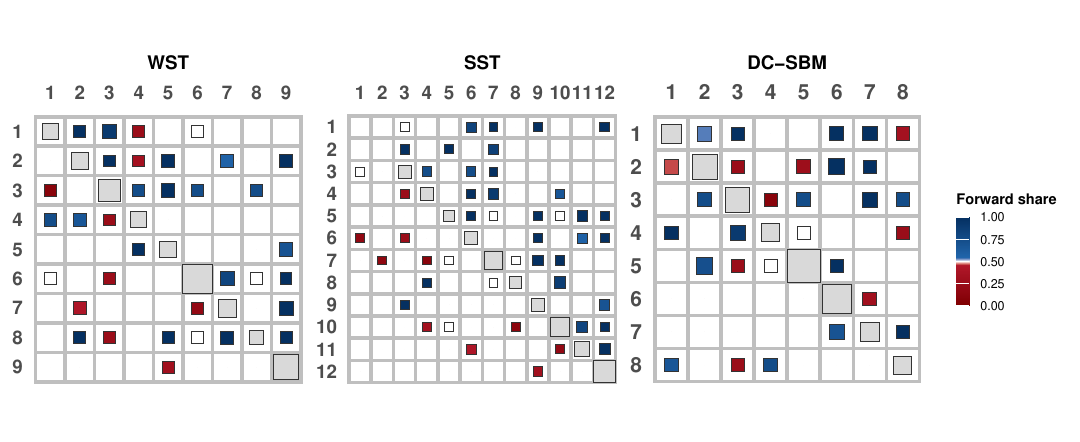}
\caption{Empirical forward share matrices for the high school network \citep{kunegisKONECTKoblenzNetwork2013, coleman1964introduction}. The DC--SBM model achieves an unordered partition incompatible with a global ordering. For a full description, see the caption of Figure~\ref{fig:block-flow-animal}.}
\label{fig:block-flow-human}
\end{figure}

To illustrate this point, we report two block-flow plots in Figures~\ref{fig:block-flow-animal} and~\ref{fig:block-flow-human} that refer to the VI point estimate of the partition. In Figure~\ref{fig:block-flow-animal} (hyenas network), both TSBM specifications recover a partition exhibiting a global order among blocks. By contrast, the DC--SBM orders blocks from 1 to 5, but block 6 wins against block 1, creating a cycle that is incompatible with the inferred ordering. This shows that the total number of backward edges, reported in the last column of Table~\ref{tab:model-selection}, should be interpreted with caution. A model may have fewer backward edges overall, while those reversals may still generate block-level cycles.

In Figure~\ref{fig:block-flow-human} (high school network), the comparison is even clearer. The DC--SBM partition is not consistent with an ordering, and appears to reflect community structure rather than hierarchy. The SST--TSBM, instead, recovers a block-level majority relation compatible with a complete ordering of all 12 blocks of the VI point estimate.
Appendix~\ref{app:application-details} reports additional cycle diagnostics for all applications.

\section{Discussion}
\label{sec:discussion}

This paper investigates when a directed weighted network is better described as a set
of \emph{ordered groups}, rather than as an unordered collection of communities
or a ranking of individual nodes. To address this question, it introduces two Transitive Stochastic Block Models (TSBMs) that combine two forms of structure. Nodes are clustered into
stochastically equivalent groups, as in a standard DC--SBM, while
the block-level directional probabilities are constrained to satisfy a form of
stochastic transitivity that induces a hierarchy among the groups. By jointly modelling grouping and hierarchy, the TSBMs treat both structures as inferential targets, rather than estimating one and imposing the other in a subsequent step.

The simulation studies and empirical applications show that jointly modelling clustering and ordering can improve both predictive performance and statistical inference. In particular, for sparse networks with a weak directional signal, TSBMs improve prediction and partition recovery by regularising the directional
component towards transitivity. The empirical applications reveal that the impact of this constraint depends on the network under study. In animal dominance networks such as Example~\ref{mot:animal}, TSBMs tend to recover more strongly ordered partitions and often achieve better predictive performance than the DC--SBM.
In the macaque network, the hierarchy is already highly pronounced and is almost compatible with a
complete node-level ordering. In this case, the DC--SBM achieves slightly better predictive performance, as accurate prediction
depends not only on capturing the direction of dominance but also on modelling the volume of interactions between blocks.
By contrast, in citation networks such as Example~\ref{mot:citations}, although a prestige gradient is apparent, the directional structure is not cleanly transitive. Consequently, imposing an ordered block hierarchy yields more limited gains in predictive performance.

The paper also contributes to Bayesian nonparametric inference for ordered partitions in latent variable models.
Because block labels encode rank, the predictive schemes associated with exchangeable Gibbs-type random partitions cannot be directly used to construct a collapsed Gibbs sampler for posterior inference. Instead, we exploit age-ordered random partitions, originally developed in population genetics, which naturally accommodate ordered clusters and allow prior beliefs about rank sizes to be incorporated. Age-ordered random partitions are a key ingredient in the derivation of a Gibbs sampler that provides a tractable approach to estimating the number of ordered groups, \(K\). Inference on \(K\) can be performed within the sampler itself, while avoiding information-criterion-based model selection or more complex trans-dimensional Metropolis--Hastings algorithms.

Several natural extensions of the proposed framework are possible. First, covariates could be incorporated into the model. In many applications, hierarchy is not only something to estimate but also something to explain: age, sex, size, field, seniority, or institutional position may help account for why some groups occupy higher ranks than others. Second, the computational strategy could be improved, as the current Gibbs sampler does not scale well to large networks. Third, more flexible specifications of strong stochastic transitivity could be developed, relaxing the Toeplitz structure adopted here.

More broadly, the Toeplitz specification illustrates how cumulative positive Gaussian increments can be used to construct structured priors while preserving tractable P\'olya--Gamma updates and allowing the number of blocks to be estimated within the model. The same principle could be adapted to encode other meso-scale network structures, such as core--periphery organisation, assortative mixing, or more complex forms of hierarchy, within a similar Bayesian nonparametric framework.

\paragraph*{Acknowledgements}

This publication has emanated from research conducted with the financial support of Taighde Éireann -- Research Ireland under Grant number 18/CRT/6049. The Insight Centre for Data Analytics is supported by Science Foundation Ireland under Grant Number 12/RC/2289$\_$P2.

For the purpose of Open Access, the author has applied a CC BY public copyright license to any Author Accepted Manuscript (AAM) version arising from this submission.

\paragraph{Reproducibility.}
Code and supporting materials to reproduce the simulation studies and empirical analyses reported in this paper will be made available at
\href{https://github.com/laposanti/Reproducibility-support-for-Ordered-SBMs}
{\texttt{GitHub Reproducibility Support}}.
The repository will contain the scripts required to fit the proposed ordered stochastic block models, reproduce the posterior summaries, and generate the figures and tables presented in the paper.

\printbibliography

\newpage 
\appendix

\paragraph*{Supplementary material to the Manuscript Entitled \comillas{Ordering Stochastic Block Models via prior transitivity}}

\section{Full conditional derivations}
\label{app:full-conditionals}

This appendix derives the full conditional updates used by the TSBM Gibbs sampler. Recall that:
\[
\mathcal I=\{(i,j):1\le i<j\le n\},
\qquad
N_{ij}=A_{ij}+A_{ji},
\qquad
\Abar_{ij}=A_{ij}-\frac12 N_{ij},
\qquad (i,j)\in\mathcal I.
\]
Thus \(\Abar_{ij}\) is a signed version of the directed count in the canonical dyad orientation \(i<j\). It is useful to keep this convention explicit, because the allocation update later considers dyads involving a fixed node \(i\), and that node need not be the first index of the unordered dyad.

Throughout, we write
\[
\Theta=(\bm\eta,\kappa,\psi),
\qquad
\mathcal M\in\{\mathrm{WST},\mathrm{SST}\},
\]
where \(\mathrm{SST}\) denotes the Toeplitz strong-stochastic-transitivity specification. For each \((i,j)\in\mathcal I\), define
\[
s_{ij}=\operatorname{sgn}(z_j-z_i),
\]
and
\begin{equation}
\label{eq:app-phi-def}
\phi_{ij}
=
\begin{cases}
0, & z_i=z_j,\\[3pt]
s_{ij}\psi_{z_i\wedge z_j,z_i\vee z_j}, & z_i\neq z_j,\ \mathcal M=\mathrm{WST},\\[3pt]
s_{ij}\psi_{|z_i-z_j|}, & z_i\neq z_j,\ \mathcal M=\mathrm{SST}.
\end{cases}
\end{equation}
Here \(a\wedge b=\min(a,b)\) and \(a\vee b=\max(a,b)\). The quantity \(\phi_{ij}\) is the directional log-odds that an interaction in the unordered dyad \((i,j)\), written in the canonical orientation \(i<j\), is oriented from \(i\) to \(j\). Hence
\[
A_{ij}\mid N_{ij},\mathbf{z},\psi
\sim
\operatorname{Binomial}\{N_{ij},\sigm(\phi_{ij})\},
\qquad
\sigm(x)=\operatorname{logit}^{-1}(x).
\]

The priors used below are
\[
\kappa_{k\ell}\overset{\mathrm{ind}}{\sim}
\operatorname{Gamma}(a_\kappa,b_\kappa),
\qquad k\le \ell,
\]
\[
\eta_i\overset{\mathrm{ind}}{\sim}
\operatorname{Gamma}(a_\eta,b_\eta),
\qquad i=1,\ldots,n,
\]
where Gamma distributions are parameterised by shape and rate. Under WST,
\[
\psi_{k\ell}\overset{\mathrm{ind}}{\sim}
\mathcal N^+(\mu_0,\sigma_0^2),
\qquad 1\le k<\ell\le K,
\]
with truncation to \([0,\infty)\). Under SST,
\[
\psi_d=\sum_{r=1}^d\delta_r,
\qquad d=1,\ldots,K-1,
\]
with
\[
\delta_r\overset{\mathrm{ind}}{\sim}
\mathcal N^+(0,\tau_0^2),
\qquad r=1,\ldots,K-1.
\]
This increment parameterisation enforces
\[
0\le \psi_1\le \psi_2\le\cdots\le\psi_{K-1}.
\]

\subsection{Augmented posterior kernel}

The Binomial directional likelihood can be written as
\[
p(A_{ij}\mid N_{ij},\phi_{ij})
\propto
\frac{\exp(A_{ij}\phi_{ij})}{\{1+\exp(\phi_{ij})\}^{N_{ij}}},
\]
where constants depending only on \(A_{ij}\) and \(N_{ij}\) have been omitted. The Pólya--Gamma identity gives
\[
\frac{\exp(A_{ij}\phi_{ij})}{\{1+\exp(\phi_{ij})\}^{N_{ij}}}
\propto
\exp\!\left\{\Abar_{ij}\phi_{ij}\right\}
\int_0^\infty
\exp\!\left\{-\frac12\omega_{ij}\phi_{ij}^2\right\}
p_{\mathrm{PG}}(\omega_{ij}\mid N_{ij},0)\,d\omega_{ij},
\]
where
\[
\Abar_{ij}=A_{ij}-\frac12N_{ij}.
\]
After augmentation, the joint posterior distribution is
\begin{align}
\label{eq:app-augmented-posterior}
p(\Theta,\mathbf{z},\omega\mid A,N,\mathcal M)
&\propto
\prod_{(i,j)\in\mathcal I}
\exp\!\left\{
-\eta_i\eta_j\kappa_{z_i z_j}
+
N_{ij}\bigl(\log\eta_i+\log\eta_j+\log\kappa_{z_i z_j}\bigr)
\right\}
\nonumber\\
&\quad\times
\prod_{(i,j)\in\mathcal I}
\exp\!\left\{
\Abar_{ij}\phi_{ij}
-\frac12\omega_{ij}\phi_{ij}^2
\right\}
p_{\mathrm{PG}}(\omega_{ij}\mid N_{ij},0)
\nonumber\\
&\quad\times
\left[
\prod_{1\le k\le \ell\le K}
\kappa_{k\ell}^{a_\kappa-1}
\exp(-b_\kappa\kappa_{k\ell})
\right]
\left[
\prod_{i=1}^n
\eta_i^{a_\eta-1}
\exp(-b_\eta\eta_i)
\right]
p_{\mathcal M}(\psi)
p_\vartheta(\mathbf{z}).
\end{align}
The normalising constants in the Gamma priors are omitted in \eqref{eq:app-augmented-posterior} because they do not affect ordinary full conditionals. They will be restored later when we integrate \(\kappa\) out in the partially collapsed allocation update, where the number of integrated block-pair parameters can change.

\subsection{Full conditional for the Pólya--Gamma variables}

Conditional on the current value of \(\phi_{ij}\), the Pólya--Gamma full conditional is
\begin{equation}
\label{eq:app-pg-update}
\omega_{ij}\mid\text{rest}
\sim
\operatorname{PG}(N_{ij},\phi_{ij}),
\qquad (i,j)\in\mathcal I.
\end{equation}
Since \(\operatorname{PG}(b,c)\) is symmetric in \(c\), this is equivalently written as
\[
\omega_{ij}\mid\text{rest}
\sim
\operatorname{PG}(N_{ij},|\phi_{ij}|).
\]
The absolute value is a computational convenience rather than a change in the model.

\subsection{Full conditional for the volume intensities \texorpdfstring{\(\kappa\)}{kappa}}

The block-pair sufficient statistics for the Poisson volume component in
\eqref{eq:dyadic-volume-direction} are
\[
R_{k\ell}
=
\sum_{(i,j)\in\mathcal I_{k\ell}(\mathbf{z})}N_{ij},
\qquad
T_{k\ell}
=
\sum_{(i,j)\in\mathcal I_{k\ell}(\mathbf{z})}\eta_i\eta_j.
\]
Retaining only the terms depending on \(\kappa_{k\ell}\), the kernel is
\[
\kappa_{k\ell}^{R_{k\ell}}
\exp(-T_{k\ell}\kappa_{k\ell})
\kappa_{k\ell}^{a_\kappa-1}
\exp(-b_\kappa\kappa_{k\ell}).
\]
Therefore
\begin{equation}
\label{eq:app-kappa-update}
\kappa_{k\ell}\mid\text{rest}
\sim
\operatorname{Gamma}
\left(
a_\kappa+R_{k\ell},
\;
b_\kappa+T_{k\ell}
\right),
\qquad 1\le k\le \ell\le K.
\end{equation}
This update is unaffected by the WST or SST choice, because the ordering restriction acts only on the directional component.

\subsection{Full conditional for the degree propensities \texorpdfstring{\(\eta_i\)}{eta-i}}

For node \(i\), the terms in the volume likelihood involving \(\eta_i\) are
\[
\prod_{j\ne i}
\exp\!\left\{-\eta_i\eta_j\kappa_{z_i z_j}\right\}
\eta_i^{N_{ij}},
\]
where \(N_{ij}=N_{ji}=A_{ij}+A_{ji}\). Combining this with the Gamma prior gives
\[
\eta_i^{a_\eta-1+\sum_{j\ne i}N_{ij}}
\exp\!\left\{
-\eta_i
\left(
b_\eta+\sum_{j\ne i}\eta_j\kappa_{z_i z_j}
\right)
\right\}.
\]
Hence
\begin{equation}
\label{eq:app-eta-update}
\eta_i\mid\text{rest}
\sim
\operatorname{Gamma}
\left(
a_\eta+G_i,
\;
b_\eta+H_i
\right),
\end{equation}
where
\[
G_i=\sum_{j\ne i}N_{ij},
\qquad
H_i=\sum_{j\ne i}\eta_j\kappa_{z_i z_j}.
\]
Again, this update is common to WST and SST.

\subsection{Directional log-odds update under WST}

For \((i,j)\in\mathcal I_{k\ell}(\mathbf{z})\), the WST directional predictor is
\[
\phi_{ij}
=
\operatorname{sgn}(z_j-z_i)\psi_{k\ell}.
\]
Since \(\operatorname{sgn}(z_j-z_i)^2=1\) for cross-block dyads, the part of the augmented log-likelihood \eqref{eq:aug-dyad} involving \(\psi_{k\ell}\) is
\[
\bar y_{k\ell}\psi_{k\ell}
-
\frac12\bar\omega_{k\ell}\psi_{k\ell}^2,
\]
where
\begin{equation}
\label{eq:app-wst-suff-stats}
\bar y_{k\ell}
=
\sum_{(i,j)\in\mathcal I_{k\ell}(\mathbf{z})}
\operatorname{sgn}(z_j-z_i)\Abar_{ij},
\qquad
\bar\omega_{k\ell}
=
\sum_{(i,j)\in\mathcal I_{k\ell}(\mathbf{z})}
\omega_{ij}.
\end{equation}
Combining this likelihood kernel with the Gaussian prior gives
\begin{align}
\log p(\psi_{k\ell}\mid\text{rest})
&\propto
\bar y_{k\ell}\psi_{k\ell}
-
\frac12\bar\omega_{k\ell}\psi_{k\ell}^2
-
\frac{(\psi_{k\ell}-\mu_0)^2}{2\sigma_0^2}
+
\mathrm{const}
\nonumber\\
&=
-\frac12
\left(
\bar\omega_{k\ell}+\sigma_0^{-2}
\right)
\psi_{k\ell}^2
+
\left(
\bar y_{k\ell}+\sigma_0^{-2}\mu_0
\right)
\psi_{k\ell}
+
\mathrm{const}.
\end{align}
Completing the square, define
\[
v_{k\ell}
=
\left(
\bar\omega_{k\ell}+\sigma_0^{-2}
\right)^{-1},
\qquad
m_{k\ell}
=
v_{k\ell}
\left(
\bar y_{k\ell}+\sigma_0^{-2}\mu_0
\right).
\]
The WST constraint is \(\psi_{k\ell}\ge0\). Hence
\begin{equation}
\label{eq:app-wst-psi-update}
\psi_{k\ell}\mid\text{rest},\mathcal M=\mathrm{WST}
\sim
\mathcal N^+(m_{k\ell},v_{k\ell}),
\qquad k<\ell.
\end{equation}
If \(\mathcal I_{k\ell}(\mathbf{z})\) is empty, then \(\bar y_{k\ell}=0\) and \(\bar\omega_{k\ell}=0\), so \eqref{eq:app-wst-psi-update} reduces to the prior \(\mathcal N^+(\mu_0,\sigma_0^2)\), as it should.

\subsection{Directional log-odds update under SST}
\label{app:toep-directional-update}

We now derive the full conditional update for the directional parameters under the
SST specification. This is the detailed version of the update summarised in
the directional-parameter step of Algorithm~\ref{alg:TSBM-gibbs-compact}. The
notation follows the volume/direction decomposition in
Section~\ref{sec:likelihood-decomp}, especially
\eqref{eq:dyadic-volume-direction}, and the SST prior introduced in
Section~\ref{sec:psi_prior}. The starting point is the Pólya--Gamma augmented
directional likelihood in \eqref{eq:aug-dyad}.

Recall that, for each unordered dyad \((i,j)\in\mathcal I=\{(i,j):1\le i<j\le n\}\),
the directional likelihood is written in terms of the centred directed count
\[
\Abar_{ij}
=
A_{ij}-\frac{N_{ij}}{2},
\qquad
N_{ij}=A_{ij}+A_{ji}.
\]
After introducing the Pólya--Gamma variable \(\omega_{ij}\),
\eqref{eq:aug-dyad} gives the augmented directional kernel
\[
\exp\left\{
\Abar_{ij}\varphi_{ij}
-
\frac12\omega_{ij}\varphi_{ij}^2
\right\},
\]
up to factors that do not depend on the directional log-odds. Under the SST
model, defined in \eqref{eq:sst-rho-distance}, the linear predictor is
\[
\phi_{ij}
=
\operatorname{sgn}(z_j-z_i)\psi_{|z_i-z_j|},
\qquad z_i\ne z_j.
\]
The sign term records whether the direction \(i\to j\) agrees with the current
ordering of the blocks. Since the SST model assigns a common log-odds
\(\psi_d\) to all block pairs separated by rank distance \(d\), observations can be
collected by distance.

For \(d=1,\ldots,K-1\), define
\[
\mathcal I_d(\mathbf{z})
=
\left\{
(i,j)\in\mathcal I:
|z_i-z_j|=d
\right\}.
\]
The corresponding distance-wise sufficient statistics are
\begin{equation}
\label{eq:app-toep-suff-stats}
\bar y_d
=
\sum_{(i,j)\in\mathcal I_d(\mathbf{z})}
\operatorname{sgn}(z_j-z_i)\Abar_{ij},
\qquad
\bar\omega_d
=
\sum_{(i,j)\in\mathcal I_d(\mathbf{z})}
\omega_{ij}.
\end{equation}
The statistic \(\bar y_d\) is a signed directional imbalance at distance \(d\). It is
positive when, after centring by \(N_{ij}/2\), the observed directions tend to follow
the block order, and negative when they tend to go against it. The statistic
\(\bar\omega_d\) is the Pólya--Gamma precision accumulated over all dyads at that
distance. Notice that the quadratic term does not retain the sign, because
\(\operatorname{sgn}(z_j-z_i)^2=1\).

Summing \eqref{eq:aug-dyad} over dyads and collecting terms by distance gives
the SST directional kernel
\begin{equation}
\label{eq:app-toep-psi-kernel}
\sum_{d=1}^{K-1}
\left[
\bar y_d\psi_d
-
\frac12\bar\omega_d\psi_d^2
\right].
\end{equation}
This is the key simplification: conditionally on the Pólya--Gamma variables and on
the allocation \(\mathbf{z}\), the logistic likelihood has become a Gaussian quadratic
kernel in the distance-level log-odds \(\psi_1,\ldots,\psi_{K-1}\).

The SST prior in \eqref{eq:sst-monotone-psi} imposes
\[
0\le \psi_1\le \psi_2\le\cdots\le \psi_{K-1}.
\]
Using the increments defined in \eqref{eq:sst-prior2}, we enforce this monotonicity by writing the ordered
log-odds as cumulative sums of non-negative increments. Let
\[
\delta=(\delta_1,\ldots,\delta_{K-1})^\top,
\qquad
\delta_r\ge 0,
\]
and let \(S\) be the \((K-1)\times(K-1)\) lower-triangular cumulative-sum matrix
\[
S_{dr}=\ind\{r\le d\}.
\]
Then
\[
\bm\psi=S\delta,
\qquad
\psi_d=\sum_{r=1}^d\delta_r.
\]
Thus \(\delta_r=\psi_r-\psi_{r-1}\), with the convention \(\psi_0=0\). The increment
\(\delta_r\) measures how much extra forward log-odds is gained when moving from
rank distance \(r-1\) to rank distance \(r\). Now define
\[
\bar{\bm y}
=
(\bar y_1,\ldots,\bar y_{K-1})^\top,
\qquad
\bm\Omega
=
\operatorname{diag}(\bar\omega_1,\ldots,\bar\omega_{K-1}).
\]
Using \(\bm\psi=S\delta\), the likelihood kernel in
\eqref{eq:app-toep-psi-kernel} becomes
\[
\bar{\bm y}^{\top}S\delta
-
\frac12
\delta^\top S^\top\bm\Omega S\delta.
\]
Equivalently,
\[
-\frac12\delta^\top Q\delta
+
g^\top\delta,
\qquad
Q=S^\top\bm\Omega S,
\qquad
g=S^\top\bar{\bm y}.
\]
The entries of \(Q\) and \(g\) have a simple cumulative interpretation. Since
\(S_{dr}=\ind\{r\le d\}\),
\[
Q_{rq}
=
\sum_{d=1}^{K-1}
\bar\omega_d S_{dr}S_{dq}
=
\sum_{d=\max(r,q)}^{K-1}
\bar\omega_d,
\]
and
\[
g_r
=
\sum_{d=1}^{K-1}
S_{dr}\bar y_d
=
\sum_{d=r}^{K-1}
\bar y_d.
\]
Hence the \(r\)-th increment affects all distances \(d\ge r\), but none of the
distances \(d<r\). This is the only point in the derivation where the cumulative-sum
parametrisation really matters. It is also the reason why the conditional update for
\(\delta_r\) depends on all larger distances, not only on distance \(r\).

The prior on the increments is independent half-Normal,
\[
\delta_r\stackrel{\mathrm{ind}}{\sim}\mathcal N^+(0,\tau_0^2),
\qquad r=1,\ldots,K-1,
\]
as stated in \eqref{eq:sst-prior2}. Ignoring the normalising constants of the
truncated priors, this contributes
\[
-\frac12\tau_0^{-2}\delta^\top\delta,
\qquad
\delta_r\ge 0,\quad r=1,\ldots,K-1.
\]
Combining the likelihood kernel and the prior kernel gives
\begin{equation}
\label{eq:app-toep-delta-joint}
p(\delta\mid\text{rest},\mathcal M=\mathrm{SST})
\propto
\exp\!\left\{
-\frac12\delta^\top Q^\ast\delta
+
g^\top\delta
\right\}
\ind\{\delta_1,\ldots,\delta_{K-1}\ge0\},
\end{equation}
where
\[
Q^\ast
=
Q+\tau_0^{-2}I_{K-1}.
\]
Thus the posterior for the increment vector is a multivariate Gaussian kernel
restricted to the positive orthant. Directly sampling from this truncated
multivariate distribution is possible, but a component-wise Gibbs update is simpler
and preserves the full conditional tractability mentioned in
Section~\ref{sec:psi_prior}.

To update a single increment \(\delta_r\), keep \(\delta_{-r}\) fixed and isolate the
terms in \eqref{eq:app-toep-delta-joint} that depend on \(\delta_r\):
\[
-\frac12Q^\ast_{rr}\delta_r^2
-
\delta_r\sum_{q\ne r}Q^\ast_{rq}\delta_q
+
g_r\delta_r
+
\mathrm{const}.
\]
Since the prior contribution is diagonal, \(Q^\ast_{rq}=Q_{rq}\) whenever \(q\ne r\).
Moreover,
\[
Q^\ast_{rr}
=
\tau_0^{-2}+Q_{rr}
=
\tau_0^{-2}
+
\sum_{d=r}^{K-1}\bar\omega_d,
\]
and
\[
g_r
=
\sum_{d=r}^{K-1}\bar y_d.
\]
It remains to rewrite the cross-product term. For fixed \(\delta_{-r}\), define the
distance-\(d\) log-odds with the \(r\)-th increment removed as
\[
\psi_d^{(-r)}
=
\sum_{\substack{q=1\\ q\ne r}}^d\delta_q
=
\psi_d-\delta_r\ind\{r\le d\}.
\]
For \(d\ge r\), this is simply \(\psi_d-\delta_r\); for \(d<r\), the increment
\(\delta_r\) does not enter \(\psi_d\). Using the cumulative form of \(Q\),
\[
\sum_{q\ne r}Q^\ast_{rq}\delta_q
=
\sum_{q\ne r}
\delta_q
\sum_{d=\max(r,q)}^{K-1}\bar\omega_d
=
\sum_{d=r}^{K-1}
\bar\omega_d
\sum_{\substack{q=1\\q\ne r}}^d\delta_q
=
\sum_{d=r}^{K-1}
\bar\omega_d\psi_d^{(-r)}.
\]
This expression is often the cleanest way to implement the update, because it removes the current contribution of \(\delta_r\) from every affected distance, computes
the residual directional imbalance, and then samples the new increment from a
one-dimensional truncated Gaussian.

Define
\[
c_r
=
Q^\ast_{rr}
=
\tau_0^{-2}
+
\sum_{d=r}^{K-1}\bar\omega_d,
\]
and
\[
b_r
=
g_r-\sum_{q\ne r}Q^\ast_{rq}\delta_q
=
\sum_{d=r}^{K-1}
\left(
\bar y_d-\bar\omega_d\psi_d^{(-r)}
\right).
\]
Then the terms involving \(\delta_r\) can be written as
\[
-\frac12c_r\delta_r^2+b_r\delta_r
=
-\frac12c_r
\left(
\delta_r-\frac{b_r}{c_r}
\right)^2
+
\mathrm{const}.
\]
Therefore,
\[
\mu_r=\frac{b_r}{c_r},
\qquad
v_r=c_r^{-1},
\]
and the full conditional distribution is
\begin{equation}
\label{eq:app-toep-delta-update}
\delta_r
\mid
\delta_{-r},\text{rest},\mathcal M=\mathrm{SST}
\sim
\mathcal N^+(\mu_r,v_r),
\qquad r=1,\ldots,K-1.
\end{equation}
Here \(\mathcal N^+(\mu_r,v_r)\) denotes a Gaussian distribution with mean parameter
\(\mu_r\) and variance \(v_r\), truncated to \((0,\infty)\). 

Finally, consider the degenerate case in which no directional information is available
for any distance affected by \(\delta_r\). If
\[
\sum_{d=r}^{K-1}\bar\omega_d=0,
\]
then all dyads contributing to distances \(d\ge r\) have \(N_{ij}=0\), hence also
\(\bar y_d=0\) for those distances. Consequently,
\[
c_r=\tau_0^{-2},
\qquad
b_r=0,
\]
and the update reduces to
\[
\delta_r\mid\text{rest}\sim\mathcal N^+(0,\tau_0^2).
\]
In words, if the data provide no information about the distances affected by the
increment \(\delta_r\), the sampler correctly falls back to the half-Normal prior.

\subsection{Ordered allocation update: exact probabilities up to normalisation}
\label{app:z-full-conditional}

This subsection explains how the probabilities in
\eqref{eq:z-update-existing-weight} and \eqref{eq:z-update-new-weight} are
computed. Throughout, the comparison is always made against the state obtained after temporarily removing node \(i\). We reserve \(i,j,u,v\) for node indices and
\(k,\ell,r,a,b\) for ordered-block labels. This separation is useful because the update
is node-wise, but the resulting likelihood ratios are most naturally written by
aggregating dyads at the block level.

\subsubsection{The network after removing node \texorpdfstring{$i$}{i} and the set of candidate allocations}

After node \(i\) has been removed, let \(K_{-i}\) be the number of occupied blocks
where the ordered-CRP prior term is evaluated after removing node $i$. There are then two classes of probabilities to
evaluate:
\[
\Pr(z_i=k^{\mathrm{old}}\mid z_{-i},\text{rest}),
\qquad k\in[K_{-i}],
\]
for assignment to an existing block, and
\[
\Pr(z_i=r^{\mathrm{new}}\mid z_{-i},\text{rest}),
\qquad r\in[K_{-i}+1],
\]
for creation of a new singleton block at insertion slot \(r\).

If removing node \(i\) empties its current block \(h\), that block is deleted
before any probability is evaluated and the surviving blocks are relabelled.
Under WST, rows and columns of \(\kappa\) and \(\psi\) involving the deleted
block are removed and the surviving pair-specific entries are carried to their
new labels. Under SST, the volume parameters are contracted in the same way,
but the directional parameters remain indexed by distance in the order. Hence deleting a
block changes the distance of old block pairs that straddled \(h\), and the
largest Toeplitz distance disappears together with its terminal increment.
Accordingly, under SST the distance-indexed pooled quantities are recomputed
after deletion, or equivalently updated by moving straddling old-old pairs to
their new distances.

If a singleton is inserted in slot \(r\), an old block \(\ell\) moves to
\[
h_r(\ell)
=
\ell+\ind\{\ell\ge r\},
\qquad
\ell=1,\ldots,K_{-i}.
\]
The map \(h_r\) is needed only in the new-block case: it records how the old
blocks are relabelled after the singleton is inserted.

For quick reference, the quantities that appear repeatedly below are
\[
\begin{array}{ll}
K_{-i}: & \text{number of occupied blocks after removing node } i,\\
h_r(\ell): & \text{new label of old block } \ell \text{ after insertion at slot } r,\\
N_{i\ell}: & \text{total dyadic volume between node } i \text{ and block } \ell,\\
\eta_{-i,\ell}: & \text{sum of degree propensities in block } \ell,\\
R_{ab}^{(-i)},\,T_{ab}^{(-i)}: & \text{old--old volume totals and exposures for block pair } (a,b),\\
\Abar_{i\ell},\,\bar\omega_{i\ell}: & \text{pooled directional imbalance and P\'olya--Gamma precision for node } i \text{ versus block } \ell,\\
\bar A_{id}^{(k)},\,\bar\omega_{id}^{(k)}: & \text{SST versions after pooling by distance } d \text{ from candidate block } k.
\end{array}
\]

\subsubsection{Probability of assigning node \texorpdfstring{$i$}{i} to an existing block}

For an existing block \(k\in[K_{-i}]\), the probability in
\eqref{eq:z-update-full-conditional} can be written as
\begin{equation}
\Pr(z_i=k^{\mathrm{old}}\mid z_{-i},\text{rest})
\propto
\Pr(z_i=k^{\mathrm{old}}\mid z_{-i})
\exp\!\left\{
\ell_{i,\mathrm{vol}}^{\mathrm{old}}(k)
+
\ell_{i,\mathrm{dir}}^{\mathrm{old}}(k;\mathcal M)
\right\}.
\label{eq:app-existing-probability}
\end{equation}
The two terms in the exponent come from the volume and directional parts of the
likelihood ratio. We derive them in turn.

\paragraph{Volume contribution.}
Only the block pairs involving the candidate block \(k\) can differ between the
numerator and denominator. This is why it is enough to keep track of
total amount of edges between existing blocks (after excluding node $i$) and of the additional contribution of node \(i\) to
each existing block. For \(1\le a\le b\le K_{-i}\), define
\begin{align}
R_{ab}^{(-i)}
&:=
\sum_{\substack{u<v,\;u,v\ne i\\ z_u\wedge z_v=a,\;z_u\vee z_v=b}}N_{uv},
&
T_{ab}^{(-i)}
&:=
\sum_{\substack{u<v,\;u,v\ne i\\ z_u\wedge z_v=a,\;z_u\vee z_v=b}}\eta_u\eta_v,
\nonumber\\
N_{i\ell}
&:=
\sum_{\substack{j\ne i\\ z_j=\ell}}N_{ij},
&
\eta_{-i,\ell}
&:=
\sum_{\substack{j\ne i\\ z_j=\ell}}\eta_j.
\label{eq:app-volume-local-sufficient-statistics}
\end{align}
Here \(R_{ab}^{(-i)}\) and \(T_{ab}^{(-i)}\) denote the block-pair sufficient statistics computed under the ordered partition obtained after removing node \(i\). In particular, \(R_{ab}^{(-i)}\) is the total observed interaction volume currently allocated to block pair \((a,b)\), while \(T_{ab}^{(-i)}\) is the corresponding total degree-correction exposure. Dyads involving node \(i\) are excluded from both quantities and are added separately when evaluating each candidate reassignment of \(z_i\). The pair \((N_{i\ell},\eta_i\eta_{-i,\ell})\) is what node \(i\) would add if it were attached to block \(\ell\).

For one block pair, integrating \(\kappa\) against its Gamma prior gives
\begin{align}
M_\kappa(R,T)
&:=
\int_0^\infty
\kappa^R e^{-T\kappa}
\frac{b_\kappa^{a_\kappa}}{\Gamma(a_\kappa)}
\kappa^{a_\kappa-1}e^{-b_\kappa\kappa}\,d\kappa
\nonumber\\
&=
\frac{b_\kappa^{a_\kappa}}{\Gamma(a_\kappa)}
\frac{\Gamma(a_\kappa+R)}{(b_\kappa+T)^{a_\kappa+R}},
\qquad
M_\kappa(0,0)=1.
\label{eq:app-M-kappa}
\end{align}
Thus \(M_\kappa(R,T)\) is the collapsed Poisson--Gamma contribution of one block
pair with total count \(R\) and total exposure \(T\).

With these definitions, every block-pair factor not involving \(k\) cancels
between numerator and denominator in \eqref{eq:z-update-full-conditional}. The
dyad-specific factors
\(\prod_{j\ne i}(\eta_i\eta_j)^{N_{ij}}/N_{ij}!\) also cancel because they do
not depend on the candidate block. The surviving volume ratio is therefore
\begin{equation}
\prod_{\ell=1}^{K_{-i}}
\frac{
M_\kappa\!\left(
R_{k\wedge\ell,k\vee\ell}^{(-i)}+N_{i\ell},
T_{k\wedge\ell,k\vee\ell}^{(-i)}+\eta_i\eta_{-i,\ell}
\right)
}{
M_\kappa\!\left(
R_{k\wedge\ell,k\vee\ell}^{(-i)},
T_{k\wedge\ell,k\vee\ell}^{(-i)}
\right)
}.
\label{eq:app-volume-existing-ratio}
\end{equation}
Taking logs gives \(\ell_{i,\mathrm{vol}}^{\mathrm{old}}(k)\).

\paragraph{Directional contribution under WST.}
For the directional part, the basic quantity \(\Abar_{ij}\) was defined in the
canonical dyad orientation \(i<j\), because the likelihood is written on
unordered dyads. In the allocation update, however, node \(i\) is fixed and the
affected dyads can appear in either order: some have \(i<j\), others have
\(j<i\). If we kept only the notation \(\Abar_{ij}\), every pooled sum over a
block would have to be split into these two cases. It is therefore convenient
to re-express the same centred count in the common orientation ``from node
\(i\) to node \(j\)'' and write
\begin{align}
\Abar_{i\to j}
&=
\begin{cases}
\Abar_{ij}, & i<j,\\[2pt]
-\Abar_{ji}, & j<i,
\end{cases}
\nonumber\\
\Abar_{i\ell}
&:=
\sum_{\substack{j\ne i\\ z_j=\ell}}\Abar_{i\to j},
\qquad
\bar\omega_{i\ell}
:=
\sum_{\substack{j\ne i\\ z_j=\ell}}\omega_{i\wedge j,i\vee j},
\qquad
s_{k\ell}
:=
\operatorname{sgn}(\ell-k).
\label{eq:app-node-to-rank-dir-stats}
\end{align}
This arrow notation does not introduce a new observable: it is only a
bookkeeping device for the already-defined centred dyadic count. A positive
\(\Abar_{i\to j}\) means that, after centring by \(N_{ij}/2\), the observed
interaction in dyad \(\{i,j\}\) points more from \(i\) to \(j\) than from \(j\)
to \(i\); a negative value means the reverse. Its role is to keep the focal
node \(i\) in the sender position throughout the update, so that all
contributions from node \(i\) to block \(\ell\) can be pooled in the single
quantity \(\Abar_{i\ell}\), and under SST can then be pooled again by distance.
The quantity \(\Abar_{i\ell}\) is the pooled directional signal from node \(i\)
toward block \(\ell\), whereas \(\bar\omega_{i\ell}\) is the corresponding
P\'olya--Gamma precision. The sign \(s_{k\ell}\) records whether block \(\ell\)
is above or below the candidate block \(k\) in the ordering.

Under WST, dyads from node \(i\) to the same block have \(\phi_{ij}=0\), so only
\(\ell\ne k\) contribute. Hence
\begin{equation}
\ell_{i,\mathrm{dir}}^{\mathrm{old}}(k;\mathrm{WST})
=
\sum_{\substack{\ell=1\\\ell\ne k}}^{K_{-i}}
\left[
s_{k\ell}\Abar_{i\ell}\psi_{k\wedge\ell,k\vee\ell}
-
\frac12\bar\omega_{i\ell}\psi_{k\wedge\ell,k\vee\ell}^{\,2}
\right].
\label{eq:app-dir-old-wst-log}
\end{equation}

\paragraph{Directional contribution under SST.}
Under SST, the directional parameters are indexed by distance in the ordering
rather than by block pair. We therefore need to pool the same node-to-block
contributions by their distance from \(k\):
\begin{equation}
\bar A_{id}^{(k)}
:=
\sum_{\substack{\ell=1\\|\ell-k|=d}}^{K_{-i}}
s_{k\ell}\Abar_{i\ell},
\qquad
\bar\omega_{id}^{(k)}
:=
\sum_{\substack{\ell=1\\|\ell-k|=d}}^{K_{-i}}
\bar\omega_{i\ell}.
\label{eq:app-dir-old-toep-pooled}
\end{equation}
Here \(\bar A_{id}^{(k)}\) is the total signed imbalance between node \(i\) and
all blocks at distance \(d\) from \(k\), and \(\bar\omega_{id}^{(k)}\) is the
matching precision. No old-old distance changes occur in the existing-block
case, so these are the only SST quantities that are needed:
\begin{equation}
\ell_{i,\mathrm{dir}}^{\mathrm{old}}(k;\mathrm{SST})
=
\sum_{d=1}^{K_{-i}-1}
\left[
\bar A_{id}^{(k)}\psi_d
-
\frac12\bar\omega_{id}^{(k)}\psi_d^{\,2}
\right].
\label{eq:app-dir-old-toep-log}
\end{equation}

\subsubsection{New-block insertion probability}

For a new block inserted at slot \(r\in[K_{-i}+1]\), the probability in
\eqref{eq:z-update-full-conditional} can be written as
\begin{equation}
\Pr(z_i=r^{\mathrm{new}}\mid z_{-i},\text{rest})
\propto
\Pr(z_i=r^{\mathrm{new}}\mid z_{-i})
\exp\!\left\{
\ell_{i,\mathrm{vol}}^{\mathrm{new}}(r)
+
\ell_{i,\mathrm{dir}}^{\mathrm{new}}(r;\mathcal M)
\right\}.
\label{eq:app-birth-probability}
\end{equation}
The structure is parallel to the existing-block case, but now all parameters
attached to the newborn block must be integrated out, and under SST the
insertion can also change distances between old blocks.

\paragraph{Volume contribution.}
The same node-to-block quantities \(N_{i\ell}\) and \(\eta_{-i,\ell}\) are
needed here, because the volume part still depends only on how much node \(i\)
interacts with each old block. There is no within-new-block dyad, so the missing
\((r,r)\) term contributes \(M_\kappa(0,0)=1\). All old-old block-pair factors
cancel because the insertion only relabels them. What remains is
\begin{equation}
\prod_{\ell=1}^{K_{-i}}
M_\kappa\!\left(
N_{i\ell},
\eta_i\eta_{-i,\ell}
\right).
\label{eq:app-volume-birth-ratio}
\end{equation}
Taking logs of \eqref{eq:app-volume-birth-ratio} gives
\(\ell_{i,\mathrm{vol}}^{\mathrm{new}}(r)\). It does not depend on the
slot \(r\): the volume model can detect that node \(i\) is forming a singleton
block, but not where that singleton is inserted in the ordering.

\paragraph{Directional contribution under WST.}
Under WST, the old-old directional terms are merely relabelled and therefore
cancel. The only non-cancelling terms are the new comparisons between the
singleton block \(r\) and each old block \(h_r(\ell)\). For each old block
\(\ell\), define
\[
\Abar_{i\ell}^{(r)}
:=
\operatorname{sgn}\{h_r(\ell)-r\}\Abar_{i\ell},
\]
which is the pooled directional signal from node \(i\) toward block \(\ell\)
after assigning the sign implied by insertion at slot \(r\). The associated
probability contribution is
\begin{equation}
\ell_{i,\mathrm{dir}}^{\mathrm{new}}(r;\mathrm{WST})
=
\sum_{\ell=1}^{K_{-i}}
\log I_+\!\left(
\Abar_{i\ell}^{(r)},
\bar\omega_{i\ell};
\mu_0,\sigma_0^2
\right),
\label{eq:app-dir-new-wst-log}
\end{equation}
where \(\pi_+(\cdot;\mu,\sigma^2)\) denotes the truncated--normal density on
\([0,\infty)\) with location \(\mu\) and variance \(\sigma^2\), and
\begin{align}
I_+(y,w;\mu_0,\sigma_0^2)
&:=
\int_0^\infty
\exp\left[
y\psi-\frac12w\psi^2
\right]
\pi_+(\psi;\mu_0,\sigma_0^2)\,d\psi
\nonumber\\
&=
\exp\left[
-\frac12\log\{1+\sigma_0^2w\}
-\frac{\mu_0^2}{2\sigma_0^2}
+
\frac{(y+\sigma_0^{-2}\mu_0)^2}
{2(w+\sigma_0^{-2})}
\right]
\frac{
\Phi\!\left(
m_+(y,w)/\sqrt{v_+(y,w)}
\right)
}{
\Phi(\mu_0/\sigma_0)
},
\label{eq:app-I-plus}
\\
v_+(y,w)
&:=
(w+\sigma_0^{-2})^{-1},
\qquad
m_+(y,w)
:=
v_+(y,w)\{y+\sigma_0^{-2}\mu_0\}.
\nonumber
\end{align}
Thus \(I_+\) is the collapsed contribution of one new positive WST log-odds.
Its normalising constant must be retained, because birth moves integrate
parameters that are absent in the existing-block case. If the new-block move is
selected, that log-odds is then sampled from
\(\mathcal N^+\{m_+(y,w),v_+(y,w)\}\) with
\((y,w)=(\Abar_{i\ell}^{(r)},\bar\omega_{i\ell})\).

\paragraph{Directional contribution under SST.}
Under SST, the insertion slot \(r\) matters for two reasons. First, it fixes
the distance from the singleton block to each old block. Second, it can change
the distance between two old blocks that straddle the insertion point. This is
why the new-block probability requires two kinds of pooled quantities: those
involving node \(i\) and those involving old-old block pairs.

The singleton-to-old-block distance is
\[
d_{r\ell}
:=
|h_r(\ell)-r|.
\]
We adopt the boundary convention \(\psi_0=0\), so when \(K_{-i}=1\) the new
terminal parameter is simply \(\psi_{K_{-i}}=\delta_{K_{-i}}\).
For the singleton-old dyads we reuse the same signed node-to-block quantity
\[
\Abar_{i\ell}^{(r)}
:=
\operatorname{sgn}\{h_r(\ell)-r\}\Abar_{i\ell},
\]
now interpreted together with the distance \(d_{r\ell}\). To account for the
old-old pairs whose distance can change, let \(a<b\) denote two old blocks and
define
\begin{align}
\bar A_{ab}^{(-i)}
&:=
\sum_{\substack{u<v,\;u,v\ne i\\\{z_u,z_v\}=\{a,b\}}}
\operatorname{sgn}(z_v-z_u)\Abar_{uv},
\qquad
\bar\omega_{ab}^{(-i)}
:=
\sum_{\substack{u<v,\;u,v\ne i\\\{z_u,z_v\}=\{a,b\}}}
\omega_{uv}.
\label{eq:app-toep-birth-basic-defs}
\end{align}
Here \(a\) and \(b\) are block labels, whereas \(u\) and \(v\) are node
indices. The pair
\((\bar A_{ab}^{(-i)},\bar\omega_{ab}^{(-i)})\) summarises the directional
information already attached to the old block pair \((a,b)\) after excluding node $i$. These quantities are needed only because inserting a new block can push
some old block pairs one step farther apart.

Using
\[
B(y,w;x)=yx-\frac12wx^2,
\]
the SST directional contribution to
\(\Pr(z_i=r^{\mathrm{new}}\mid z_{-i},\text{rest})\) is
\begin{align}
\ell_{i,\mathrm{dir}}^{\mathrm{new}}(r;\mathrm{SST})
&=
\sum_{\substack{\ell=1\\d_{r\ell}\le K_{-i}-1}}^{K_{-i}}
B\!\left(
\Abar_{i\ell}^{(r)},
\bar\omega_{i\ell};
\psi_{d_{r\ell}}
\right)
\nonumber\\
&\quad+
\sum_{\substack{1\le a<b\le K_{-i}\\a<r\le b\\b-a\le K_{-i}-2}}
\left[
B\!\left(\bar A_{ab}^{(-i)},\bar\omega_{ab}^{(-i)};\psi_{b-a+1}\right)
-
B\!\left(\bar A_{ab}^{(-i)},\bar\omega_{ab}^{(-i)};\psi_{b-a}\right)
\right]
\nonumber\\
&\quad-
\sum_{\substack{1\le a<b\le K_{-i}\\a<r\le b\\b-a=K_{-i}-1}}
B\!\left(\bar A_{ab}^{(-i)},\bar\omega_{ab}^{(-i)};\psi_{K_{-i}-1}\right)
\nonumber\\
&\quad+
\log I_{\mathrm{term}}\!\left(
\bar A_{\mathrm{term}}^{(r)},
\bar\omega_{\mathrm{term}}^{(r)};
\psi_{K_{-i}-1},\tau_0^2
\right).
\label{eq:app-dir-new-toep-log}
\end{align}
The first line keeps the singleton-old dyads whose post-insertion distance is
already represented by an existing parameter. The second line corrects the
old-old pairs with \(a<r\le b\) and \(b-a\le K_{-i}-2\), whose distance changes
from \(b-a\) to \(b-a+1\). The third line removes the previous contribution of
the straddling pairs that were already at the largest old distance \(K_{-i}-1\).
The final term adds back, in integrated form, everything that lands at the new
terminal distance \(K_{-i}\).

The terminal pooled quantities are
\begin{align}
\bar A_{\mathrm{term}}^{(r)}
&:= 
\sum_{\substack{\ell=1\\d_{r\ell}=K_{-i}}}^{K_{-i}}
\Abar_{i\ell}^{(r)}
+
\sum_{\substack{1\le a<b\le K_{-i}\\a<r\le b\\b-a=K_{-i}-1}}
\bar A_{ab}^{(-i)},
\nonumber\\
\bar\omega_{\mathrm{term}}^{(r)}
&:=
\sum_{\substack{\ell=1\\d_{r\ell}=K_{-i}}}^{K_{-i}}
\bar\omega_{i\ell}
+
\sum_{\substack{1\le a<b\le K_{-i}\\a<r\le b\\b-a=K_{-i}-1}}
\bar\omega_{ab}^{(-i)}.
\nonumber
\end{align}
They collect exactly the directional information that must be attached to the
new farthest distance: singleton-old dyads with \(d_{r\ell}=K_{-i}\) and old-old
straddling pairs that move from distance \(K_{-i}-1\) to \(K_{-i}\).

Writing the new endpoint as
\(\psi_{K_{-i}}=\psi_{K_{-i}-1}+\delta_{K_{-i}}\) with \(\delta_{K_{-i}}>0\), the integrated terminal
contribution is
\begin{align}
I_{\mathrm{term}}(y,w;\psi_T,\tau_0^2)
&:=
\int_0^\infty
\exp\left[
y(\psi_T+\delta)-\frac12w(\psi_T+\delta)^2
\right]
\pi_+(\delta;0,\tau_0^2)\,d\delta
\nonumber\\
&=
\exp\left[
y\psi_T-\frac12w\psi_T^2
-\frac12\log(1+\tau_0^2w)
+
\frac{(y-w\psi_T)^2}{2(w+\tau_0^{-2})}
\right]
\nonumber\\
&\quad\times
2\Phi\left(
\frac{y-w\psi_T}{\sqrt{w+\tau_0^{-2}}}
\right),
\label{eq:app-I-terminal}
\end{align}
which is the collapsed contribution of the new positive terminal increment.
Again the normalising constant must be retained, because it is specific to the
new-block move. If the move is selected, then
\[
\delta_{K_{-i}}
\mid z_i=r^{\mathrm{new}},\text{rest}
\sim
\mathcal N^+\!\left(
\frac{y-w\psi_{K_{-i}-1}}{w+\tau_0^{-2}},
(w+\tau_0^{-2})^{-1}
\right),
\]
evaluated at
\((y,w)=(\bar A_{\mathrm{term}}^{(r)},\bar\omega_{\mathrm{term}}^{(r)})\).

\subsubsection{Normalisation, parameter instantiation, and retained constants}

After evaluating
\(\Pr(z_i=k^{\mathrm{old}}\mid z_{-i},\text{rest})\) for \(k=1,\ldots,K_{-i}\) and
\(\Pr(z_i=r^{\mathrm{new}}\mid z_{-i},\text{rest})\) for
\(r=1,\ldots,K_{-i}+1\), we normalise across all \(2K_{-i}+1\) possibilities and
sample \(z_i\). If an existing block is selected, only \(z_i\) changes relative
to the previous state. If a new block is selected, old blocks
\(r,\ldots,K_{-i}\) shift to \(r+1,\ldots,K_{-i}+1\), node \(i\) becomes the
singleton block at slot \(r\),
and the parameters integrated out in the probability calculation are then
instantiated. Concretely, the new cross-block terms
\(\kappa_{r,h_r(\ell)}\), \(\ell=1,\ldots,K_{-i}\), are drawn from their Gamma
full conditionals, while the unused within-block term \(\kappa_{rr}\) is drawn
from its prior because the newborn singleton contributes no within-block dyad.
Under WST, the new pair-specific log-odds are drawn from the corresponding
truncated--normal full conditionals; under SST, the new terminal increment
\(\delta_{K_{-i}}\), equivalently the new endpoint \(\psi_{K_{-i}}\), is
drawn from the truncated--normal full conditional above.

Constants common to all possibilities may be dropped from the intermediate
calculations, but constants specific to birth moves or to a particular
integrated parameter must be retained. This is why the normalising constants
inside \(M_\kappa\), \(I_+\), and \(I_{\mathrm{term}}\) are kept explicitly:
existing-block and new-block probabilities integrate different numbers and types
of parameters, so those constants do affect the final normalised probabilities.

\section{DC--SBM full conditionals under a CRP partition prior}
\label{app:DCSBM_gnedin}

This appendix describes the directed degree-corrected stochastic block model
(DC--SBM) used as an unconstrained benchmark for the TSBM.  The purpose is not
to introduce a new model, but to specify precisely the benchmark fitted in the
simulation and application studies, and to record the full conditional updates
used in the sampler.

The benchmark is deliberately aligned with the directed DC--SBM formulation used
in the main text.  It keeps the same Poisson edge model,
node-specific degree propensities, and asymmetric block-pair intensities, but it
removes the TSBM volume/direction decomposition and all stochastic-transitivity
constraints.  The block labels are therefore unordered, and the partition prior
is the ordinary Chinese Restaurant Process (CRP), rather than the ordered CRP
used for the TSBM.

\subsection{Motivation and relation to existing implementations}

At the time of writing, no readily available \textsf{R} implementation provided
exactly the benchmark required here: a fully Bayesian directed Poisson
DC--SBM, with node-specific degree propensities, asymmetric block-pair
interaction rates, Gamma--Poisson conjugate parameter updates, and a CRP prior
allowing the number of occupied blocks to vary.  This motivated a custom
implementation.  The custom implementation is useful mainly for comparability:
the benchmark should differ from the TSBM by the absence of ordering
constraints, not by a different likelihood or a different inferential target.

Several related tools are nevertheless close to the present model.  The
\texttt{sbm} package \citep{sbmPackage} provides likelihood-based inference for
stochastic block models with Bernoulli, Poisson, and Gaussian edges, including
directed networks, but its inference is variational or ICL-based rather than
fully Bayesian, and it does not target the CRP-based posterior sampler used
here.  The \texttt{SBMSplitMerge} package \citep{sbmSplitMerge} implements
Bayesian MCMC for stochastic block models with Dirichlet-process-type priors and
split--merge moves, but it does not provide the directed Poisson
Gamma-conjugate DC--SBM benchmark required for the present comparison.
Optimisation-based tools such as \texttt{greed}, and related methods in the
\texttt{randnet} ecosystem, are useful for model selection and
classification-type summaries, but they do not provide the posterior sampling
scheme needed here.

Peixoto's nonparametric degree-corrected SBM, implemented in
\texttt{graph-tool}, is the closest Bayesian alternative
\citep{peixoto2014hierarchical}.  It supports directed weighted networks and a
variable number of blocks, but it uses a different hierarchical prior and
inference framework, is implemented through a Python/C++ ecosystem, and is not a
drop-in replacement for the conjugate CRP benchmark used in this paper.
Related research code for exchangeable SBMs under Gibbs-type priors, such as
that associated with \citet{Legramanti_2022}, is also close in spirit but does
not provide the exact directed Poisson degree-corrected sampler described below.

\subsection{Model}
\label{app:dcsbm-model}

Let \(A=(A_{ij})_{i,j=1}^n\) be a directed count adjacency matrix with
\(A_{ii}\equiv0\).  The entry \(A_{ij}\) is the number of directed interactions
from node \(i\) to node \(j\).  Each node has a latent block label
\(z_i\in\mathbb N\), and \(K=K(\mathbf{z})\) denotes the number of occupied blocks in
\(\mathbf{z}=(z_1,\ldots,z_n)\).  Conditional on \((\mathbf{z},\bm\eta,\lambda)\), the
edges are mutually independent and
\begin{equation}
A_{ij}\mid \mathbf{z},\bm\eta,\lambda
\sim
\operatorname{Poisson}\!\bigl(\eta_i\eta_j\lambda_{z_i z_j}\bigr),
\qquad i\ne j.
\label{eq:dir-dcsbm-gibbs}
\end{equation}
Here \(\eta_i>0\) is a node-specific, non-directional degree propensity, and
\(\lambda=(\lambda_{k\ell})\) is a generally asymmetric matrix of block-pair
interaction intensities.  Thus \(\lambda_{k\ell}\) controls the expected
interaction rate from block \(k\) to block \(\ell\), after degree correction.
This is the unconstrained directed DC--SBM appearing in the main text before the
TSBM reparametrisation into volume and direction.

The model is not invariant to permutations of node indices, because the observed
network fixes the nodes, but it is invariant to permutations of block labels.
Unlike the TSBM, the labels \(1,\ldots,K\) carry no hierarchical meaning.

\subsection{Priors}
\label{app:dcsbm-priors}

All Gamma distributions are parametrised by shape and rate.  The node
propensities and block-pair intensities are assigned independent priors,
\begin{align}
\eta_i
&\sim
\operatorname{Gamma}(a_\eta,b_\eta),
&& i=1,\ldots,n,
\label{eq:dcsbm-prior-eta}\\
\lambda_{k\ell}
&\sim
\operatorname{Gamma}(a_\lambda,b_\lambda),
&& k,\ell\in[K].
\label{eq:dcsbm-prior-lambda}
\end{align}
No finite-dimensional mixing vector is introduced.  Instead, the random
partition induced by \(\mathbf{z}\) follows a CRP prior with concentration
\(\alpha>0\).  After removing node \(i\), let
\(K_{-i}=K(\bm z_{-i})\) and let
\(n_k^-=\#\{j\ne i:z_j=k\}\) for \(k=1,\ldots,K_{-i}\).  The CRP predictive
probabilities are
\begin{equation}
\Pr(z_i=k^{\mathrm{old}}\mid \bm z_{-i},\alpha)
=
\frac{n_k^-}{\alpha+n-1},
\qquad
\Pr(z_i=\mathrm{new}\mid \bm z_{-i},\alpha)
=
\frac{\alpha}{\alpha+n-1}.
\label{eq:dcsbm-crp-predictive}
\end{equation}
For single-site allocation weights, the common denominator \(\alpha+n-1\) can
be dropped.  The induced prior mean of the number of occupied blocks is
\[
\mathbb E(K_n)
=
\sum_{r=1}^n \frac{\alpha}{\alpha+r-1}
=
\alpha\{\psi(\alpha+n)-\psi(\alpha)\},
\]
where \(\psi\) denotes the digamma function.  This identity is used to calibrate
\(\alpha\).  In the simulations and applications in the main text, the DC--SBM
uses the ordinary CRP with \(\alpha=0.5\), whereas the ordered TSBM uses the
ordered-CRP prior with concentration \(\vartheta=0.5\).

\subsection{Posterior kernel}
\label{app:dcsbm-posterior}

The likelihood is
\begin{equation}
L(A\mid \mathbf{z},\bm\eta,\lambda)
=
\prod_{i\ne j}
\frac{\{\eta_i\eta_j\lambda_{z_i z_j}\}^{A_{ij}}}{A_{ij}!}
\exp\{-\eta_i\eta_j\lambda_{z_i z_j}\}.
\label{eq:dcsbm-likelihood}
\end{equation}
Let \(\varphi=(a_\eta,b_\eta,a_\lambda,b_\lambda,\alpha)\) collect the
hyperparameters.  Up to the normalising constant, the posterior density is
\begin{align}
p(\bm\eta,\lambda,\mathbf{z}\mid A,\varphi)
&\propto
L(A\mid \mathbf{z},\bm\eta,\lambda)
\left\{\prod_{i=1}^n p(\eta_i)\right\}
\left\{\prod_{k=1}^{K(\mathbf{z})}\prod_{\ell=1}^{K(\mathbf{z})}p(\lambda_{k\ell})\right\}
p(\mathbf{z}\mid\alpha).
\label{eq:dcsbm-posterior-kernel}
\end{align}
The sampler cycles through updates of \(\lambda\), \(\bm\eta\), and \(\mathbf{z}\).
When the number of occupied blocks changes, the corresponding rows and columns
of \(\lambda\) are created or removed as described below.

\subsection{Full conditional for the block-pair intensities}
\label{app:dcsbm-lambda-update}

For a fixed allocation \(\mathbf{z}\), define the sufficient statistics
\begin{align}
R_{k\ell}
&=
\sum_{i:z_i=k}\sum_{\substack{j:z_j=\ell\\ j\ne i}} A_{ij},
\label{eq:dcsbm-Rkl}\\
T_{k\ell}(\bm\eta)
&=
\sum_{i:z_i=k}\sum_{\substack{j:z_j=\ell\\ j\ne i}} \eta_i\eta_j.
\label{eq:dcsbm-Tkl-def}
\end{align}
Equivalently, if \(S_k=\sum_{i:z_i=k}\eta_i\), then
\begin{equation}
T_{k\ell}(\bm\eta)
=
S_kS_\ell
-\mathbb I(k=\ell)\sum_{i:z_i=k}\eta_i^2.
\label{eq:dcsbm-Tkl-mass}
\end{equation}
The subtraction in the diagonal case is necessary because self-edges
\(A_{ii}\) are excluded.  Notice that \(R_{k\ell}\) and \(R_{\ell k}\) are
usually different, and so are \(\lambda_{k\ell}\) and \(\lambda_{\ell k}\).

The part of the likelihood involving \(\lambda_{k\ell}\) is
\[
\lambda_{k\ell}^{R_{k\ell}}
\exp\{-\lambda_{k\ell}T_{k\ell}(\bm\eta)\}.
\]
Combining this factor with the Gamma prior gives
\begin{equation}
\lambda_{k\ell}\mid \text{rest}
\sim
\operatorname{Gamma}\{a_\lambda+R_{k\ell},\;
 b_\lambda+T_{k\ell}(\bm\eta)\},
\qquad k,\ell=1,\ldots,K.
\label{eq:dcsbm-lambda-fullcond}
\end{equation}
Equivalently, conditional on \((\mathbf{z},\bm\eta,\lambda)\), the aggregated count
\(R_{k\ell}\) has a Poisson likelihood with mean
\(\lambda_{k\ell}T_{k\ell}(\bm\eta)\), up to factors not involving
\(\lambda_{k\ell}\).

\subsection{Full conditional for the node propensities}
\label{app:dcsbm-eta-update}

Fix a node \(i\), and write \(k=z_i\).  Define its total incident count,
\begin{equation}
G_i
=
\sum_{j\ne i}(A_{ij}+A_{ji}),
\label{eq:dcsbm-Gi}
\end{equation}
and, for each block \(\ell\), the block mass excluding node \(i\),
\begin{equation}
S_\ell^{-i}
=
\sum_{\substack{j:z_j=\ell\\j\ne i}}\eta_j.
\label{eq:dcsbm-Sminus}
\end{equation}
The likelihood factors depending on \(\eta_i\) are the outgoing and incoming
edges incident to \(i\).  Up to factors independent of \(\eta_i\),
\begin{align}
L(A\mid \mathbf{z},\bm\eta,\lambda)
&\propto
\prod_{j\ne i}
(\eta_i\eta_j\lambda_{k,z_j})^{A_{ij}}
\exp\{-\eta_i\eta_j\lambda_{k,z_j}\}
\prod_{j\ne i}
(\eta_j\eta_i\lambda_{z_j,k})^{A_{ji}}
\exp\{-\eta_i\eta_j\lambda_{z_j,k}\}
\nonumber\\
&\propto
\eta_i^{G_i}
\exp\left[
-\eta_i\sum_{j\ne i}\eta_j\{\lambda_{k,z_j}+\lambda_{z_j,k}\}
\right]
\nonumber\\
&=
\eta_i^{G_i}
\exp\left[
-\eta_i\sum_{\ell=1}^{K}
\{\lambda_{k\ell}+\lambda_{\ell k}\}S_\ell^{-i}
\right].
\label{eq:dcsbm-eta-kernel}
\end{align}
After multiplication by the Gamma prior, the full conditional is therefore
\begin{equation}
\eta_i\mid\text{rest}
\sim
\operatorname{Gamma}\left(
 a_\eta+G_i,
 b_\eta+
 \sum_{\ell=1}^{K}
 \{\lambda_{k\ell}+\lambda_{\ell k}\}S_\ell^{-i}
\right).
\label{eq:dcsbm-eta-fullcond}
\end{equation}
The shape depends on the observed in- and out-degree mass of node \(i\).  The
rate is its total exposure to the rest of the network, with each block mass
weighted by both the outgoing and incoming interaction intensities involving
block \(k\).

\subsection{Single-site allocation update}
\label{app:dcsbm-z-update}

The allocation update is computed excluding the node we are currently evaluating. Therefore, we start by removing node
\(i\), deleting any empty block, and relabelling the remaining occupied blocks as
\(1,\ldots,K\), where now \(K=K(\bm z_{-i})\).  For each exist. block \(\ell\), set
\begin{align}
S_\ell
&=
\sum_{j:z_j=\ell}\eta_j,
&
R_{i\to\ell}
&=
\sum_{j:z_j=\ell}A_{ij},
&
R_{\ell\to i}
&=
\sum_{j:z_j=\ell}A_{ji},
\label{eq:dcsbm-node-block-stats}
\end{align}
where all sums are taken after removing node $i$.  Also define
\begin{equation}
T_{i\ell}=\eta_i S_\ell.
\label{eq:dcsbm-node-block-exposure}
\end{equation}
The quantities \(R_{i\to\ell}\) and \(R_{\ell\to i}\) are directed counts,
whereas \(T_{i\ell}\) is the common exposure for the two directions because the
benchmark uses a non-directional node propensity \(\eta_i\).

\subsubsection{Existing-block candidate}

For an existing candidate block \(k\in\{1,\ldots,K\}\), the incident-edge
likelihood contribution, up to constants independent of \(k\), is
\begin{equation}
\ell_i(k)
=
\sum_{\ell=1}^{K}
\left[
R_{i\to\ell}\log\lambda_{k\ell}
+R_{\ell\to i}\log\lambda_{\ell k}
-T_{i\ell}\{\lambda_{k\ell}+\lambda_{\ell k}\}
\right].
\label{eq:dcsbm-ell-existing}
\end{equation}
The corresponding unnormalised allocation weight is
\begin{equation}
w_i(k)
=
n_k^-\exp\{\ell_i(k)\},
\qquad k=1,\ldots,K,
\label{eq:dcsbm-z-existing-weight}
\end{equation}
where \(n_k^-\) is the size of block \(k\) obtained after removing node $i$.  The
factor \((\alpha+n-1)^{-1}\) in the CRP predictive probability is common to all
candidates and has been omitted.

\subsubsection{New-block candidate}

If node \(i\) starts a new singleton block, the new row and column of
\(\lambda\) do not yet exist.  The allocation weight is therefore evaluated by
integrating out the new interaction rates
\(\lambda_{\mathrm{new},\ell}\) and \(\lambda_{\ell,\mathrm{new}}\) under their
Gamma priors.  For a generic directed count \(R\) and exposure \(T\), define the
Gamma--Poisson marginal factor
\begin{equation}
M_\lambda(R,T)
=
\frac{b_\lambda^{a_\lambda}}{\Gamma(a_\lambda)}
\frac{\Gamma(a_\lambda+R)}{(b_\lambda+T)^{a_\lambda+R}}.
\label{eq:dcsbm-Mlambda}
\end{equation}
This is the integral of the Poisson likelihood kernel
\(\lambda^R\exp(-T\lambda)\) against the normalised
\(\operatorname{Gamma}(a_\lambda,b_\lambda)\) prior.  In particular,
\(M_\lambda(0,0)=1\), so an empty parameter contributes neither a penalty nor a
bonus.

For the new singleton block, the collapsed incident-edge log likelihood is
\begin{align}
\ell_i(\mathrm{new})
&=
\sum_{\ell=1}^{K}
\left[
\log M_\lambda(R_{i\to\ell},T_{i\ell})
+
\log M_\lambda(R_{\ell\to i},T_{i\ell})
\right]
\nonumber\\
&=
\sum_{\ell=1}^{K}
\Bigl[
2a_\lambda\log b_\lambda-2\log\Gamma(a_\lambda)
+
\log\Gamma(a_\lambda+R_{i\to\ell})
+
\log\Gamma(a_\lambda+R_{\ell\to i})
\nonumber\\
&\hspace{8em}
-(a_\lambda+R_{i\to\ell})\log(b_\lambda+T_{i\ell})
-(a_\lambda+R_{\ell\to i})\log(b_\lambda+T_{i\ell})
\Bigr].
\label{eq:dcsbm-ell-new-expanded}
\end{align}
The within-new-block parameter \(\lambda_{\mathrm{new},\mathrm{new}}\) is not
involved in any likelihood term, because the new block contains only node \(i\)
and self-edges are excluded; its integrated contribution is
\(M_\lambda(0,0)=1\).  The normalised prior constants in
\(M_\lambda\) are included because the new-block candidate changes the number of
parameters being integrated.  Dropping them in the new-block weight, while
conditioning on existing \(\lambda\)'s for existing-block candidates, would alter
the Bayes factor for creating a block.

The unnormalised new-block weight is
\begin{equation}
w_i(\mathrm{new})
=
\alpha\exp\{\ell_i(\mathrm{new})\}.
\label{eq:dcsbm-z-new-weight}
\end{equation}
Finally,
\begin{equation}
\Pr(z_i=t\mid A,\bm z_{-i},\bm\eta,\lambda,\alpha)
=
\frac{w_i(t)}{\sum_{k=1}^{K}w_i(k)+w_i(\mathrm{new})},
\qquad
 t\in\{1,\ldots,K,\mathrm{new}\}.
\label{eq:dcsbm-z-normalised}
\end{equation}
All weights should be evaluated on the log scale, subtracting the maximum
log-weight before exponentiation.

\subsection{Instantiation of new parameters and pruning}
\label{app:dcsbm-dimension-change}

If the new-block candidate is selected, the new row and column of \(\lambda\) are
instantiated from their posterior full conditionals.  For each exist. block
\(\ell=1,\ldots,K\),
\begin{align}
\lambda_{\mathrm{new},\ell}\mid\text{rest}
&\sim
\operatorname{Gamma}(a_\lambda+R_{i\to\ell},\;b_\lambda+T_{i\ell}),
\label{eq:dcsbm-new-row}\\
\lambda_{\ell,\mathrm{new}}\mid\text{rest}
&\sim
\operatorname{Gamma}(a_\lambda+R_{\ell\to i},\;b_\lambda+T_{i\ell}).
\label{eq:dcsbm-new-col}
\end{align}
The unused within-block rate for the singleton block is sampled from the prior,
\begin{equation}
\lambda_{\mathrm{new},\mathrm{new}}
\sim
\operatorname{Gamma}(a_\lambda,b_\lambda),
\label{eq:dcsbm-new-within}
\end{equation}
because no self-edge likelihood term is available at the moment of creation.
After a full sweep of allocation updates, empty blocks are removed and the
remaining labels are compactly relabelled as \(1{:}K\).  The matrix \(\lambda\)
is reduced to the occupied submatrix.

\subsection{Identifiability and optional degree normalisation}
\label{app:dcsbm-identifiability}

The likelihood in \eqref{eq:dir-dcsbm-gibbs} has a scale non-identifiability.
More generally, for positive constants \(c_1,\ldots,c_K\), the transformation
\begin{equation}
\eta_i^\star=c_{z_i}\eta_i,
\qquad
\lambda_{k\ell}^\star=\frac{\lambda_{k\ell}}{c_kc_\ell}
\label{eq:dcsbm-scale-transform}
\end{equation}
leaves every Poisson mean \(\eta_i\eta_j\lambda_{z_i z_j}\) unchanged.  A common
identifying convention is the within-block normalisation
\begin{equation}
\sum_{i:z_i=k}\eta_i=n_k,
\qquad k=1,\ldots,K.
\label{eq:dcsbm-eta-normalisation}
\end{equation}
If this convention is enforced during sampling, the rescaling in
\eqref{eq:dcsbm-scale-transform} must be applied jointly to \(\bm\eta\) and
\(\lambda\).  Equivalently, after an unconstrained update of \(\bm\eta\), set
\[
c_k=\frac{n_k}{\sum_{i:z_i=k}\eta_i},
\qquad
\eta_i\leftarrow c_{z_i}\eta_i,
\qquad
\lambda_{k\ell}\leftarrow \frac{\lambda_{k\ell}}{c_kc_\ell}.
\]
This deterministic projection leaves the likelihood unchanged and fixes a
representative of each likelihood-equivalent parameterisation.  If one wants the
literal posterior induced by the independent Gamma priors in
\eqref{eq:dcsbm-prior-eta}--\eqref{eq:dcsbm-prior-lambda}, the projection should
instead be omitted and degree-normalised quantities should be reported only in
post-processing.  This distinction is worth making explicit, since otherwise one
is silently changing the prior geometry, always an excellent way to make a
sampler look more mysterious than it is.

\section{Supplementary simulation results}
\label{app:simulation-supplement}

This section collects the simulation details and checks referenced in the main
text.  The main paper reports the headline tables and figures; here we retain
the design details, implementation sanity checks, robustness summaries, and
misspecification check needed to interpret those results.

\subsection{Minimal sanity-check simulation for the DC--SBM sampler}
\label{app:sim_dcsbm_gnedin}

This section reports a small simulation study whose purpose is only to check the
implementation of the Gibbs sampler above under controlled conditions.  It is
separate from the main simulation study, where the data are generated from the
TSBM and the DC--SBM is used as an unconstrained benchmark.

\subsubsection{Data-generating mechanism}

Fix \(n\) and a ground-truth number of blocks
\(K_{\mathrm{true}}\in\{3,5,7\}\).  A balanced ground-truth allocation
\(\mathbf{z}^\star\) is obtained by assigning nodes cyclically to the
\(K_{\mathrm{true}}\) labels.  Node propensities are sampled independently as
\[
\eta_i^\star\sim\operatorname{Unif}(0.8,1.2),
\qquad i=1,\ldots,n,
\]
and then normalised within each true block so that
\[
\sum_{i:z_i^\star=k}\eta_i^\star=n_k^\star,
\qquad k=1,\ldots,K_{\mathrm{true}},
\]
where \(n_k^\star=\#\{i:z_i^\star=k\}\).  The block interaction matrix
\(\lambda^\star=(\lambda_{k\ell}^\star)\) is constructed with stronger diagonal
entries, weaker off-diagonal entries, and mild directed asymmetry.  For example,
for \(K_{\mathrm{true}}=3\), one such matrix is
\[
\lambda^\star
=
\begin{bmatrix}
2.146 & 0.258 & 0.053\\
0.055 & 2.190 & 0.219\\
0.198 & 0.045 & 2.156
\end{bmatrix}.
\]
Finally, conditional on \((\mathbf{z}^\star,\bm\eta^\star,\lambda^\star)\), the
adjacency matrix is generated from
\[
A_{ij}\mid \mathbf{z}^\star,\bm\eta^\star,\lambda^\star
\sim
\operatorname{Poisson}\!\bigl(
\eta_i^\star\eta_j^\star\lambda^\star_{z_i^\star z_j^\star}
\bigr),
\qquad i\ne j,
\qquad A_{ii}=0.
\]
This simulation uses the same non-directional degree correction as the sampler.
The asymmetry of the directed network is carried by \(\lambda^\star\), not by
separate sending and receiving propensities.

\subsubsection{Inference setup}

For each \(K_{\mathrm{true}}\in\{3,5,7\}\), five independent datasets are
generated and fitted with the Gibbs sampler described above.  The allocation is
initialised from an over-specified partition with \(K_{\mathrm{init}}=n\).  The
initial values of \((\bm\eta,\lambda)\) are drawn from the corresponding Gamma
priors, followed by the optional within-block normalisation described in
Section~\ref{app:dcsbm-identifiability}.  Each chain is run for
\(T=2500\) iterations, with the first \(B=1250\) iterations discarded as burn-in
and every second post-burn-in draw retained.

\subsubsection{Posterior summaries}

Let \(\{\mathbf{z}^{(s)}\}_{s=1}^S\) denote the retained allocation draws.  A point
estimate \(\hat{\mathbf{z}}\) is obtained by minimising posterior expected Variation
of Information loss.  Since the truth is known in this simulation, performance
is summarised by the Adjusted Rand Index \(\operatorname{ARI}(\mathbf{z}^\star,
\hat{\mathbf{z}})\), the Variation of Information distance
\(\operatorname{VI}(\mathbf{z}^\star,\hat{\mathbf{z}})\), and posterior summaries of the
number of occupied blocks.  We write \(\hat K\) for the posterior mean of
\(K(\mathbf{z}^{(s)})\) across retained draws.

Table~\ref{tab:sim_dcsbm_gnedin_summary} reports performance aggregated over the
five replications for each value of \(K_{\mathrm{true}}\).  In this deliberately
well-specified check, the sampler recovers both the partition and the number of
occupied blocks in all replications.

\begin{table}[t]
\centering
\caption{Simulation sanity check for the directed DC--SBM with a CRP prior on the number of occupied blocks.  Results are aggregated over five replicated datasets for each value of \(K_{\mathrm{true}}\).}
\label{tab:sim_dcsbm_gnedin_summary}
\begin{tabular}{rrrrr}
\toprule
\(K_{\mathrm{true}}\) & mean ARI & sd(ARI) & \(\hat K\) & sd\((\hat K)\)\\
\midrule
3 & 1 & 0 & 3 & 0\\
5 & 1 & 0 & 5 & 0\\
7 & 1 & 0 & 7 & 0\\
\bottomrule
\end{tabular}
\end{table}

\subsection{Simulation design details for the ordered data-generating models}
\label{app:sim-design-details}

The main simulation uses \(n=60\), \(K^\star\in\{3,5,8\}\),
\(\kappa^\star\in\{0.75,1.5,3,6\}\), and three hierarchy-strength settings
\(\psi^\star\in\{0.2,0.7,1.3\}\).  The block sizes are balanced.  The symmetric
volume matrix is drawn from a Gamma law with mean \(\kappa^\star\) and coefficient
of variation \(0.6\), then symmetrised.  Node propensities are drawn independently
from \(\mathrm{Unif}(0.8,1.2)\).

For WST-generated data, each upper-triangular block-pair log-odds
\(\psi^\star_{k\ell}\), \(k<\ell\), is drawn independently from
\(\mathcal N^+(\psi^\star,0.3^2)\), and the lower triangle is filled by
skew-symmetry.  For SST-generated data, the distance log-odds are
cumulative sums \(\psi^\star_d=\sum_{r\le d}\delta^\star_r\), where
\[
\delta^\star_r
\overset{\mathrm{ind}}{\sim}
\mathcal N^+\!\left(
\frac{\psi^\star}{K^\star-1},
\frac{0.3^2}{(K^\star-1)^2}
\right),
\qquad r=1,\ldots,K^\star-1.
\]
Thus \(\psi^\star\) controls the approximate end-to-end directional strength on
the log-odds scale in both generating regimes, while WST permits pair-specific
heterogeneity and SST pools by rank distance.

\subsection{Supplementary partition-recovery summaries for the main simulation}
\label{app:sim-vi-ari}

\begin{figure}[htbp]
\centering
\includegraphics[width=\textwidth]{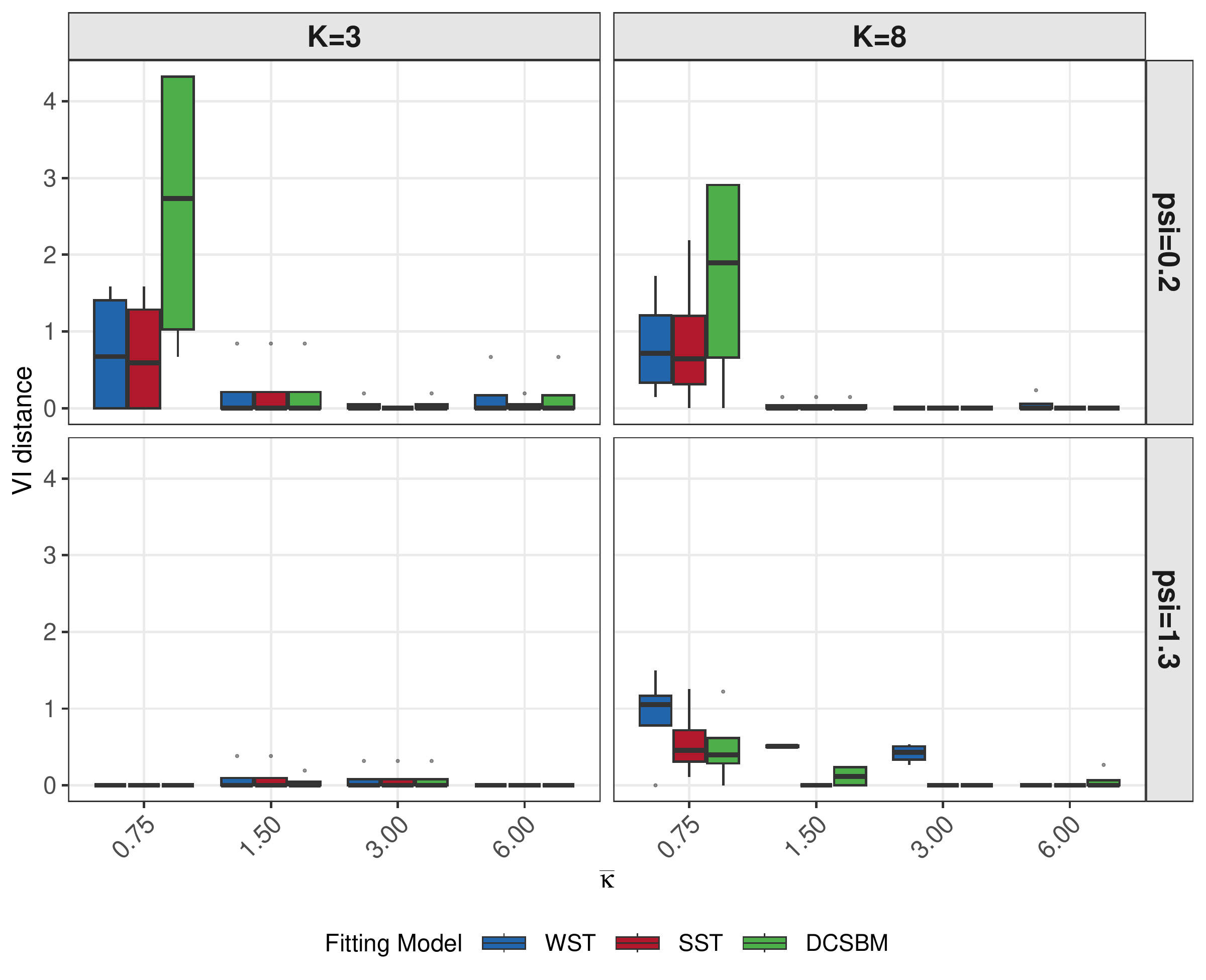}
\caption{VI distance between the minimum-VI partition estimate and the true partition across SST-generated scenarios, stratified by density \((\kappa^\star)\) and hierarchy strength \((\psi^\star)\).  Lower values indicate better recovery.  The ordered models improve over the unconstrained DC--SBM most clearly in sparse regimes, especially when the hierarchical signal is strong.}
\label{fig:vi-SST-gen}
\end{figure}

\begin{figure}[htbp]
\centering
\includegraphics[width=\textwidth]{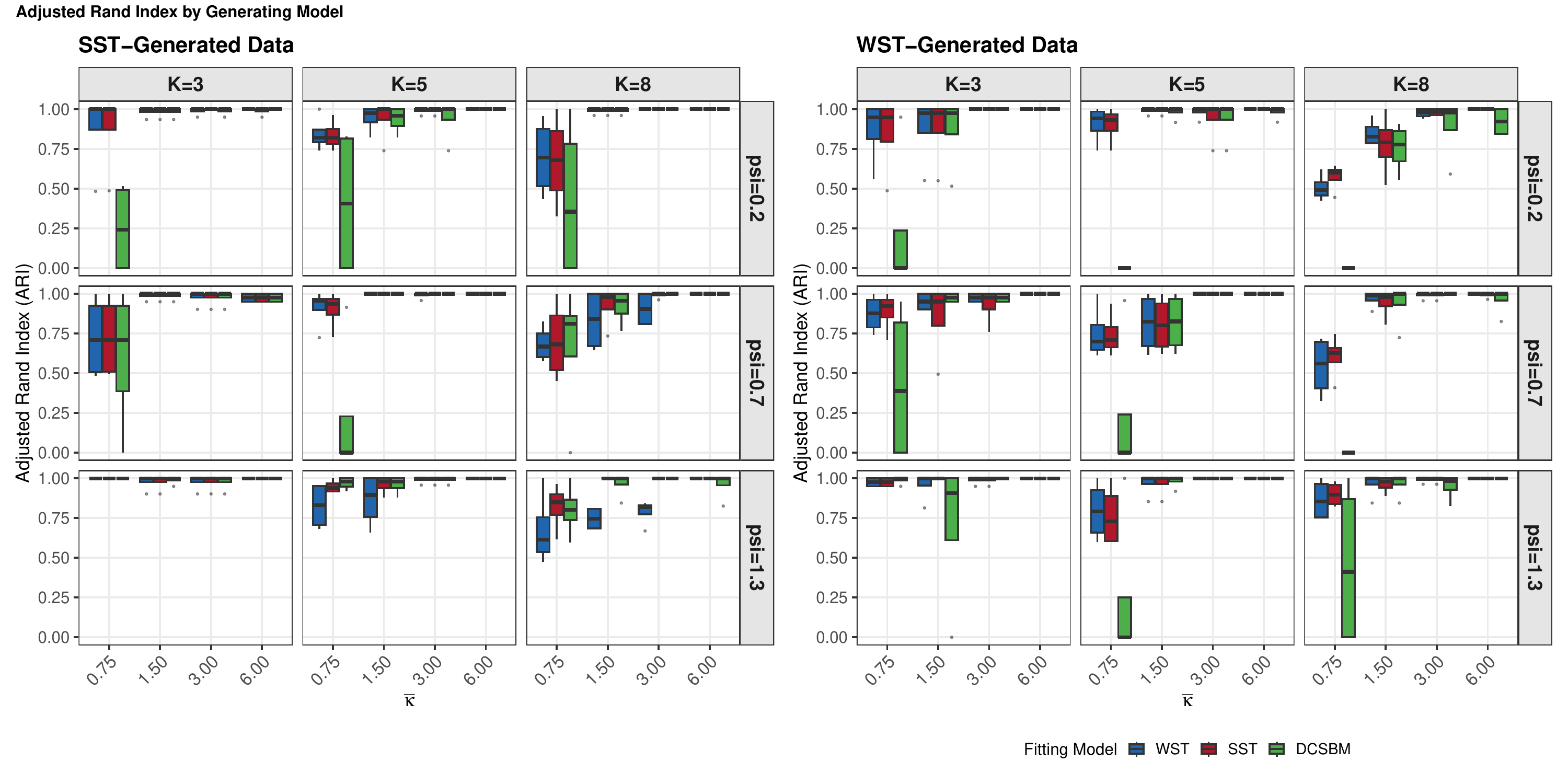}
\caption{Adjusted Rand Index between the minimum-VI partition estimate and the true partition for SST-generated scenarios and WST-generated scenarios, by density \((\kappa^\star)\) and hierarchy strength \((\psi^\star)\).  Higher values indicate better recovery.}
\label{fig:ari-combined}
\end{figure}

Figures~\ref{fig:vi-SST-gen} and~\ref{fig:ari-combined} complement the
partition-recovery summaries reported in the main text.  They compare the
minimum-VI partition estimate with the true partition across scenarios varying
the generating model, network density, hierarchy strength, and true number of
blocks.

\subsection{Supplementary scenario-by-scenario summaries}
\label{app:sim-full-tables}

The full scenario-by-scenario simulation summaries report, for each combination
of generating model, hierarchy strength, true block count \(K^\star\), and
network density \(\kappa^\star\), the posterior mean number of occupied blocks,
the probability of recovering the true number of blocks exactly, and the ARI of
the minimum-VI estimate.  These tables are included in the project source as
separate table files.  The qualitative pattern is the one described in the main
text: ordered prior information is most useful in sparse networks with a real
but noisy directional signal, whereas the unconstrained DC--SBM is more robust
when the data-generating mechanism is not transitive.

\subsection{MCMC convergence diagnostics for the main simulation}
\label{app:sim-mcmc}

To keep the appendix compact, we retain one Gelman--Rubin summary for the
scalar chains that are genuinely informative here: the degree parameters
\(\bm\eta\), the volume component, and the directional component. For the
ordered models, these are the chains that determine whether the sampler has
stabilised on the likelihood scale. For the unconstrained DC--SBM, the same
summaries are computed after ex post ordering of the sampled or estimated
blocks, as described in the main text. We do not report a Gelman--Rubin
statistic for \(K\), because in the DC--SBM the occupied block count is weakly
identified once degree heterogeneity is allowed to absorb part of the
structure; a poor value there would reflect identifiability rather than
sampler failure.

The table below shows that the ordered models mix well on all three scalar
summaries. Values close to 1 indicate that the chains have reached a common
stationary regime, so the posterior comparisons in the paper are not driven by
obvious Monte Carlo instability. The largest upper confidence limit is only
noticeably inflated in the strong-prior SST case, where the volume component
mixes more slowly. That flags the one regime where Monte Carlo uncertainty is
less tight, but it does not change the overall conclusion that the scalar
chains are well behaved.

\begin{table}[H]
\caption{Convergence summary for the scalar chains that are genuinely informative here. Weak, default, and strong correspond to $(\vartheta_{\mathrm{ocrp}},\alpha_{\mathrm{crp}},b_{\mathrm{scale}},d_{\mathrm{scale}})=(1.0,1.0,0.5,1.5)$, $(0.5,0.5,1.0,1.0)$, and $(0.2,0.2,2.0,0.67)$. We report Gelman--Rubin for $\bm\eta$, volume, and direction summaries, plus the largest upper 97.5\% confidence limit across those three quantities. Values near 1 indicate good mixing.}
\label{tab:sim-mcmc-convergence}
\centering
\begin{tabular}{llrrrr}
\toprule
Dataset & Setting & $\hat R_{\eta}$ & $\hat R_{\mathrm{vol}}$ & $\hat R_{\mathrm{dir}}$ & max upper CI \\
\midrule
Simulated SST dataset & Weak prior & 1.006 & 1.021 & 1.010 & 1.078 \\
Simulated SST dataset & Default prior & 1.011 & 1.133 & 1.022 & 1.694 \\
Simulated SST dataset & Strong prior & 1.028 & 1.252 & 1.037 & 15.707 \\
Bighorn sheep & Weak prior & 1.020 & 1.008 & 1.011 & 1.108 \\
Bighorn sheep & Default prior & 1.004 & 1.105 & 1.016 & 1.502 \\
Bighorn sheep & Strong prior & 1.007 & 1.050 & 1.010 & 1.170 \\
\bottomrule
\end{tabular}
\end{table}

\subsection{Prior sensitivity check}
\label{app:sim-sensitivity}

We ran a compact sensitivity check for the ordered-CRP and volume-prior scales.
The setting labels are shorthand for
\[
\begin{array}{c|cccc}
&\vartheta_{\mathrm{ocrp}}&\alpha_{\mathrm{crp}}&b_{\mathrm{scale}}&d_{\mathrm{scale}}\\
\hline
\text{weak} & 1.0&1.0&0.5&1.5\\
\text{default} & 0.5&0.5&1.0&1.0\\
\text{strong} & 0.2&0.2&2.0&0.67 
\end{array}
\]
Here \(\vartheta_{\mathrm{ocrp}}\) is the ordered-CRP concentration used by the
TSBM, and \(\alpha_{\mathrm{crp}}\) is the ordinary CRP concentration used by the
DC--SBM.  Table~\ref{tab:supp-small-sim-sensitivity} shows that the simulated
SST example is most stable under the default prior in terms of the VI summary,
while the sheep example shows a similar pattern: the posterior block count
changes more than the violation rate when the prior is tightened. The
regenerated version of the same table also reports pooled LOOIC and paired
\(\Delta\)LOOIC (SE) as a predictive cross-check.

\begin{table}[htbp]
\centering
\caption{Sensitivity on one SST-generated dataset ($n=60$, $K^\star=5$, moderate density and strong hierarchy). For each model and prior setting we pool all retained draws across chains, report the posterior block-count summary $\hat K[2.5\%,97.5\%]$, the VI loss of the pooled \texttt{minVI} partition against the truth, the pooled LOOIC and paired $\Delta$LOOIC (SE) against the best model within each prior setting, and the largest Gelman--Rubin upper confidence limit across the scalar chains. Lower VI and LOOIC are better.\label{tab:supp-small-sim-sensitivity}}
\begin{tabular}{llccccc}
\toprule
Setting & Model & $\hat K[2.5\%,97.5\%]$ & VI & LOOIC & $\Delta$LOOIC (SE) & max upper CI \\
\midrule
Weak prior & WST--TSBM & $4\;[3,6]$ & 0.380 & $8623.4$ & $38.4 (5.8)$ & 1.125 \\
Weak prior & Toeplitz SST--TSBM & $5\;[4,6]$ & 0.380 & $\mathbf{8585.0}$ & $\mathbf{-0.0 (0.0)}$ & 1.099 \\
Weak prior & DC--SBM & $3\;[3,4]$ & 0.405 & $8669.8$ & $84.8 (11.2)$ & 4.050 \\
Default prior & WST--TSBM & $4\;[3,5]$ & 0.425 & $8658.3$ & $80.1 (11.9)$ & 1.694 \\
Default prior & Toeplitz SST--TSBM & $4\;[4,5]$ & 0.335 & $\mathbf{8578.3}$ & $\mathbf{-0.0 (0.0)}$ & 1.033 \\
Default prior & DC--SBM & $3\;[3,4]$ & 0.519 & $8648.0$ & $69.8 (12.1)$ & 4.633 \\
Strong prior & WST--TSBM & $3\;[2,3]$ & 0.519 & $8723.3$ & $75.1 (9.5)$ & 10.938 \\
Strong prior & Toeplitz SST--TSBM & $3\;[3,4]$ & 0.415 & $8662.2$ & $14.0 (8.8)$ & 15.707 \\
Strong prior & DC--SBM & $3\;[3,4]$ & 0.519 & $\mathbf{8648.2}$ & $\mathbf{-0.0 (0.0)}$ & 4.454 \\
\bottomrule
\end{tabular}
\end{table}

The practical conclusion is that the default prior gives the most balanced
behaviour: weak priors make births too cheap and encourage extra blocks, while
strong priors can over-regularise the ordered structure and move posterior mass
between nearby values of \(K_n\). The appendix table makes this visible in the
simulated SST example through the shift in \(\hat K\) and the modest movement in
VI, and in the sheep example through the much smaller change in the
violation rate relative to the change in \(\hat K\). The predictive
cross-check is consistent with that same warning: the Toeplitz SST fit wins the
simulated example under the weak and default priors, while the strongest prior
can tip the predictive ranking toward DC--SBM.

\subsection{Misspecified non-hierarchical simulation}
\label{app:sim-misspec}

We also checked a deliberately misspecified setting in which the data-generating
block structure has no compatible global hierarchy.  The simulation uses
\(K^\star=5\), \(n=60\), \(\kappa_{kk}=8\), \(\kappa_{k\ell}=3\) for
\(k\ne\ell\), and a heterogeneous directional matrix with mixed signs.  Hence
no total order can represent all cross-block directions.  Table~\ref{tab:sim-misspec-app}
reports normalised VI, LOOIC, and the violation rate diagnostic.

\begin{table}[htbp]
\centering
\caption{Misspecification experiment with non-hierarchical directional
asymmetry. Lower LOOIC and NVI are better.}
\label{tab:sim-misspec-app}
\begin{tabular}{lccccc}
\toprule
Model & \(\hat K\) & \(\bar K_{\mathrm{post}}\) & NVI & LOOIC & \(\hat\zeta_z\) \\
\midrule
WST--TSBM & 2 & 2.00 & 0.33 & 13\,058 & 0.49 \\
SST--TSBM & 2 & 2.00 & 0.33 & 13\,003 & 0.49 \\
DC--SBM & 5 & 5.00 & 0.00 & 11\,867 & 0.49 \\
\bottomrule
\end{tabular}
\end{table}

Both ordered models collapse to two blocks and have NVI \(=0.33\), while the
unconstrained DC--SBM recovers \(K=5\) with NVI \(=0\).  The predictive gap is
large, with LOOIC lower by roughly \(1{,}100\) for the DC--SBM.  This is the
expected failure mode: when the data contain incompatible directional cycles,
the ordered models pool across incompatible relations, and the unconstrained
benchmark is preferred.

\section{Geometry and support surfaces of transitivity constraints}
\label{app:geometry-supports}

This section derives the areas of the geometric supports for the SST and Linear Stochastic Transitivity (LST) models in the $\psi$-space. The comparison of these supports, visualised in Figure~\ref{fig:psi-geometries} in the main text for $K=3$, provides geometric intuition for the relative flexibility of different transitivity assumptions.

For $K=3$, the $\psi$-space is three-dimensional with coordinates $(x,y,z)=(\psi_{12},\psi_{23},\psi_{13})$, which represent the upper-triangular entries of the distance log-odds matrix. The support of Weak Stochastic Transitivity (WST) is the entire positive octant $[0,B]^3$. The support of Strong Stochastic Transitivity (SST) is a pyramid defined by the monotonicity constraints $z\ge x$ and $z\ge y$. The SST and LST supports are both two-dimensional surfaces embedded in $\mathbb{R}^3$, and hence have zero volume but non-zero Hausdorff area.

\subsection{Surface area of \texorpdfstring{$\mathcal{C}_{\mathrm{Toep}}$}{C-Toep}}

The Toeplitz SST support is defined by the constraint $x=y$ within the SST pyramid, yielding the surface
\[
\mathcal{C}_{\mathrm{Toep}} = \{(x,x,z) : 0\le x\le z\le B\}.
\]

To compute the area, we parametrise this surface by $(u,v)$ with $u,v\in[0,B]$ and $u\le v$:
\[
\mathbf{r}(u,v) = (u,u,v).
\]

The tangent vectors are \(\mathbf r_u=(1,1,0)\) and
\(\mathbf r_v=(0,0,1)\), so
\(\mathbf r_u\times\mathbf r_v=(1,-1,0)\) and
\(\|\mathbf r_u\times\mathbf r_v\|=\sqrt{2}\). Hence
\begin{align*}
\mathrm{Area}(\mathcal{C}_{\mathrm{Toep}}\cap[0,B]^3)
&=
\int_0^B\int_u^B\sqrt{2}\,\mathrm d v\,\mathrm d u\\
&=
\sqrt{2}\int_0^B(B-u)\,\mathrm d u\\
&=
\frac{B^2}{\sqrt{2}}.
\end{align*}

\subsection{Surface area of \texorpdfstring{$\mathcal{C}_{\mathrm{LST}}$}{C-LST}}

The Linear Stochastic Transitivity (LST) support is defined by the constraint $z=x+y$ within the SST pyramid, with the additional constraint $x,y\ge 0$, yielding the surface
\[
\mathcal{C}_{\mathrm{LST}} = \{(x,y,x+y) : x,y\ge 0,\; x+y\le B\}.
\]

To compute the area, we parametrise this surface by $(s,t)$ with $s,t\ge 0$ and $s+t\le B$:
\[
\mathbf{r}(s,t) = (s,t,s+t).
\]

The tangent vectors are \(\mathbf r_s=(1,0,1)\) and
\(\mathbf r_t=(0,1,1)\), so
\(\mathbf r_s\times\mathbf r_t=(-1,-1,1)\) and
\(\|\mathbf r_s\times\mathbf r_t\|=\sqrt{3}\). The integration domain is
the triangle \(\{(s,t):s,t\ge0,\ s+t\le B\}\), giving
\begin{align*}
\mathrm{Area}(\mathcal{C}_{\mathrm{LST}}\cap[0,B]^3)
&=
\int_0^B\int_0^{B-s}\sqrt{3}\,\mathrm d t\,\mathrm d s\\
&=
\sqrt{3}\int_0^B(B-s)\,\mathrm d s\\
&=
\frac{\sqrt{3}\,B^2}{2}.
\end{align*}

\subsection{Comparison of surface areas}

The ratio of the LST area to the Toeplitz SST area is:
\[
\frac{\mathrm{Area}(\mathcal{C}_{\mathrm{LST}})}{\mathrm{Area}(\mathcal{C}_{\mathrm{Toep}})} = 
\frac{\sqrt{3}\, B^2 / 2}{B^2 / \sqrt{2}} = 
\frac{\sqrt{3}}{2} \cdot \sqrt{2} = 
\frac{\sqrt{6}}{2} = 
\sqrt{\frac{3}{2}} \approx 1.225.
\]

Thus, the LST surface subtends approximately $22.5\%$ more area than the Toeplitz SST surface within the cube $[0,B]^3$. This reflects the fact that LST allows for more flexible path-additivity patterns, whereas Toeplitz SST restricts to constant distance-based log-odds; geometrically, this difference in flexibility translates to a proportionally larger surface area.

\newpage

\section{Age-ordered prior: derivations and prior sensitivity}
\label{app:ocrp-derivation}

This appendix derives the age-ordered partition prior used in Section~\ref{sec:ocrp-prior} and summarises its prior sensitivity to \(\vartheta\). The concentration parameter is denoted by \(\vartheta>0\); \(K_n\) denotes the number of occupied blocks among \(n\) nodes; and an ordered composition is written as
\[
(n_1,\ldots,n_K),
\qquad n_k\ge 1,
\qquad \sum_{k=1}^K n_k=n.
\]
Ranks are read from top to bottom: rank \(1\) is the upper end of the hierarchy and rank \(K\) is the lower end.  We write
\[
m_k=\sum_{\ell=1}^k n_\ell,
\qquad k=1,\ldots,K,
\qquad m_0=0,
\]
for the prefix sums.  This is the convention used in the ordered allocation predictive probabilities \eqref{eq:ocrp-old}--\eqref{eq:ocrp-new}.  The role of this appendix is to derive those formulae directly in the age-ordered orientation used in the paper and to show how \(\vartheta\) affects the induced prior on the hierarchy.

\subsection{The age-ordered composition law}
\label{app:ocrp-derivation-ocrp}

Let \(\Pi_n=(A_1,\ldots,A_K)\) be an age-ordered partition of \([n]\), where the block \(A_k\) has size \(n_k=|A_k|\).  In the TSBM the labels are not arbitrary names: they encode the hierarchy.  Therefore, unlike in an exchangeable partition model, the ordered vector \((A_1,\ldots,A_K)\) is part of the latent state.

The age-ordered partition prior used in the main text is
\begin{equation}
\label{eq:app-ocrp-allocation}
p_{\mathrm{ord}}(A_1,\ldots,A_K)
=
\frac{\vartheta^K}{\vartheta_{(n)}}
\frac{\prod_{k=1}^K n_k!}{m_1m_2\cdots m_K},
\qquad
\vartheta_{(n)}=\frac{\Gamma(\vartheta+n)}{\Gamma(\vartheta)}.
\end{equation}
If only the ordered composition \((n_1,\ldots,n_K)\) is retained, and the identities of the nodes inside the blocks are ignored, then there are \(n!/\prod_{k=1}^K n_k!\) ordered partitions with those block sizes.  Hence the induced law on ordered compositions is
\begin{equation}
\label{eq:app-ocrp-composition}
p_{\mathrm{ord}}(n_1,\ldots,n_K)
=
\frac{\vartheta^K}{\vartheta_{(n)}}
\frac{n!}{m_1m_2\cdots m_K}.
\end{equation}

The prefix sums in~\eqref{eq:app-ocrp-allocation} are the defining feature of the age-ordered prior.  Because \(m_k\) counts the cumulative mass already assigned to the first \(k\) ranks, large upper blocks are penalised more strongly than large lower blocks.  This is exactly the pyramidal tilt used in the main paper: upper ranks are encouraged to be relatively sparse a priori, while lower ranks may accumulate more nodes.

We now derive the predictive probabilities.  Suppose an ordered composition \((n_1,\ldots,n_K)\) of \(n\) nodes is given, with allocation vector \(\mathbf z=(z_1,\ldots,z_n)\).  If node \(n+1\) is added to an existing rank \(k\), the new composition is
\[
(n_1,\ldots,n_{k-1},n_k+1,n_{k+1},\ldots,n_K),
\]
and the prefix sums \(m_\ell\) change only for \(\ell\ge k\).  Taking the ratio of~\eqref{eq:app-ocrp-allocation} after and before the addition gives
\begin{align}
\Pr(z_{n+1}=k^{\text old}\mid \mathbf z)
&=
\frac{n_k+1}{\vartheta+n}
\prod_{\ell=k}^{K}\frac{m_\ell}{m_\ell+1},
\qquad k=1,\ldots,K.
\label{eq:app-ocrp-old}
\end{align}
If instead node \(n+1\) creates a new singleton block inserted at slot \(r\in\{1,\ldots,K+1\}\), the new composition is
\[
(n_1,\ldots,n_{r-1},1,n_r,\ldots,n_K),
\]
with the convention that insertion at \(K+1\) appends the singleton at the bottom.  The same ratio calculation gives
\begin{align}
\Pr(z_{n+1}=r^{\mathrm{new}}\mid \mathbf z)
&=
\frac{\vartheta}{\vartheta+n}
\frac{1}{m_{r-1}+1}
\prod_{\ell=r}^{K}\frac{m_\ell}{m_\ell+1},
\qquad r=1,\ldots,K+1,
\label{eq:app-ocrp-new}
\end{align}
where \(m_0=0\) and an empty product is equal to one.  Equations~\eqref{eq:app-ocrp-old}--\eqref{eq:app-ocrp-new} are exactly the predictive probabilities stated in the main text in~(9)--(10).

For the single-site Gibbs update used in the sampler, the same formulae are applied after temorarily removing node $i$ from the network.  After removing node \(i\), let \(K_{-i}=K(z_{-i})\), let \(n_k^-=n_k(z_{-i})\), and let \(m_k^-=\sum_{\ell\le k} n_\ell^-\), so that \(m_{K_{-i}}^-=n-1\).  Then
\begin{align}
\Pr(z_i=k\mid z_{-i},\vartheta)
&=
\frac{n_k^-+1}{\vartheta+n-1}
\prod_{\ell=k}^{K_{-i}}\frac{m_\ell^-}{m_\ell^-+1},
\qquad k=1,\ldots,K_{-i},
\label{eq:app-ocrp-loo-old}\\
\Pr(z_i=r^{\mathrm{new}}\mid z_{-i},\vartheta)
&=
\frac{\vartheta}{\vartheta+n-1}
\frac{1}{m_{r-1}^-+1}
\prod_{\ell=r}^{K_{-i}}\frac{m_\ell^-}{m_\ell^-+1},
\qquad r=1,\ldots,K_{-i}+1.
\label{eq:app-ocrp-loo-new}
\end{align}
These are exactly the versions of~\eqref{eq:app-ocrp-old}--\eqref{eq:app-ocrp-new}, obtained by temporarily excluding node $i$ from the network, and replacing \(n\) with \(n-1\), \(K\) with \(K_{-i}\), \(n_k\) with \(n_k^-\), and \(m_k\) with \(m_k^-\).  The common denominator \(\vartheta+n-1\) may be dropped when these probabilities are used as unnormalised allocation weights.

\subsection{Regenerative representation}
\label{app:ocrp-derivation-regenerative}

The ordered composition law in~\eqref{eq:app-ocrp-composition} is a finite-dimensional member of the regenerative composition structures of~\citet{Gne:Pit:05aop}.  For the present paper, the predictive probabilities above are the quantities required for computation, but the regenerative representation is useful because it explains why the formula is consistent as \(n\) varies.

In the age-ordered orientation, the composition law factorises directly in prefix-sum form as
\[
p_{\mathrm{ord}}(n_1,\ldots,n_K)
=
\prod_{j=1}^K q(m_j:n_j),
\qquad m_j=\sum_{\ell=1}^j n_\ell,
\]
where the decrement matrix is
\begin{equation}
\label{eq:app-ocrp-decrement}
q(n:m)
=
\vartheta\,
\frac{\Gamma(n)}{\Gamma(n-m+1)}
\frac{\Gamma(\vartheta+n-m)}{\Gamma(\vartheta+n)},
\qquad 1\le m\le n.
\end{equation}
Indeed, using \(m_j-n_j=m_{j-1}\) with \(m_0=0\), one obtains
\[
\prod_{j=1}^K q(m_j:n_j)
=
\frac{\vartheta^K}{\vartheta_{(n)}}
\frac{n!}{m_1\cdots m_K}.
\]
This is the sense in which the prior is regenerative: after the lowest retained block is removed, the remaining upper composition again has the same form, with a smaller sample size.  In the subordinator construction of regenerative compositions, one obtains such laws by sampling iid uniform points and counting them in the interval components cut out by a multiplicatively regenerative random closed set.  We do not use this representation algorithmically, but it gives a principled route to richer ordered partition priors, such as two-parameter Ewens--Pitman analogues, should a more flexible prior on \(K_n\) or on the size profile be required.

\subsection{Prior sensitivity to \texorpdfstring{\(\vartheta\)}{vartheta}}
\label{app:ocrp-derivation-sensitivity}

The likelihood is not invariant to permutations of the block labels, because the labels encode hierarchical positions.  The age-ordered prior is therefore a prior on ordered partitions, not merely on unordered clusters.  Under the prefix-sum law in~\eqref{eq:app-ocrp-composition}, upper ranks receive stronger cumulative-size penalties than lower ranks.  This induces, a priori, a pyramidal size profile: rank \(1\) is comparatively sparse, while lower ranks are allowed to contain more nodes.

For later use, it is helpful to record the ordered-composition log-ratio.  Let \((n_1,\ldots,n_K)\) and \((n'_1,\ldots,n'_{K'})\) be two ordered compositions of the same sample size \(n\), with prefix sums \(m_j\) and \(m'_j\).  From~\eqref{eq:app-ocrp-composition},
\begin{equation}
\label{eq:app-ocrp-log-ratio-fixed-n}
\log\frac{p_{\mathrm{ord}}(n'_1,\ldots,n'_{K'})}{p_{\mathrm{ord}}(n_1,\ldots,n_K)}
=
(K'-K)\log\vartheta
-
\sum_{j=1}^{K'}\log m'_j
+
\sum_{j=1}^{K}\log m_j.
\end{equation}
The factors \(n!\) and \(\vartheta_{(n)}\) cancel because the sample size is fixed.  This is the prior ratio used by any Metropolis--Hastings move that changes the ordered composition but not the number of nodes, for example split--merge or adjacent-block proposals.  In the single-site Gibbs update, where composition of size \(n-1\) is compared with candidate compositions of size \(n\), the corresponding ratios are the predictive probabilities in~\eqref{eq:app-ocrp-loo-old}--\eqref{eq:app-ocrp-loo-new}.

The induced prior on the number of occupied blocks coincides with that of an ordinary CRP with concentration parameter \(\alpha=\vartheta\). In particular,
\begin{equation}
\label{eq:app-ocrp-prior-K}
\mathbb E[K_n]
=
\sum_{i=1}^n \frac{\vartheta}{\vartheta+i-1}
=
\vartheta\{\operatorname{digamma}(\vartheta+n)-\operatorname{digamma}(\vartheta)\}.
\end{equation}
Thus \(\vartheta\) controls the prior number of occupied ranks in exactly the same way as the CRP parameter \(\alpha\), while the age-ordering controls how the occupied block sizes are arranged along the hierarchy.

\begin{figure}[t]
\centering
\includegraphics[width=0.92\textwidth]{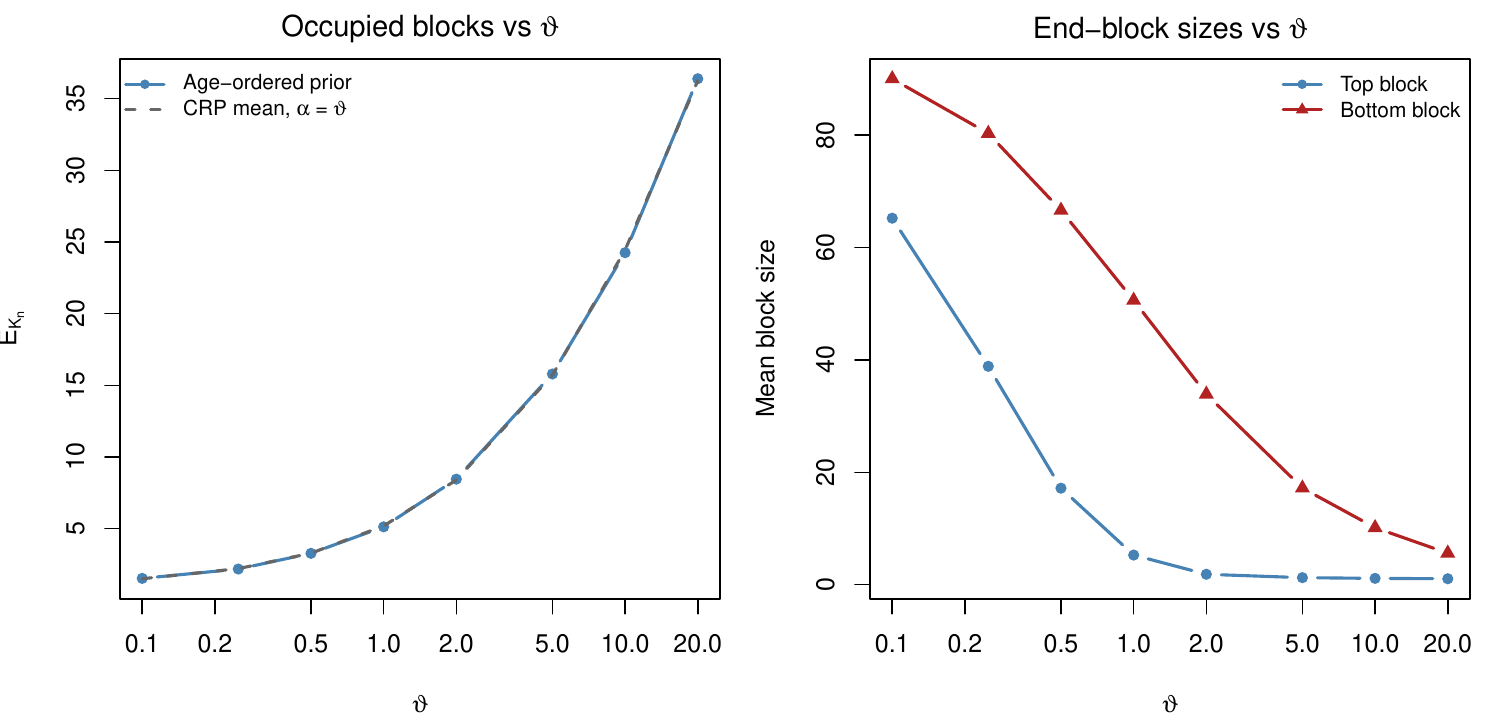}
\caption{Prior sensitivity of the age-ordered prior at \(n=100\). Left: Monte Carlo mean of \(K_n\) across \(500\) prior draws, together with the ordinary-CRP benchmark obtained by setting \(\alpha=\vartheta\). Right: Monte Carlo means of the top-block and bottom-block sizes under the same prior draws.}
\label{fig:app-age-ordered-sensitivity}
\end{figure}

Figure~\ref{fig:app-age-ordered-sensitivity} summarises the prior sensitivity to \(\vartheta\) at \(n=100\), based on \(500\) prior draws for each value of \(\vartheta\).

The left panel shows that the age-ordered prior leaves the ordinary-CRP law of \(K_n\) unchanged: the Monte Carlo averages track the benchmark~\eqref{eq:app-ocrp-prior-K} almost exactly once \(\alpha=\vartheta\).  The right panel shows the aspect of the prior that matters for the hierarchy itself.  Small \(\vartheta\) values produce few blocks and a steep pyramid, with a small top rank and a much larger bottom rank.  At the default \(\vartheta=0.5\), the prior mean is \(E[K_{100}]\approx 3.3\), the expected top-block size is about \(17\), and the expected bottom-block size is about \(67\).  As \(\vartheta\) increases, the prior spreads mass over more occupied ranks and both end blocks shrink, but the bottom block remains substantially larger than the top block throughout the range shown.  This is the intended behaviour: \(\vartheta\) tunes overall complexity, while the age-ordering preserves a clear pyramidal tilt.

\section{Additional application details}
\label{app:application-details}

\begin{table}[htbp]
\centering
\small
\setlength{\tabcolsep}{5pt}
\begin{tabular}{@{}llrcrrr@{}}
\toprule
Dataset & Point estimate & $K$ & Any cycle & Cyclic triples & Share & Tied pairs \\
\midrule
\multirow{4}{*}{Bighorn sheep} & Agony & $13$ & No & $0/286$ & $0.0\%$ & $15$ \\
 & WST--TSBM & $4$ & No & $0/4$ & $0.0\%$ & $0$ \\
 & SST--TSBM & $5$ & No & $0/10$ & $0.0\%$ & $2$ \\
 & DC--SBM & $5$ & Yes & $1/10$ & $10.0\%$ & $0$ \\
\addlinespace
\multirow{4}{*}{Spotted hyenas} & Agony & $7$ & No & $0/35$ & $0.0\%$ & $0$ \\
 & WST--TSBM & $14$ & Yes & $2/364$ & $0.5\%$ & $32$ \\
 & SST--TSBM & $7$ & No & $0/35$ & $0.0\%$ & $7$ \\
 & DC--SBM & $6$ & Yes & $3/20$ & $15.0\%$ & $0$ \\
\addlinespace
\multirow{4}{*}{Mountain goats} & Agony & $26$ & Yes & $1/2{,}600$ & $<0.1\%$ & $81$ \\
 & WST--TSBM & $4$ & No & $0/4$ & $0.0\%$ & $0$ \\
 & SST--TSBM & $6$ & No & $0/20$ & $0.0\%$ & $0$ \\
 & DC--SBM & $4$ & No & $0/4$ & $0.0\%$ & $0$ \\
\addlinespace
\multirow{4}{*}{Stat.\ journals} & Agony & $4$ & No & $0/4$ & $0.0\%$ & $0$ \\
 & WST--TSBM & $13$ & Yes & $16/286$ & $5.6\%$ & $7$ \\
 & SST--TSBM & $5$ & No & $0/10$ & $0.0\%$ & $0$ \\
 & DC--SBM & $11$ & Yes & $5/165$ & $3.0\%$ & $2$ \\
\addlinespace
\multirow{4}{*}{Macaques} & Agony & $33$ & No & $0/5{,}456$ & $0.0\%$ & $65$ \\
 & WST--TSBM & $9$ & No & $0/84$ & $0.0\%$ & $0$ \\
 & SST--TSBM & $10$ & No & $0/120$ & $0.0\%$ & $0$ \\
 & DC--SBM & $11$ & No & $0/165$ & $0.0\%$ & $0$ \\
\addlinespace
\multirow{4}{*}{High school} & Agony & $8$ & Yes & $4/56$ & $7.1\%$ & $3$ \\
 & WST--TSBM & $9$ & Yes & $4/84$ & $4.8\%$ & $15$ \\
 & SST--TSBM & $12$ & Yes & $2/220$ & $0.9\%$ & $37$ \\
 & DC--SBM & $8$ & Yes & $3/56$ & $5.4\%$ & $10$ \\
\bottomrule
\end{tabular}
\caption{Cycle diagnostics for application point estimates. For each partition we aggregate observed directed mass between blocks and orient each unordered block pair toward the larger flow; exact ties are left unoriented. ``Any cycle'' reports whether this majority-flow block graph is non-acyclic, and cyclic triples counts directed 3-cycles among block triples. The share is the unweighted proportion of the \(\binom{K}{3}\) block triples that are cyclic. The WST, SST and DC--SBM point estimates use \texttt{minVI} with \texttt{method=draws}; DC--SBM draws are strength-reordered before the point estimate is formed.}
\label{tab:application-cycle-diagnostics}
\end{table}
Table~\ref{tab:application-cycle-diagnostics} checks whether the block-level majority-flow graph induced by each point estimate contains directed cycles. For a fixed partition \(\hat z\), let
\[
  C_{ab}(\hat z)=\sum_{i:\hat z_i=a}\sum_{j:\hat z_j=b} A_{ij},
  \qquad a,b\in\{1,\ldots,K\},
\]
be the observed edge mass sent from block \(a\) to block \(b\). For every unordered block pair \(a<b\), define the empirical forward share
\[
  \tilde\rho_{ab}=
  \frac{C_{ab}}{C_{ab}+C_{ba}},
  \qquad\text{when } C_{ab}+C_{ba}>0.
\]
The diagnostic constructs a directed majority graph \(G_{\mathrm{maj}}(\hat z)\) on the \(K\) blocks by adding \(a\to b\) if \(C_{ab}>C_{ba}\), and \(b\to a\) if \(C_{ba}>C_{ab}\). Exact ties, including zero-traffic pairs, are left unoriented. A block triple \(\{a,b,c\}\) is counted as cyclic when the induced three-node majority graph contains \(a\to b\to c\to a\), or the reverse orientation. The percentage reported in Table~\ref{tab:application-cycle-diagnostics} is therefore
\[
  100\,
  \binom{K}{3}^{-1}
  \sum_{1\le a<b<c\le K}
  \ind\!\left\{
    G_{\mathrm{maj}}(\hat z)[\{a,b,c\}]
    \text{ is a directed 3-cycle}
  \right\}.
\]
This is an unweighted block-triple percentage; it is not weighted by the edge mass in the triple. The WST, SST and DC--SBM point estimates are computed with the draw-restricted \texttt{minVI} summary (\texttt{method=draws}); the DC--SBM draws are strength-reordered before the summary is taken.

The empirical block-flow matrices in the main text display the same ingredients, but they answer a slightly different visual question. In a chosen order, an upper-triangular cell \(a<b\) looks forward when \(\tilde\rho_{ab}>1/2\), backward when \(\tilde\rho_{ab}<1/2\), and ambiguous when the pair is tied or has little mass. If every upper-triangular pair is forward-majority, the majority graph is acyclic in that order. The converse is more delicate: a matrix can look visually ordered because most high-mass cells are blue, while a small number of low-volume backward-majority cells still close directed 3-cycles.


\end{document}